\documentclass{elsart}
\usepackage{float,natbib,epsfig}
\usepackage{rotating}
\newcommand{\Cerenkov}{\v{C}erenkov}

\begin{document}

\begin{frontmatter}
\title{Extending the Sensitivity of 
Air {\Cerenkov} Telescopes}
\author{I. de la Calle P\'erez}
\author{ and }
\author{S. D. Biller}

\address{University of Oxford, Department of Physics, Denys Wilkinson
 Building, Oxford, OX1 3RH, UK}

\begin{abstract}

Over the last decade, the Imaging Air {\Cerenkov} technique has proven itself
to be an extremely powerful means to study very energetic gamma-radiation from
a number of astrophysical sources in a regime which is not practically
accessible to satellite-based instruments. The further development of this
approach in recent years has generally concentrated on increasing the density
of camera pixels, increasing the mirror area and using multiple telescopes.
Here we present a practical method to substantially improve the sensitivity of
Atmospheric {\Cerenkov} Telescopes using wide-field cameras with a relatively
course density of photomultiplier tubes. The 2-telescope design considered
here is predicted to be more than $\sim$3 times more sensitive than
existing/planned arrays in the regime above 300~GeV for continuously emitting
sources; up to $\sim$10 times more sensitive for hour-scale emission (relevant
for episodic sources, such as AGN); significantly more sensitive in the regime
above 10~TeV; and possessing a sky coverage which is roughly an order of
magnitude larger than existing instruments. It should be possible to extend
this approach for even further improvement in sensitivity and sky coverage.

\end{abstract}

\end{frontmatter}

\newpage

\section{Introduction}

\par The Imaging Atmospheric {\Cerenkov} Telescope (ACT) has proven to be a
highly successful technique in the detection of high energy, gamma-rays from
astrophysical sources above $\sim$250~GeV. The catalogue of sources so far
detected by this ground-based technique includes mainly supernova remnants
(SNR) (see e.g. \cite{Weekes1989}, \cite{Aharonian2001},
\cite{Aharonian2005c}) and active galaxies (AGN) (see e.g. \cite{Horan2004}
and references therein), although recently, new classes of objects such as OB
star associations \cite{Aharonian2002} and binary systems
\cite{Aharonian2005a}, have been confirmed to be sources of high energy
gamma-rays. Also, a recent survey of the galactic plane has yielded a
significant signal from the region of the galactic center
\cite{Aharonian2004a} and the discovery of a series of objects lying along the
galactic plane \cite{Aharonian2005b}. The nature of the physics addressed via
high energy gamma-ray measurements includes clues to the understanding of the
origin of cosmic rays, the nature of AGN jets and the density if the
extragalactic infrared background. In addition, several fundamental topics of
particle physics can be probed, including searches for TeV-scale WIMP
annihilations (see \cite{Buckley1998} and references therein), limits on radiative
neutrino decay \cite{Biller1998} and a fundamental test of Lorentz invariance
based on studies of the propagation of very high energy radiation over
extragalactic distances \cite{Biller1999b}. The sensitivities of such searches
are among the very best achieved by any technique.

\par Almost all current ground-based gamma-ray efforts have concentrated
primarily on improving the sensitivities to lower energies so as to overlap
more with satellite-based observations and potentially map out a larger range
of sources. The approaches to this have generally involved increasing the
density of camera pixels, increasing the mirror area and using multiple
telescopes.  However, it is of interest to note that many of the physics
topics mentioned above actually stand to benefit most from improved
measurements at higher energies. It is in this regime where AGN spectra and
short timescale flux variability most critically test acceleration mechanisms
as well as uniquely probe the extragalactic infrared background
radiation. These higher energy emissions permit the direct study of some of
the most extreme and poorly understood environments in the universe. There are
now several sources known to produce gamma-ray emission in the regime above
10~TeV, with evidence of some producing gamma-ray emission up to 40~TeV (see
e.g. \cite{Aharonian2005c}), and potentially as high as 100~TeV
\cite{Aharonian2004b}. Establishing the actual range of this emission would,
in itself, be of substantial interest. Furthermore, it is generally
acknowledged that much greater sky coverage is desired to study extended sources 
\cite{Aharonian2005c}, to better explore the gamma-ray sky and to 
more continuously monitor multiple sources to study transient behaviour. The 
approaches so far taken have led to only a modest improvement in this regard, 
causing many to suggest the need for either a `brute force' use of many tens
of telescopes or some ``new technology'' to allow larger sky coverage at high
sensitivity ({\em e.g.}~\cite{ParisConference2005},~\cite{Maccarone}).

\par Our purpose here is to explore a possible method for greatly increasing
the sensitivity of ACT instruments using an approach which is practical,
economical and does not rely on the development of any new technology
\cite{Biller2000}. This approach involves the use of cameras with a relatively
large field of view, which has the additional advantage of acting as a test-bed 
for the development of ``all-sky'' instruments. While initially motivated to improve 
the sensitivity at higher energies (as in, {\em e.g.}, \cite{Rowell}), the approach 
is seen to lead to improvements for energies at least as low as 300~GeV. In this
present paper, we will develop the basic concept and characteristics of this technique using
simulations applied to the case of both a single ``Wide Field Of View'' telescope 
(WFOV-1) and a tandem telescope design (WFOV-2). In a later paper, we will 
extend this approach to a small array and even larger fields of view to explore 
more details of the range and sensitivity of this approach.

\section{The {\Cerenkov} Lateral Distribution of Air Showers and 
``Wide-field'' Imaging}

\par
\begin{figure}[hbt]
  \begin{center}
    \epsfig{file=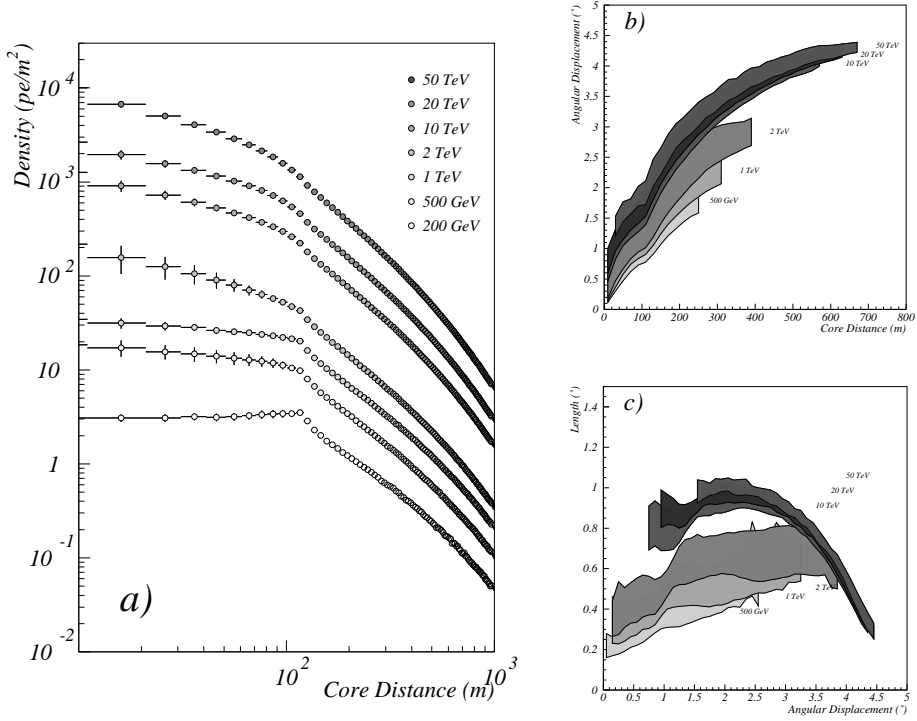,width=.9\textwidth}
    \caption{{\it a)} Simulated average {\Cerenkov} lateral distribution for
      showers initiated by gamma-rays of various energies, convolved with
      atmospheric extinction, mirror reflectivity and typical photomultiplier
      tube efficiency. All simulated showers were generated from a zenith
      angle of 0$^\circ$ with an assumed observation level for detection of
      2400~m above sea level. {\it b)} Image angular displacement
      ($\mathcal{D}$) as a function of core distance and {\it c)} image length
      ($\mathcal{L}$) as a function of angular displacement for 0.5, 1, 2, 10,
      20 and 50~TeV gamma-ray showers. The shaded bands include 68$\%$ of the
      events about the median. The observed turn-over of the image length
      distribution for higher energies and larger displacements is due to
      image truncation by the camera edge (with a radius of
      $\sim$5$^\circ$).}\label{lateral}
  \end{center}
\end{figure}

 Extensive air showers induced by primary gamma-rays in the regime near
1~TeV are known to produce a lateral distribution of {\Cerenkov} light at sea
level with a relatively constant photon density for distances within
$\sim$130~m from the shower core \cite{Hillas1996}. By a
distance of $\sim$200~m from the core, the density falls by a factor of 2-3 from
this value. Consequently, ACTs have generally concentrated on measuring
showers landing within the ``inner light pool,'' where the $\sim$constant
photon intensity is at its maximum and its relation to the primary gamma-ray
energy is not particularly dependent on a precise knowledge of the core
location. This naturally confines the effective areas of individual telescopes
to the order of a few times $10^{4}$~m$^2$. However, this form of the lateral
distribution is also known to change as a function of both detection altitude
and primary energy. As an example, figure \ref{lateral}a shows that, for an
altitude of 2400~m above sea level and for primary energies of $\sim$2~TeV or
more, there is no longer any distinctive plateau at smaller core distances and
the density falls in a nearly continuous $\sim$exponential fashion. Therefore,
the choice of a fiducial radius for analysis becomes somewhat arbitrary.

\par If one were to instead base this fiducial radius purely on the observable
light level, figure \ref{lateral}a also indicates that the ability of current
ACTs to easily trigger for energies of less than 1~TeV would imply
that a 10~TeV gamma-ray could be seen beyond a distance of $\sim$500~m. This
would correspond to an effective area more than an order of magnitude larger
than existing ACTs, and it is this attractive possibility which motivates this
present work. Owing to simple geometry, images from showers at larger core
distances tend to have an average angle in the camera which is more displaced
from the source position. As shown in figure \ref{lateral}b, an image
corresponding to a core distance of 500~m would be displaced by approximately
4~degrees.  For point-sources tracked in the center of the field of view, this
would require a camera with an angular radius of $\sim$5~degrees to contain
such images. This raises practical concerns regarding the need to balance this
with a large enough density of photomultiplier tubes (PMTs) to adequately
define shower image parameters as well as the potential impact of mirror
aberrations on such images. On the other hand, as indicated by figure
\ref{lateral}c, shower images become elongated by approximately a factor of two
at these larger core distances, which should loosen the constraints related to
these issues to some extent. Another concern is the need to not only maintain
the background rejection capability for large image displacements, but to
actually improve it, since lower intensity images will now have to ``compete''
with a much larger flux of low energy cosmic-ray background showers (mostly
due to protons) which bombard the earth from all angles.  Finally, there is
the question of how well the primary gamma-ray energies corresponding to
images with large displacements can be reconstructed.

\section{Simulated Configuration}

\par In order to address the issues previously raised, computer simulations
were used to investigate the ``wide-angle'' concept in the context of a
practical instrument. The simulation package used \cite{Biller1997} was developed
based on the EGS4 \cite{SLAC} and SHOWERSIM \cite{Wrotniak} codes. The
simulated shower development was checked against various analytic treatments
and image characteristics were compared both with results from other
simulation packages and with actual data from the Whipple imaging ACT
\cite{Calle2002}. For the purposes of this paper, a baseline telescope
configuration was chosen which was similar to that of the Whipple ACT 
({\em e.g.} \cite{Finley2000}), which has a well-documented history of performance
characteristics for a variety of different camera designs and, thus, allows
for easy comparison with this device. Accordingly, the model involved a 10~m
diameter Davies-Cotton mirror design at an altitude of 2400~m above sea level
and wavelength-dependent mirror reflectivity and PMT quantum efficiencies
similar to those measured for the Whipple 10~m ACT. However, a slightly longer
$f=1$ focal length was used (compared with $f=0.7$), which is more in line
with current telescope designs and reduces some of the optical aberrations at
large image displacements.  Each PMT was modelled with a light concentrator on
the front face, possessing a wavelength-independent reflectivity of 80\%. The
signal recorded in each channel of electronics per event was simply assumed to
consist of the earliest detected photon arrival time and the summed number of
photoelectrons (pe) for the corresponding PMT.

\newpage

\par
\begin{figure}[hbt]
  \begin{center}
    \epsfig{file=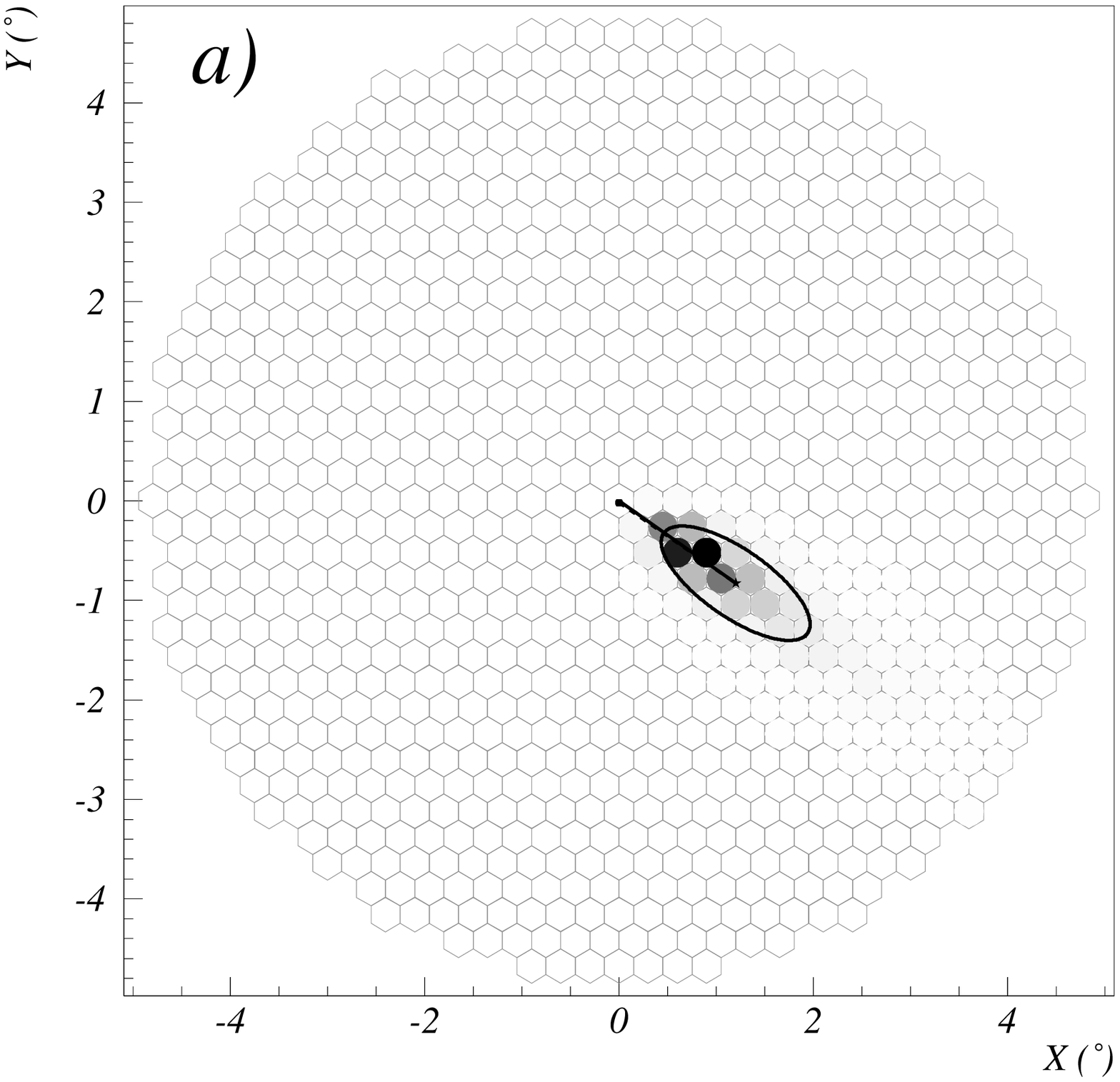,width=.45\textwidth}
    \epsfig{file=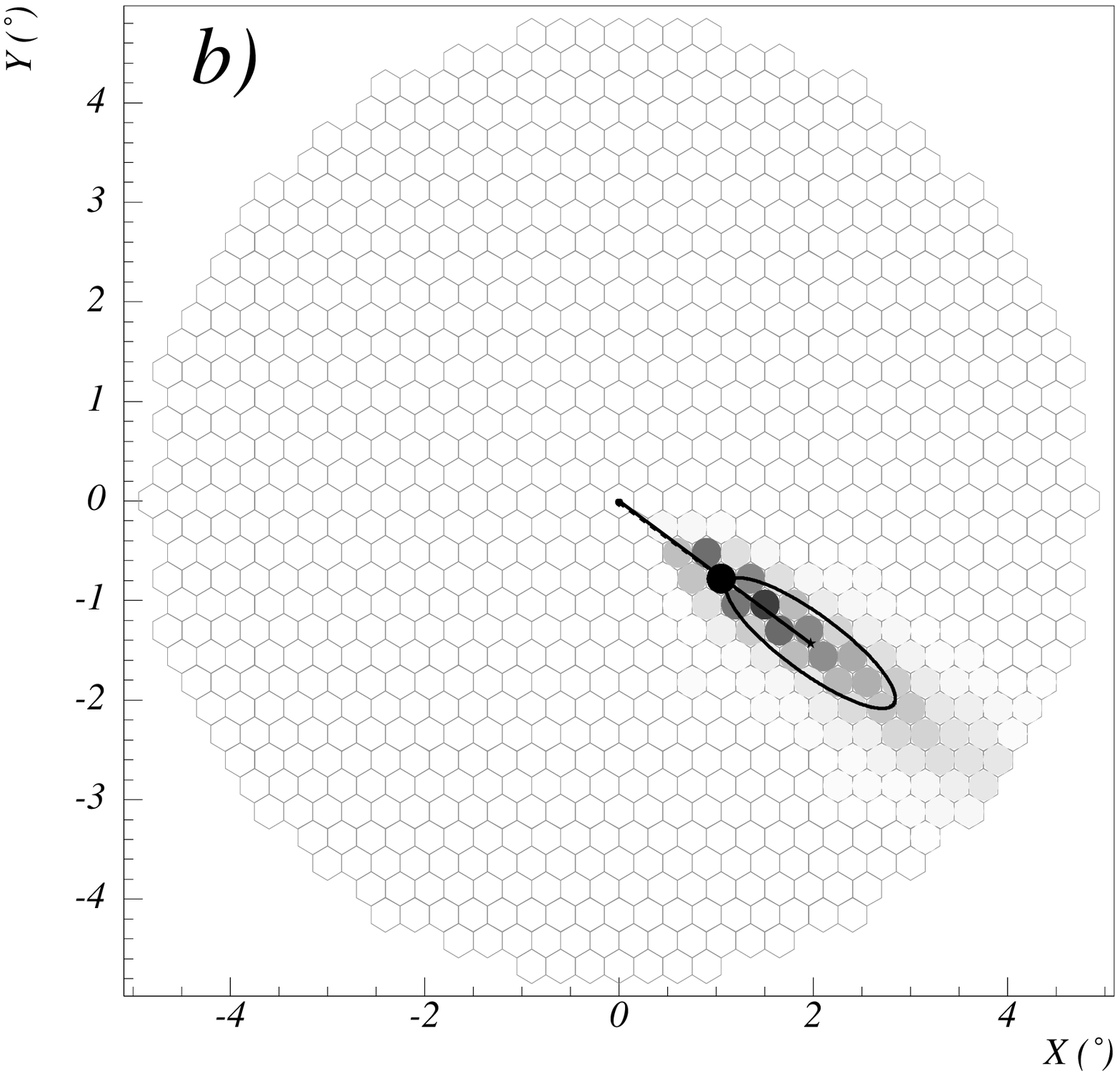,width=.45\textwidth}\\
    \epsfig{file=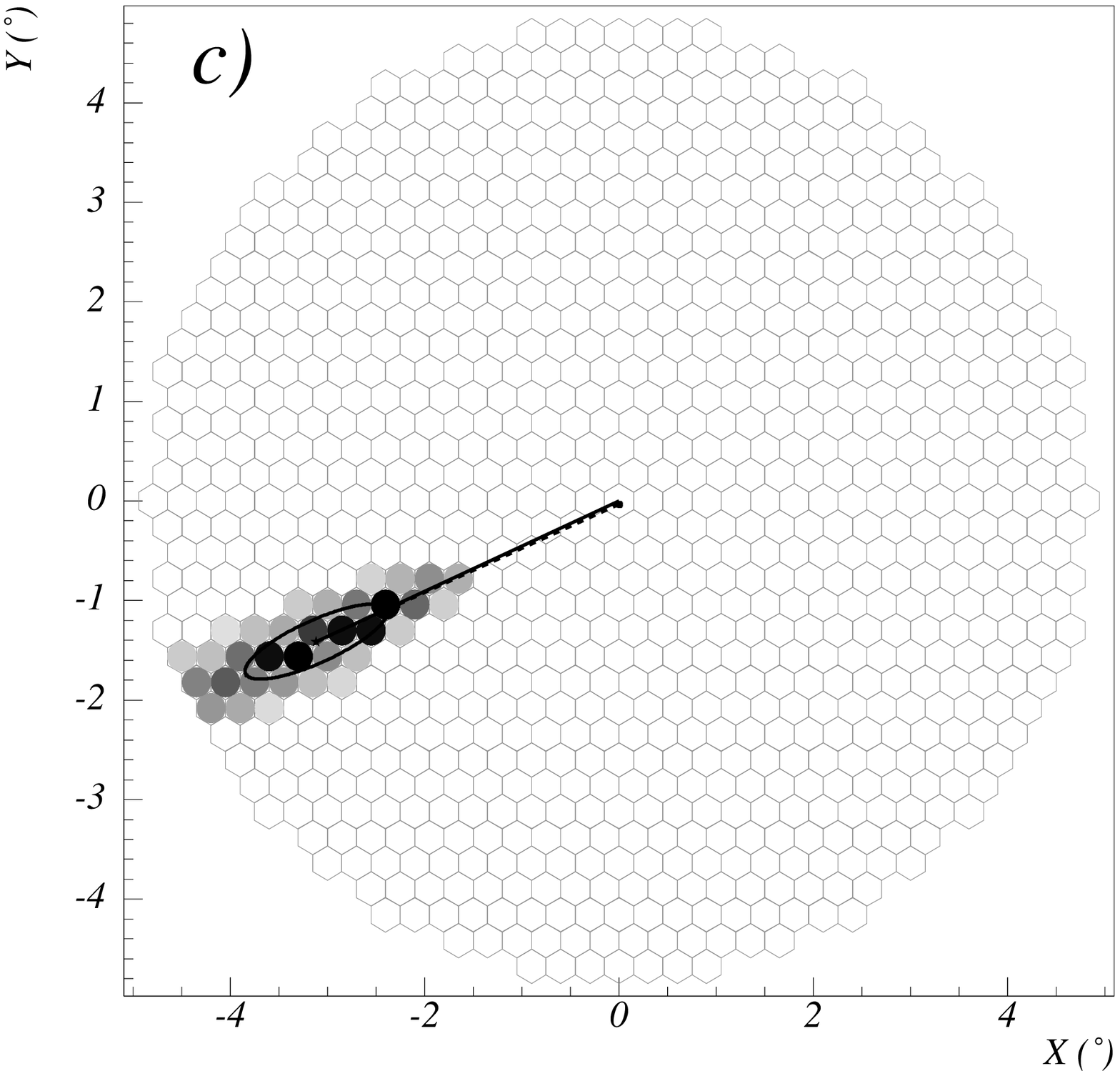,width=.45\textwidth}
    \epsfig{file=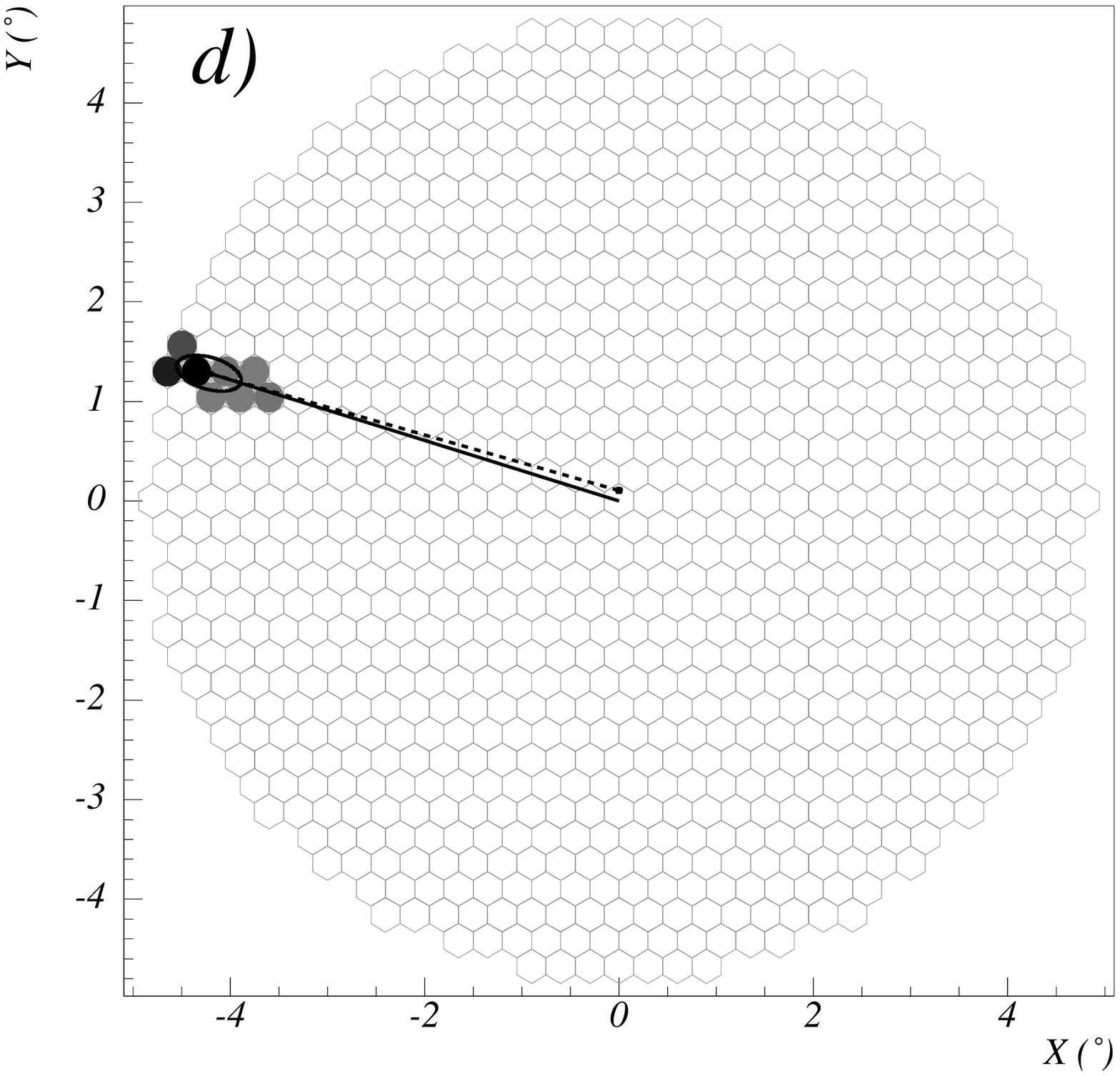,width=.45\textwidth}\\
    \caption{Representative images of simulated 20~TeV gamma-ray showers with
      increasing distance parameter, going from {\it a)}, 0.61$^\circ$ through
      {\it b)}, 2.44$^\circ$ {\it c)}, $3.42^\circ$ and {\it d)},
      4.41$^\circ$. Different shaded pixels indicate light intensity.}\label{images_distance}
  \end{center}
\end{figure}

 Based on the previous discussion, the nominal camera design simulated was
taken to have an angular radius of 5~degrees, or a full field of view angle of
10~degrees.  This is roughly twice the angular diameter of the cameras used
for the HESS ACTs \cite{Bernlohr2003} and nearly three times those of the
VERITAS project \cite{Weekes2002}. The cameras for each of these other
instruments are composed of an array of PMTs which, individually, have a full
field of view angle of $\sim$0.15~degrees. However, as indicated earlier,
gamma-ray images corresponding to shower cores which land a few hundred meters
away from the ACT are elongated by roughly a factor of two relative to those
landing within 100~m (see figure \ref{lateral}c). Hence, it is reasonable to
suspect that PMTs with twice the angular acceptance ({\em i.e.} 0.3~degrees)
might be appropriate for the approach explored here. This would also keep the
number of PMTs to less than 1000 per ACT, which is certainly in line with
current ACT camera design and couches this study in very practical
terms. Consequently, for the purposes of this paper, the nominal simulated
camera design consisted of 935 PMTs, each with an angular acceptance of
0.3~degrees, as illustrated in figure \ref{images_distance}. Subsequent
studies confirmed this PMT density to be sufficient and even suggested it
could be somewhat reduced without substantial impact (section
\ref{Dep_Pixel_Size}). We have chosen to model a camera with uniform PMT
density (as opposed to one with graded spacing) as it both simplifies image
analysis and provides a more uniform sensitivity for sources across the field
of view of the instrument. The minimum event trigger conditions required at
least any two PMTs in the camera to record a signal in excess of 20~pe above
fluctuations in the night sky background, the latter of which was scaled from
the recorded data of the Whipple 10~m camera and corresponded to a Gaussian
with an RMS of 3~pe for a PMT with a full field of view of 0.25$^\circ$
(15$\mu sr$).

\section{Sampling}

\par
\begin{figure}[hbt]
  \begin{center}
    \epsfig{file=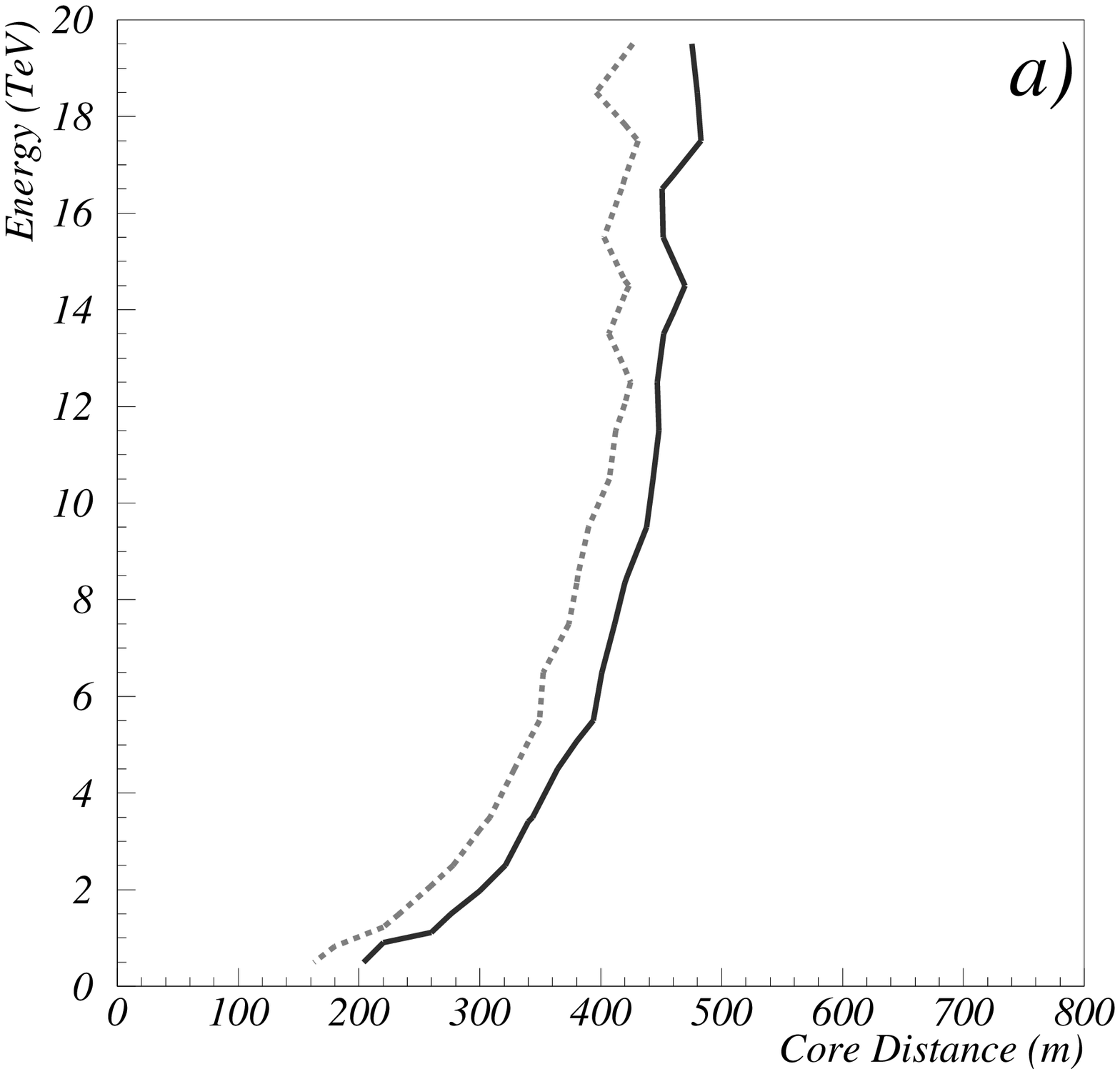,width=.33\textwidth}
    \epsfig{file=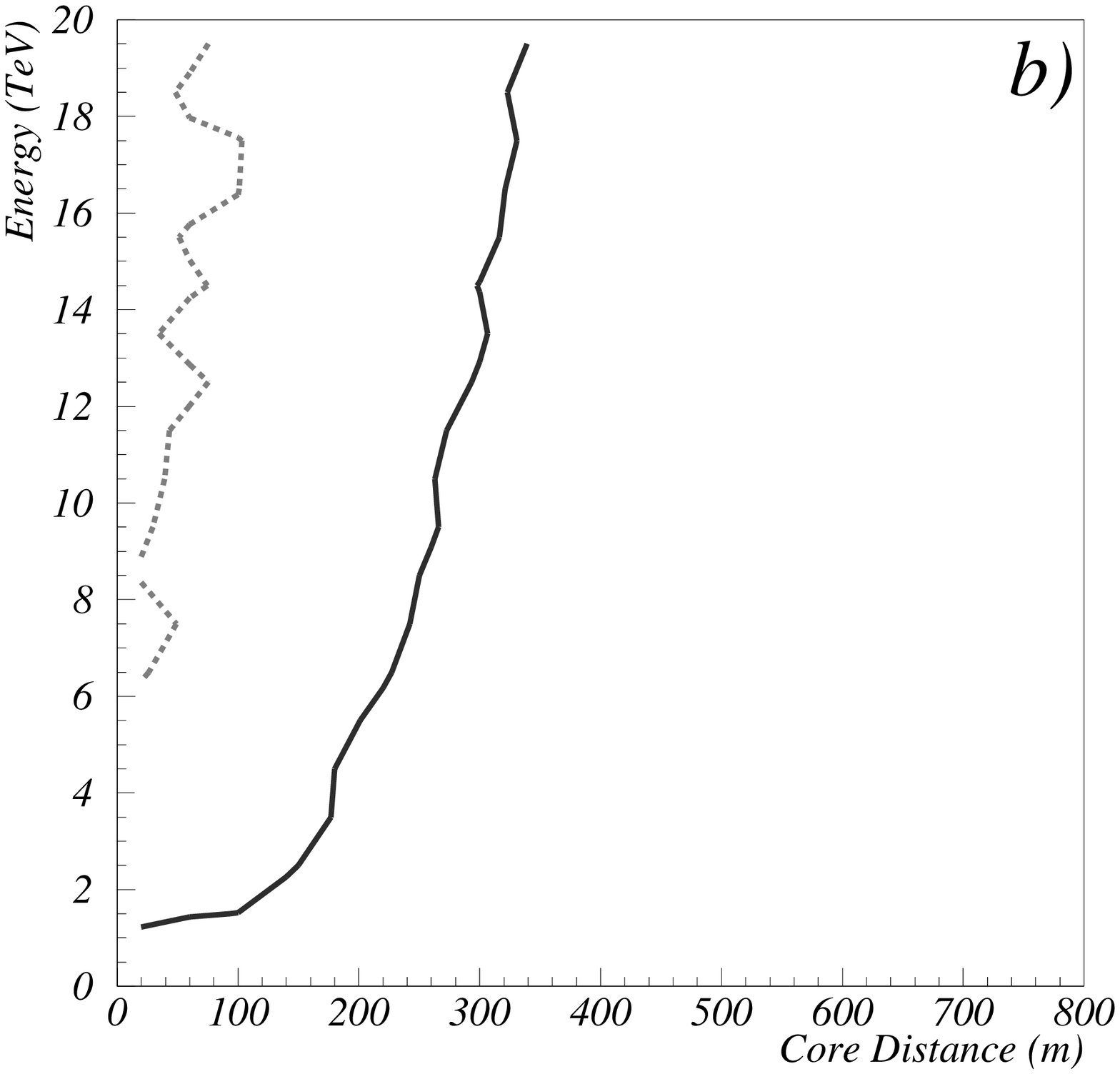,width=.33\textwidth}
    \epsfig{file=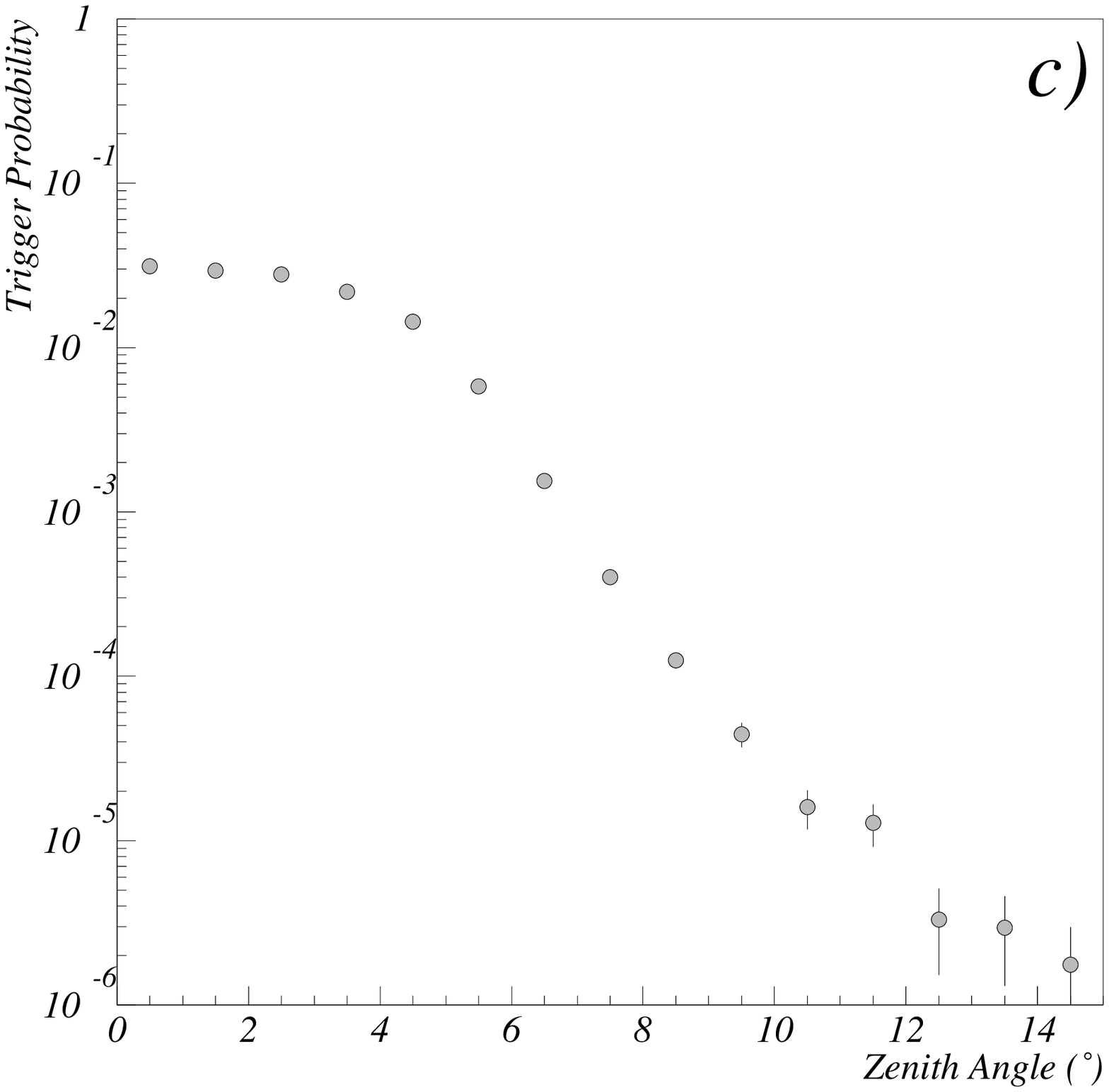,width=.33\textwidth}\\
    \caption{Trigger probability as a function of energy and core distance for
      simulated gamma-ray events {\it a)} and background events {\it b)}, where
      the solid line and dashed line indicate the 10$\%$ and 50$\%$ trigger
      probability contours, respectively, in the case of gamma events, and
      10$\%$ and 20$\%$ in the case of background events. {\it c)} Trigger
      probability for proton showers as a function of zenith angle.}\label{trigger_prob}
  \end{center}
\end{figure}

 In this work, we have concentrated mainly on assessing the performance of
the hypothetical instrument to point-sources of gamma-rays and have assumed
such a source to be tracked in the center of the field of view of the
camera. Table \ref{tab:single} summarises the primary simulation data set used
in this work.  For simplicity, gamma-ray events have been generated from the
zenith, with shower cores randomly sampled over an area on ground within 800~m
from the detector. Further increase to the effective area for gamma-rays on
the order of a factor of $\sim$3 (though this is energy-dependent) can be
achieved when viewing sources at larger zenith angles \cite{Petry2001}, but
the studies here have been limited to what might be achieved relative to the
minimum shower `footprint' from overhead sources. The largest core distance
used was chosen to ensure that the probability of triggering the ACT at the
highest gamma-ray energies falls to $\sim$zero, so that effective areas can be
accurately assessed (figure \ref{trigger_prob}a).

\par Gamma-rays with energies between 300~GeV and 20~TeV were generated based
on a power-law with a differential spectral index of 2.5, which is indicative
of many known TeV sources ({\em e.g.} \cite{Aharonian2001}, \cite{Hillas1998},
\cite{Aharonian2005d}). A smaller sample of 10 fix energies was also generated
for the study of image characteristics at those particular energies.
Background images were assumed to be predominantly due to primary proton
interactions in the atmosphere, which constitute the majority ($\sim$75\%) of
cosmic-ray showers that trigger ACTs in the TeV regime. Heavier nuclei are
expected to produce shower images which can be rejected at least as easily as
those from proton primaries. Images from isolated muons are not explicitly
considered in this study as these tend to be readily identifiable for the
energy regime considered here. The simulated proton events were
generated from a power-law with a differential spectral index of 2.7
\cite{Horandel2003} and sampled over the same area on ground as for gamma-ray
showers.  To insure a reasonable statistical sample in different energy ranges
so as to more accurately characterise the nature of these backgrounds in these
regimes, proton showers were generated over several, distinct energy intervals
with their contributions appropriately weighted to represent the full
spectrum.  Cosmic-rays are essentially isotropic with respect to the earth,
however, in practise, these events can only trigger an ACT within a restricted
angular range relative to the telescope orientation. Accordingly, simulated
proton-induced background showers were sampled within an angular range of 15
degrees from the telescope axis, which was found to be sufficient (see figure
\ref{trigger_prob}c).

\par To increase the effective statistics, both, gamma-ray and background
showers were used 100 times each by sampling the {\Cerenkov} photons over
different regions on ground. On average, local variations in the shower
properties should be relatively uncorrelated. However, within a given shower,
all such sampled densities are correlated with the first few interactions of
the primary particle, which determines the height of shower development. Thus,
in order to verify that the sampling used did not noticeably bias the results
of this study, table \ref{cuts} compares the average proton and gamma-ray
acceptances for different event selection criteria (described below) with the
case in which repeated sampling of each shower was not used. As can be seen,
there is no evidence of any bias to within the level of the statistical
uncertainties of the test. Therefore, repeated sampling was used
for increased statistics in order to better study the differential behaviour of
the ACT performance. In cases of low statistics in particular parameter
regimes, the effect of weighted sampling on the extracted differential
behaviour was also directly studied to insure there was no significant bias
due to a handful of heavily-weighted showers.

\section{Image Analysis}

\par For those images which triggered the detector, we arbitrarily chose to
use the same 2-level image cleaning and image parameterisation approach
employed by the Whipple group for a camera with a similar individual PMT
acceptance (0.25~degrees as opposed to 0.3~degrees) \cite{Reynolds1993}.  This
required a 4.25$\sigma$ signal above sky noise background for ``picture''
pixels and 2.25$\sigma$ for neighbouring ``boundary'' pixels. In addition, to
ensure that enough information was available to reconstruct the image
parameters reliably, accepted images were required to have at least 5 picture
pixels.

\par
\begin{figure}[hbt]
  \begin{center}
    \epsfig{file=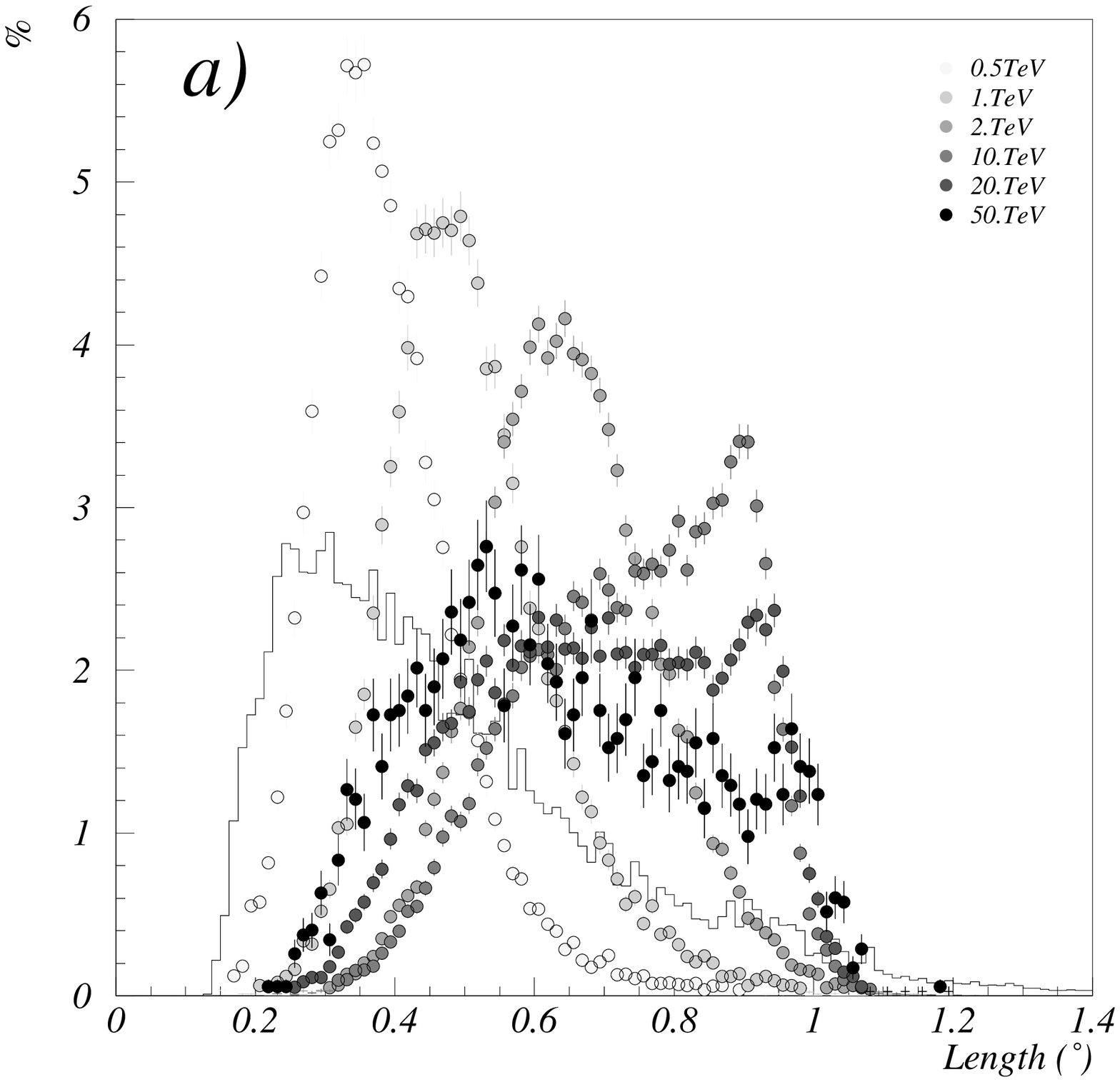,width=.45\textwidth}
    \epsfig{file=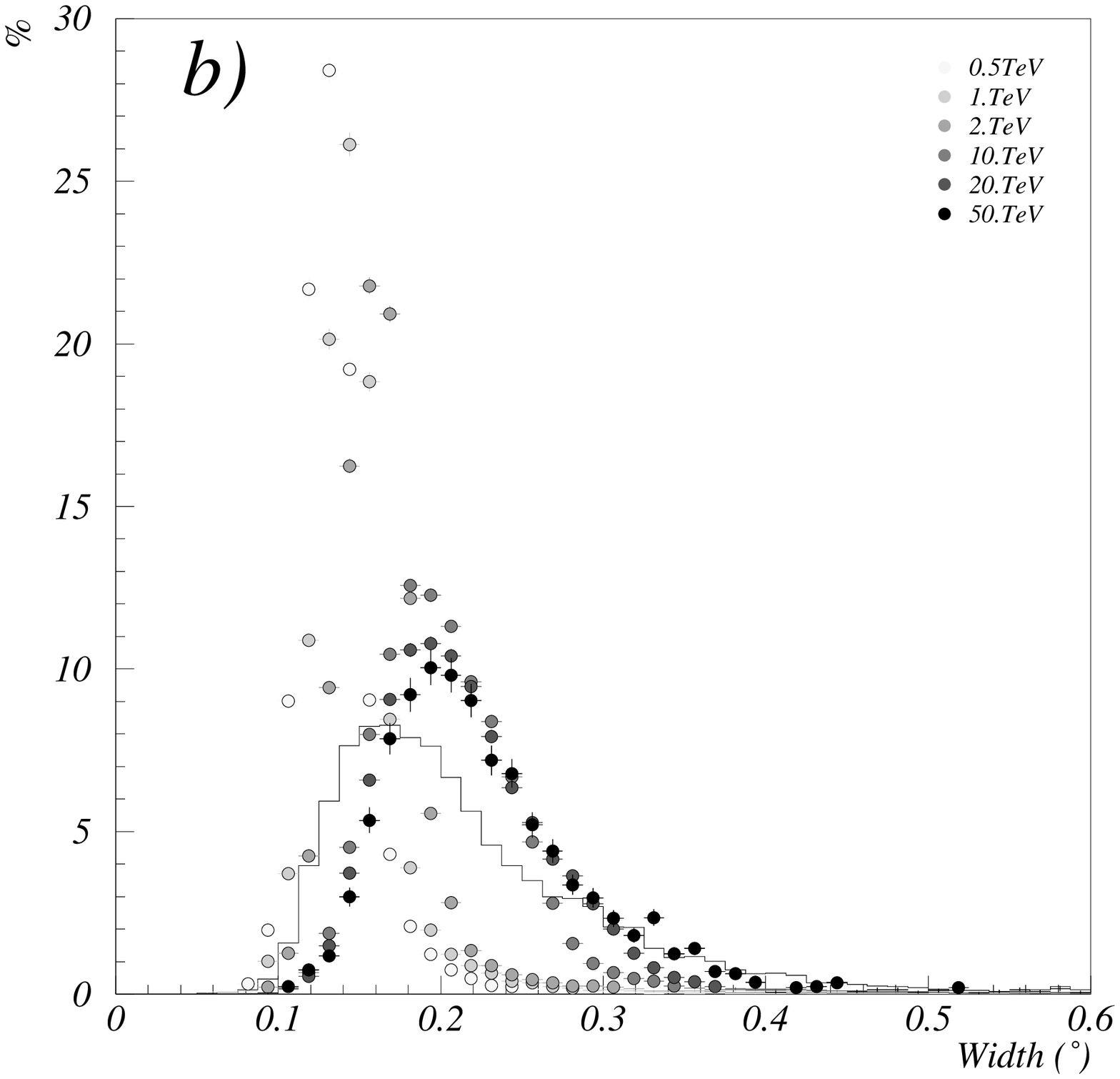,width=.45\textwidth}\\
    \epsfig{file=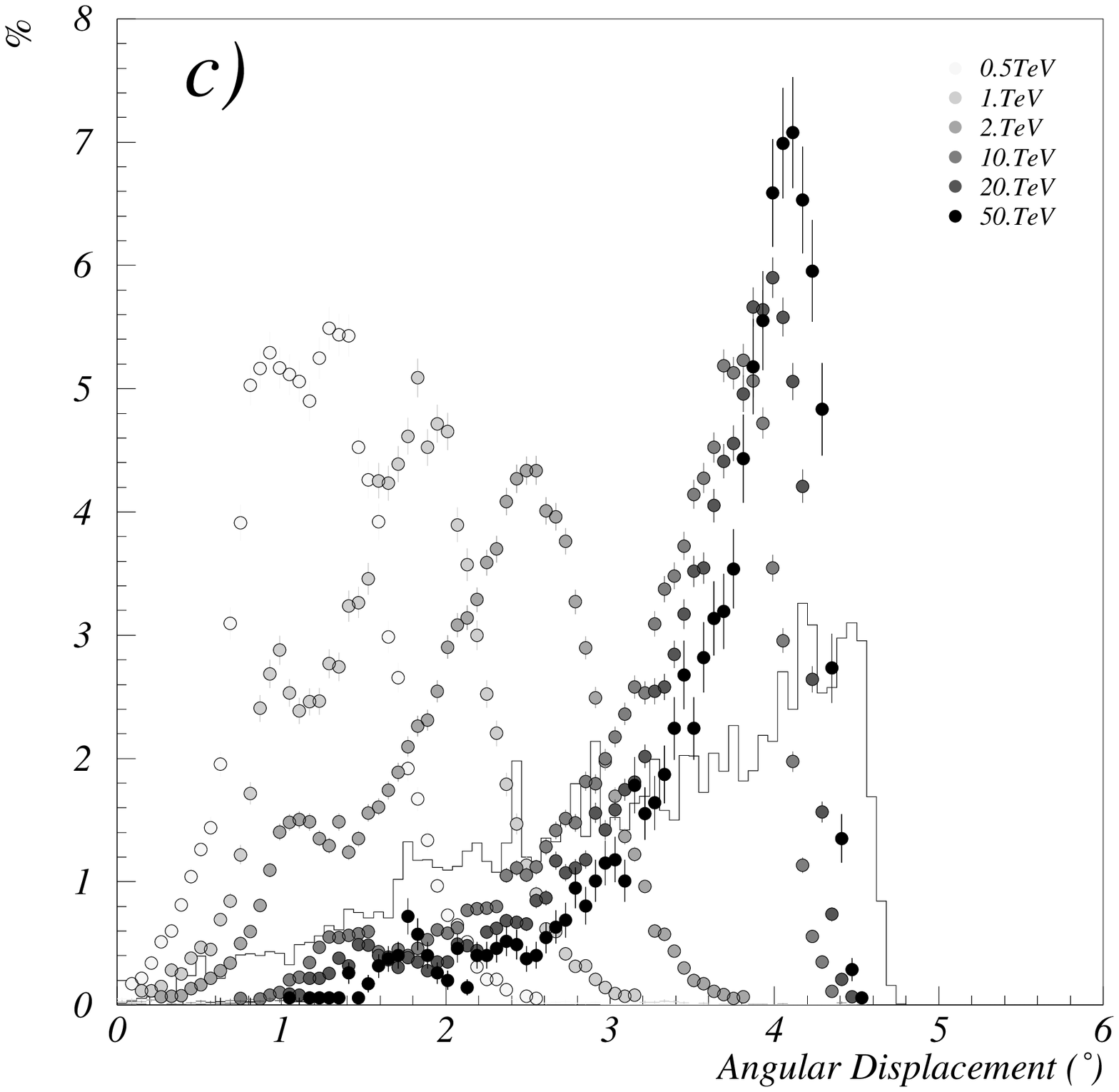,width=.45\textwidth}
    \epsfig{file=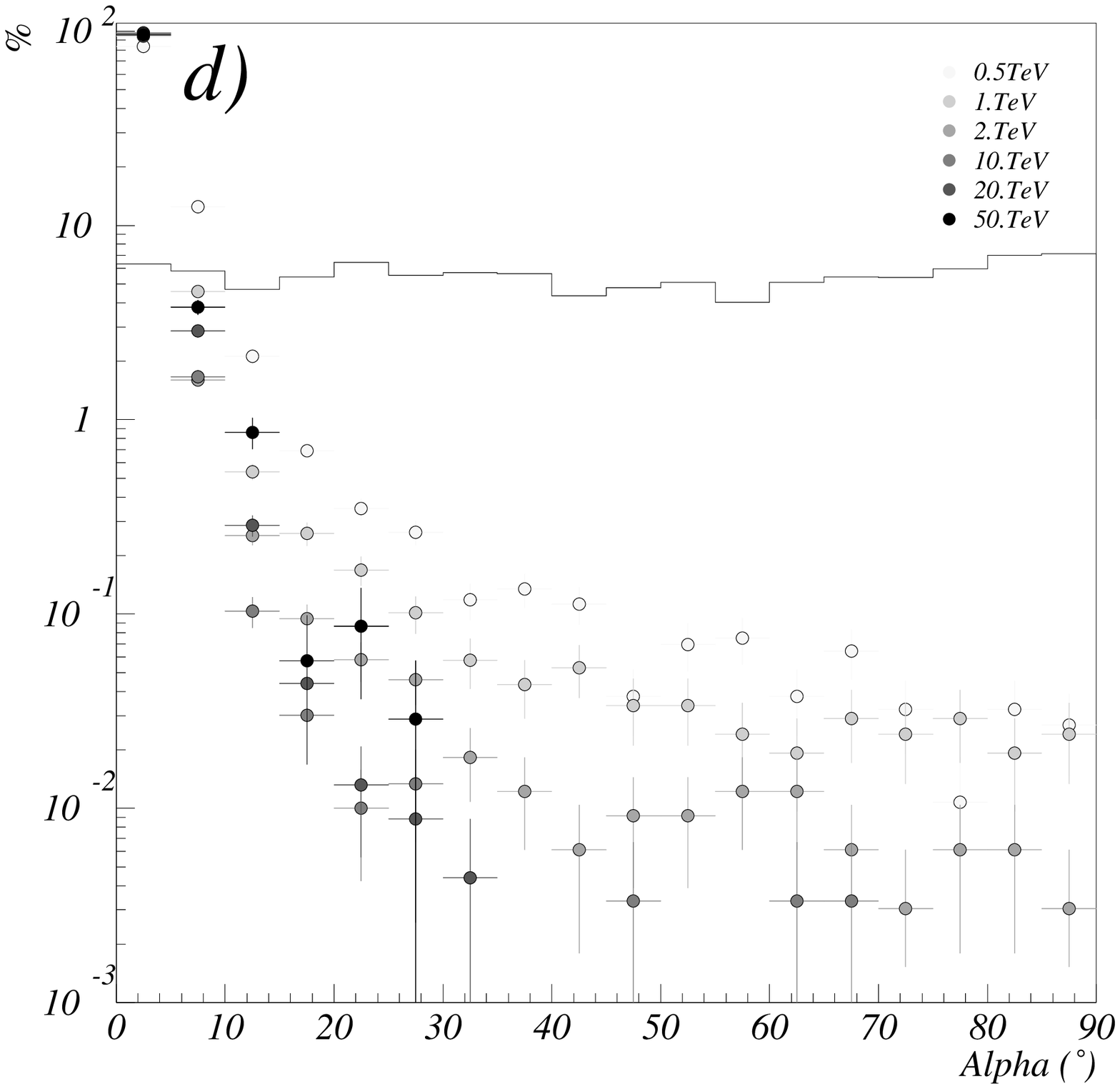,width=.45\textwidth}\\
    \caption{Image parameter distributions for simulated triggering gamma-ray
      events of 6 different energies (not normalised according to spectral
      shape) and background events summed over all energies (see table
      \ref{tab:single}).}\label{image_p_pro}
  \end{center}
\end{figure}

\begin{figure}[hbt]
  \begin{center}
    \epsfig{file=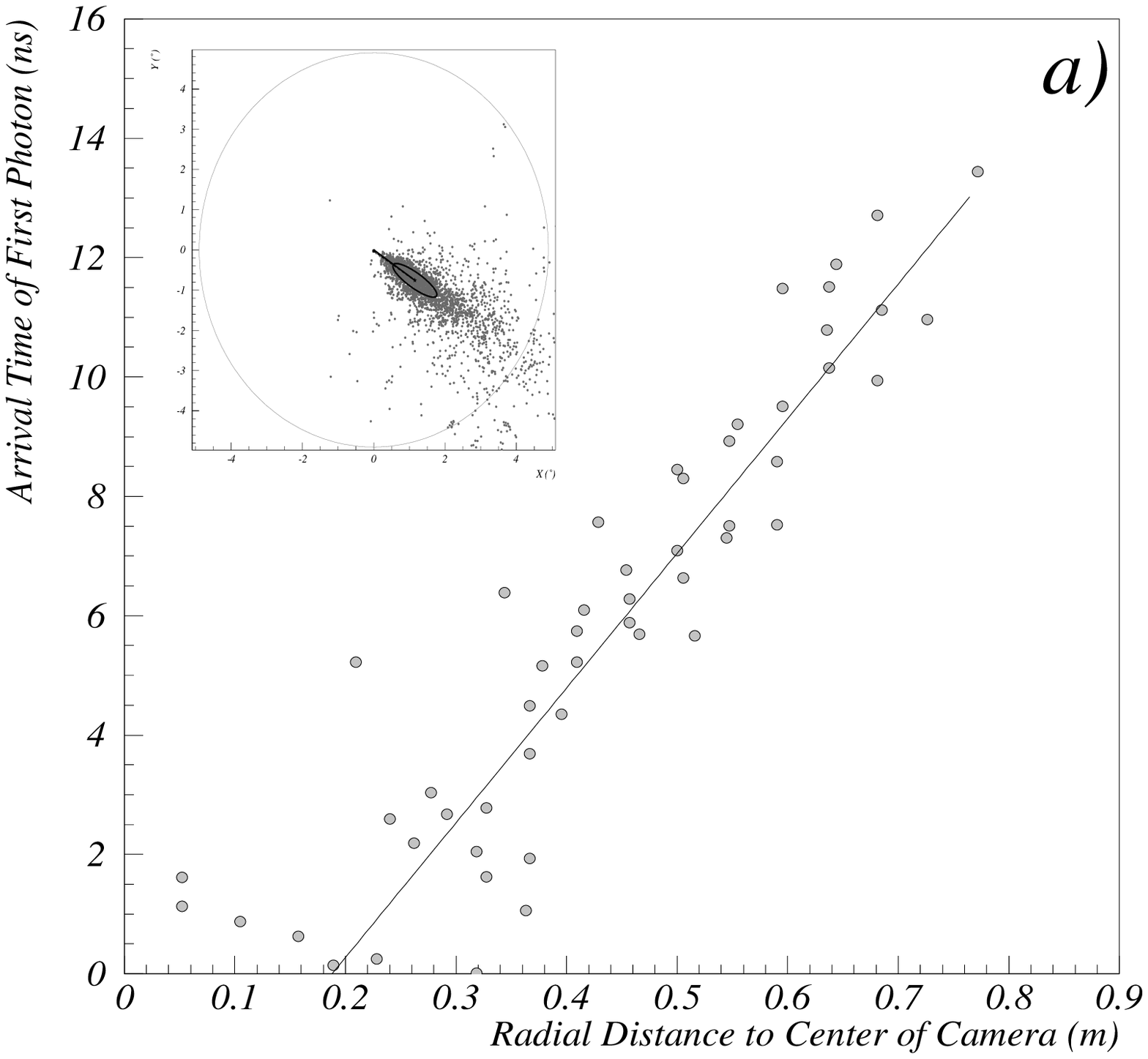,width=.45\textwidth}
    \epsfig{file=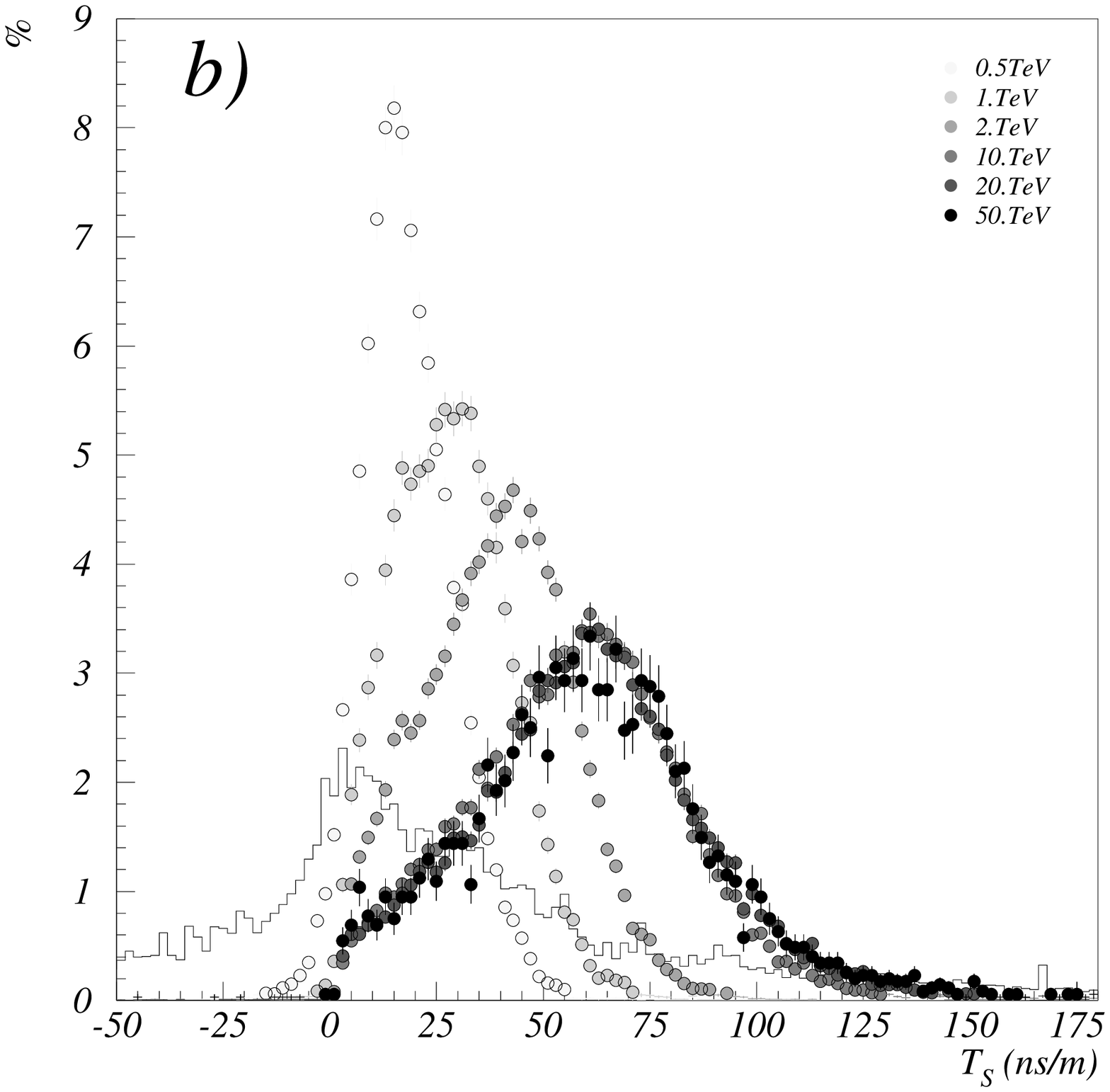,width=.45\textwidth}
    \caption{{\it a)} Arrival photon time on the camera for a simulated
      gamma-ray event. Each point in the graph corresponds to the arrival time
      of the first photon at a particular PMT. Only image pixel are used to
      derive the slope, $\mathcal{T}_S$. The corresponding gamma-ray image is
      shown in the in-set figure. {\it b)} Distribution of $\mathcal{T}_S$ for
      simulated triggering gamma-ray images of different 6 energies (not
      normalised to account for spectral shape). The continuous line is the
      equivalent distribution of $\mathcal{T}_S$ for simulated background
      events summed over all energies (see table
      \ref{tab:single}).}\label{TSlope_fig}
  \end{center}
\end{figure}

 Images were then analysed in terms of moments according to the Hillas
prescription \cite{Hillas1985}. The relevant image parameters used in this
study were: {\bf 1)} the summed number of detected photoelectrons in the
image, $\mathcal{S}$; {\bf 2)} the average angular displacement in the camera
relative to the source position, $\mathcal{D}$; {\bf 3)} the mean length of
the image, as measured along its major axis, $\mathcal{L}$; {\bf 4)} the mean
width of the image, as measured along its minor axis, $\mathcal{W}$; and {\bf
  5)} the angle of the major axis with respect to the line linking the image
centroid with the assumed source position, $\alpha$. All of these parameters
were couched in units of degrees. Figure \ref{image_p_pro} shows the
distributions of these image parameters for gamma-ray and background events.
In addition, a further image parameter, $\mathcal{T}_S$, was defined as the
slope extracted from a linear fit to the earliest photon arrival time in each
PMT as a function of the angular distance of that PMT from the assumed source
position. This characterises the projection of distances to
{\Cerenkov}-emitting regions along the axis of shower development, given the
expected primary gamma-ray direction. Figure \ref{TSlope_fig} illustrates the
definition of $\mathcal{T}_S$ for a particular event.

\section{Dependence on Pixel Size}\label{Dep_Pixel_Size}

\par
\begin{figure}[hbt]
  \begin{center}
    \epsfig{file=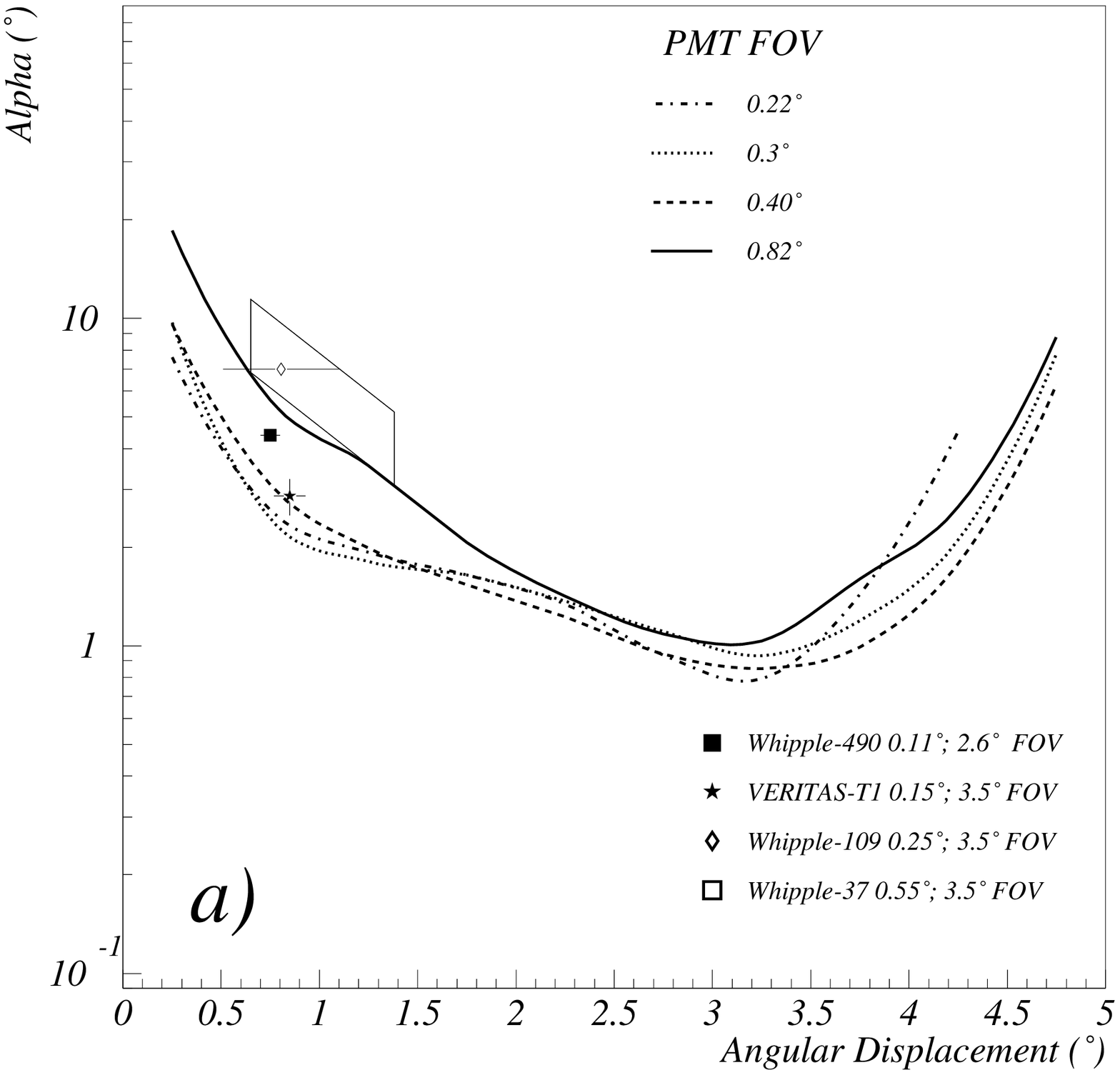,width=.45\textwidth}
    \epsfig{file=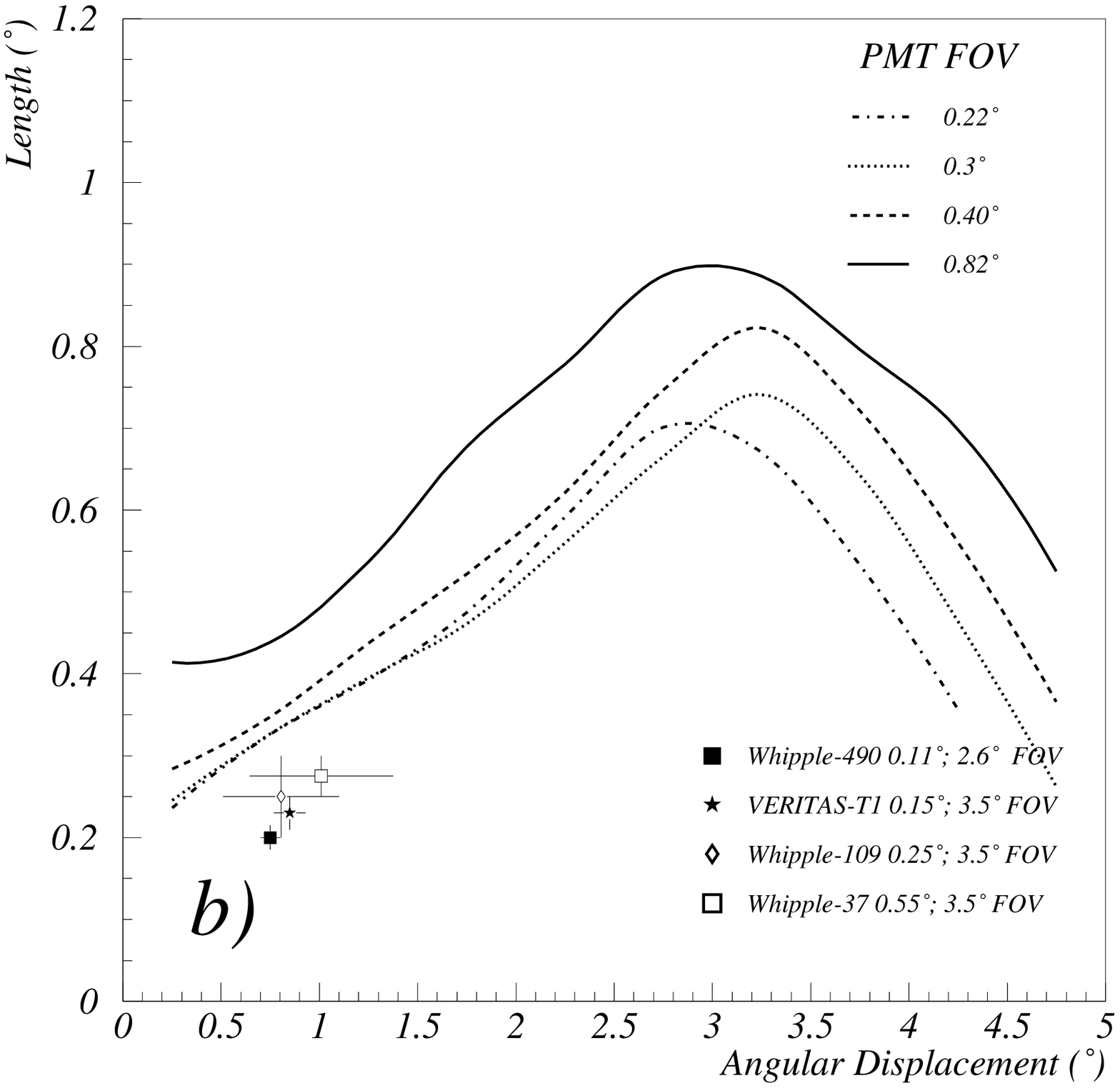,width=.45\textwidth}\\
    \epsfig{file=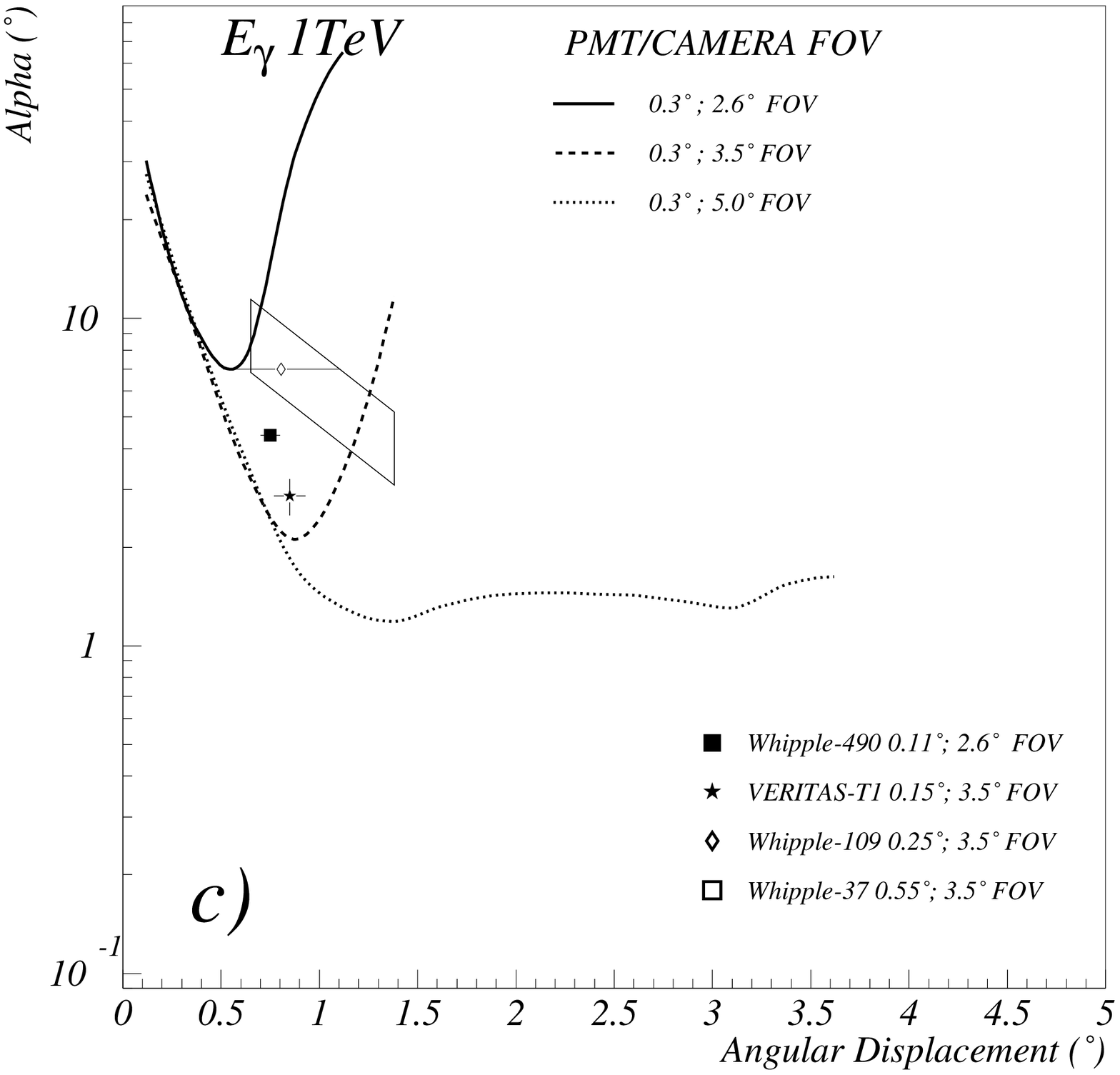,width=.45\textwidth}
    \epsfig{file=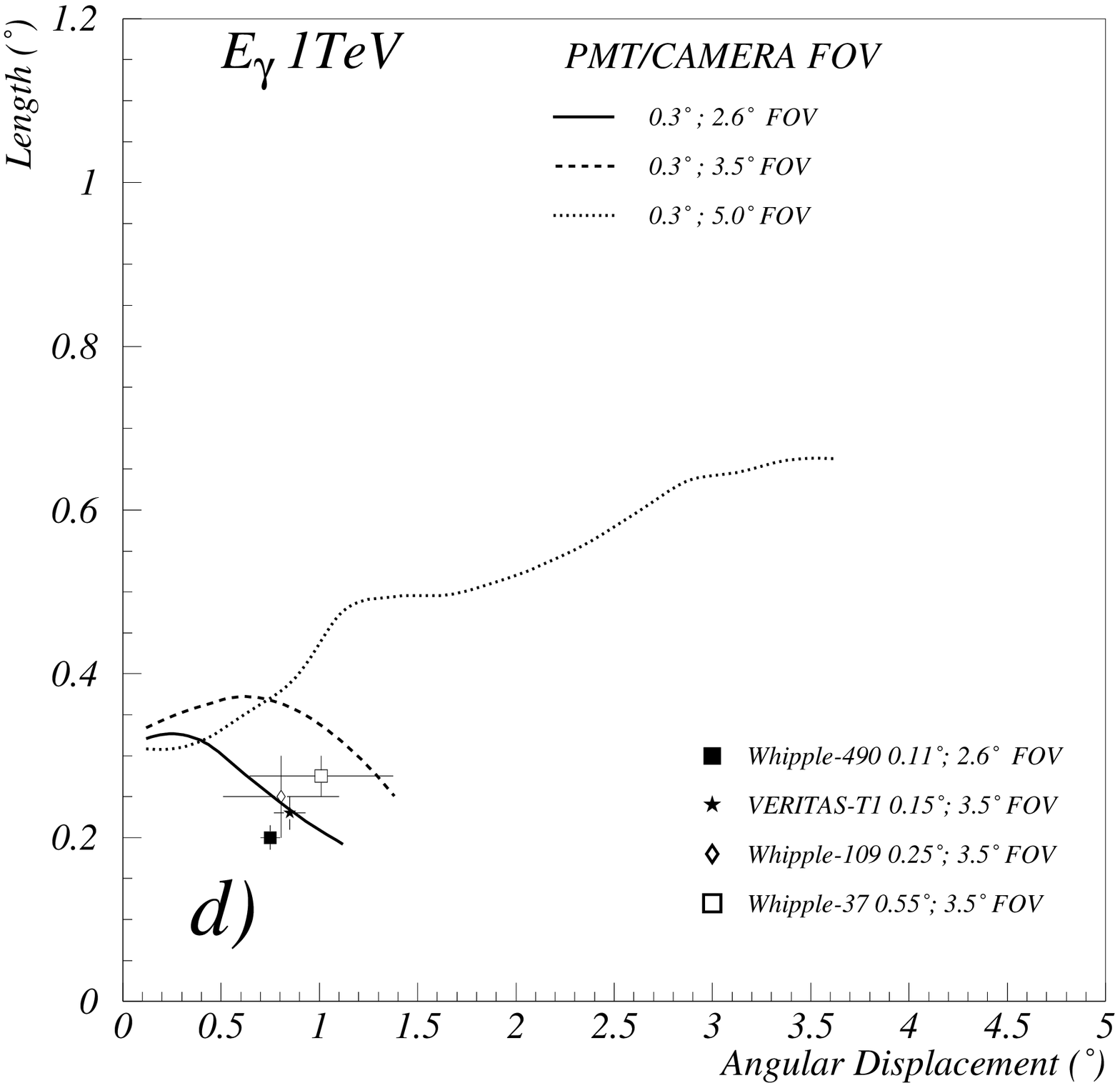,width=.45\textwidth}\\
    \caption{{\it a)} Median alpha and {\it b)} median length as a function of
      the image angular displacement for cameras with a $\sim$10.$^\circ$ FOV
      but using 4 different pixel sizes. For comparison, data from four
      different {\Cerenkov} telescope cameras, with different pixel sizes and
      FOVs, have been added to the plots, Whipple-490 \cite{Finley2000},
      VERITAS-T1 \cite{Maier2005}, Whipple-109 \cite{Mohanty1998} and
      Whipple-37 \cite{Weekes1989}. Effect of a limited FOV on {\it c)} alpha
      and {\it d)} length. For simplicity, only 1~TeV gamma-ray images are
      considered on these two bottom plots.}\label{pixel_parameters}
  \end{center}
\end{figure}

\begin{figure}[hbt]
  \begin{center}
    \epsfig{file=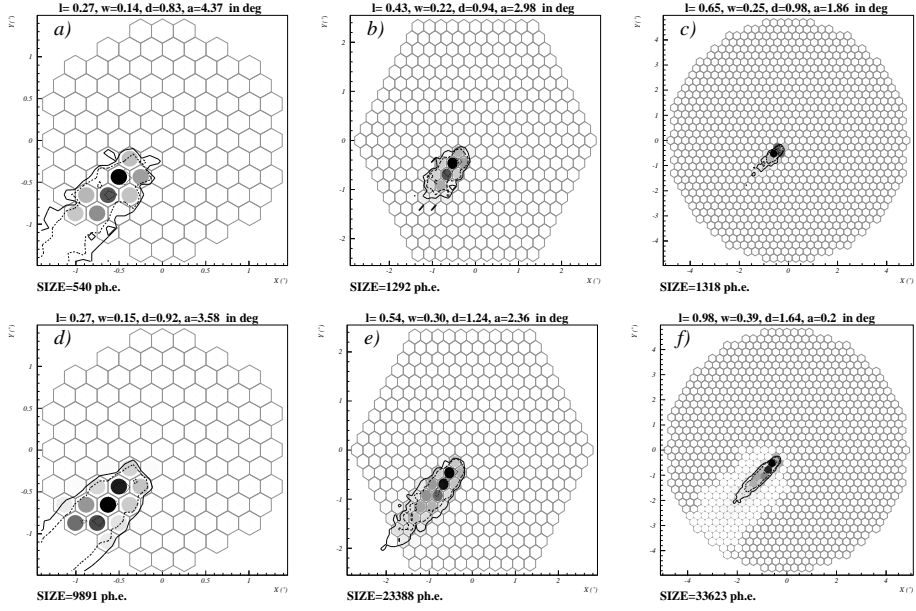,width=1.\textwidth}
    \caption{Simulated 1~TeV gamma-ray event at d$_{core}$=120~m (top panel),
      and 20~TeV gamma-ray event at d$_{core}$=120~m (bottom panel) as
      recorded by cameras with different field-of-view (FOV). From left to right,
      $\sim 3^\circ$, $\sim 5^\circ$ and $\sim 10^\circ$ FOV. The first two
      are similar in FOV to Whipple cameras and the third one,
      corresponds to the wide field of view camera used in the simulations
      presented in this work. Note as well that the pixel size is different
      for each camera (from left to right, $\sim 0.19^\circ$, $\sim
      0.23^\circ$ and $\sim 0.3^\circ$ FOV). Superimposed on top of the
      shower images are 5\% (outer line), 10\%, 50\% and 90\% photon
      density contours.}\label{camera_images}
  \end{center}
\end{figure}

 The impact of pixel size (PMT density) on ACT sensitivity enters
primarily though the ability to characterise image shape and orientation so as
to aid in background rejection. Consequently, we have explored the ability to
reconstruct the parameters $\mathcal{L}$, $\mathcal{W}$ and $\alpha$ as a
function of $\mathcal{D}$ for simulated cameras corresponding to pixel sizes
of 0.22, 0.3, 0.4 and 0.82 degrees in diameter. These were also compared with
values from actual Whipple and VERITAS data for cameras with various pixel
sizes over their limited range in $\mathcal{D}$. The results are shown in
figure \ref{pixel_parameters}. Several characteristics are worth noting: {\bf
  1)} The ability to determine image orientation ($\alpha$) is substantially
improved at larger image displacements. {\bf 2)} The ability to characterise
image parameters is noticeably improved for pixel sizes of 0.4$^\circ$
compared to 0.8$^\circ$, but any further improvement from using even smaller
pixels is relatively modest. This is consistent with the findings of Aharonian
{\em et al.} \cite{Aharonian1995}. {\bf 3)} The fact that very small pixel
sizes even appear to do worse in determining $\alpha$ at larger image
displacements is an artifact of trigger selection owing to the decreasing
photon levels in smaller PMTs and the specific trigger requirements assumed.
{\bf 4)} Data from actual ACTs seem to have generally larger values of
$\alpha$ in the angular displacement range of $\sim$1$^\circ$ than is
predicted by the simulation. Once more, this appears to be largely due to the
field of view and the fact that smaller cameras tend to truncate more of the
outer parts of these images. This effect is illustrated in figure
\ref{camera_images}, in which comparisons are made with simulations using
different fields of view.

\par
\begin{figure}[hbt]
  \begin{center}
    \epsfig{file=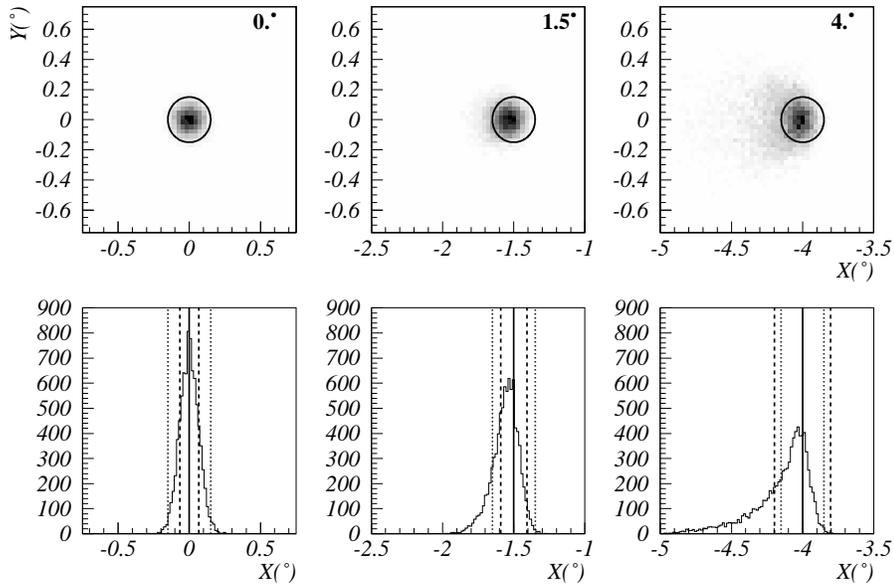,width=.9\textwidth}\\
    \caption{Distribution of photons on the camera plane for an on-axis point
      source and for two different off-set positions, 1.5$^\circ$ and
      4.$^\circ$. The circle corresponds to the individual PMT area. The
      solid lines in the three bottom plots indicate the expected peak
      position, the dashed lines are the 1~$\sigma$ region along the radial
      direction ($\sigma_r$) and the dotted lines indicate the PMT
      size.}\label{psf2}
  \end{center}
\end{figure}

\begin{figure}[hbt]
  \begin{center}
    \epsfig{file=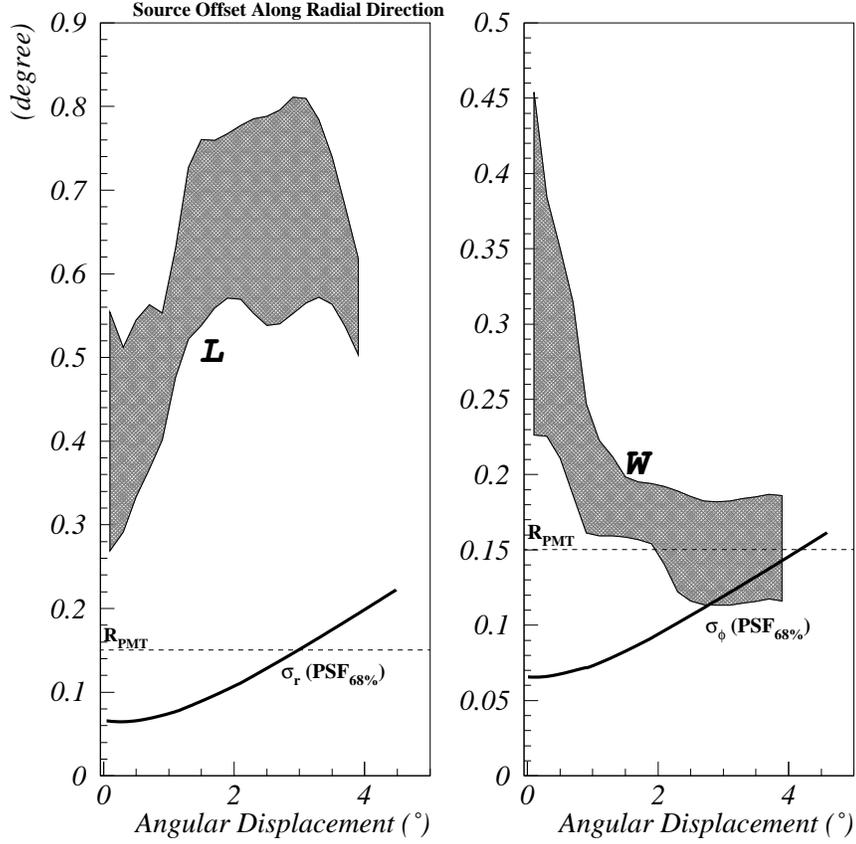,width=.9\textwidth}\\
    \caption{{\it a)} RMS of the PSF ($\sigma_{r}$) along the radial direction               
      for different source offset positions. The shaded region includes 68$\%$               
          of the events about the median length, $\mathcal{L}$, as a function of             
          radial distance for a subset of simulated 2~TeV gamma-ray showers.                 
          {\it b)} Similar plot for the RMS of                                               
      the PSF along the angular direction ($\sigma_{\phi}$), where the shaded                
          region includes 68$\%$ of the events about the median width, $\mathcal{W}$,        
          for the same showers.}\label{psf1}
  \end{center}
\end{figure}

 In addition, the Point Spread Function (PSF) of the detector due to
mirror aberrations, has been calculated as a function of the source offset
angle (figure \ref{psf2}). Figure \ref{psf1} shows the PSF together with the median
length ($\mathcal{L}$) and width ($\mathcal{W}$) as a function of the angular
displacement ($\mathcal{D}$). The figure indicates that the PSF should not
affect the ability to reconstuction the image angular size.

\par Based on the above studies, the choice of the 0.3$^\circ$ pixel size used
to explore the wide-field concept in this current paper would appear to be
more than adequate. Indeed, it is likely that a camera with even larger pixels
could be used without significantly affecting the performance of the
instrument.

\section{Single Telescope Performance}

\par We will first consider a single, isolated, wide-angle telescope and
assess its performance in terms of background rejection, energy resolution,
and effective area. Based on this, the sensitivity of the instrument to a
hypothetical point-source will be estimated for both long term (50 hours) and
short term (1 hour) observation periods, which will be compared with that of
other ACT instruments.

\subsection{Image Selection Criteria}

\par
\begin{figure}[hbt]
  \begin{center}
    \epsfig{file=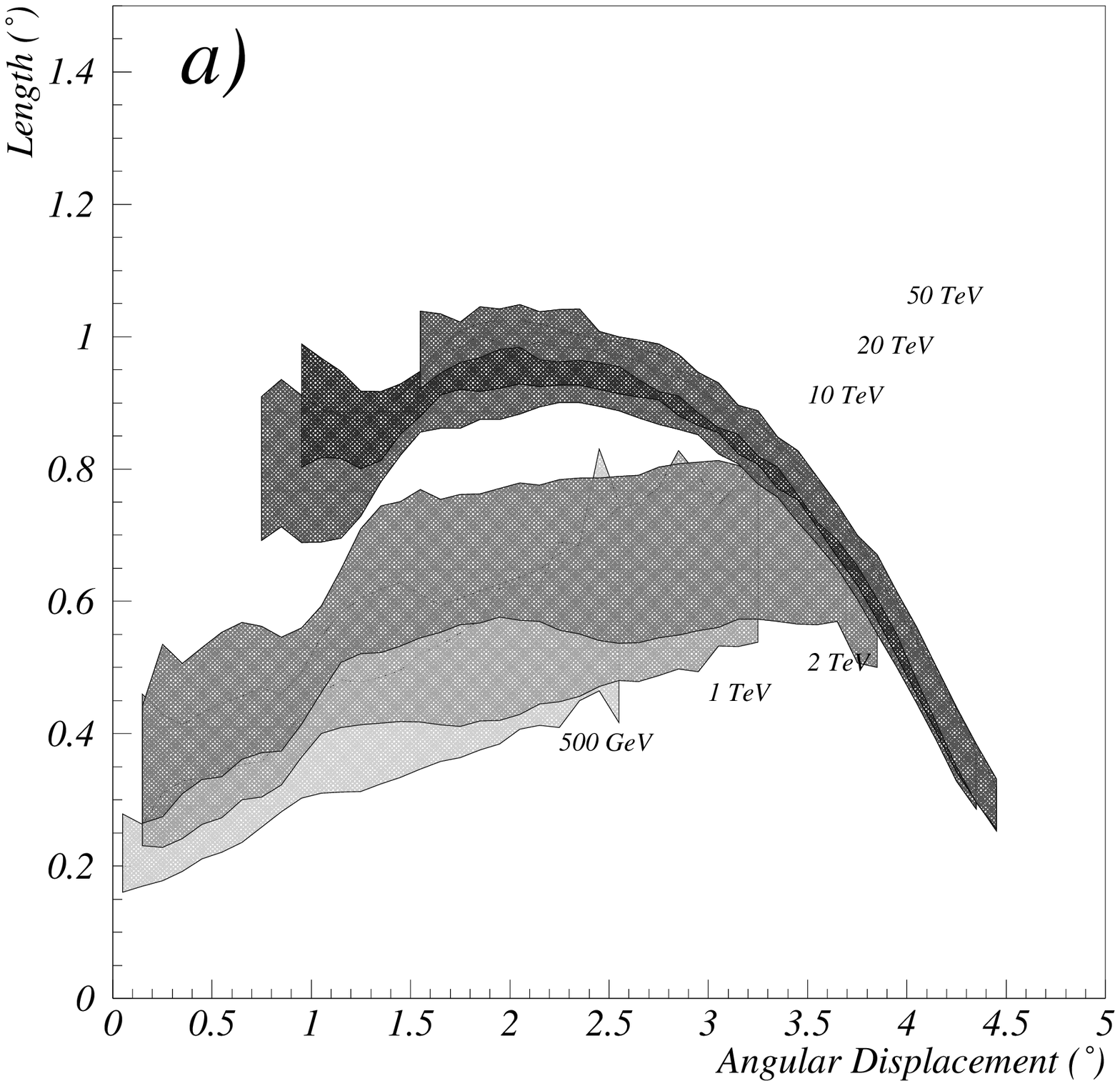,width=.45\textwidth}
    \epsfig{file=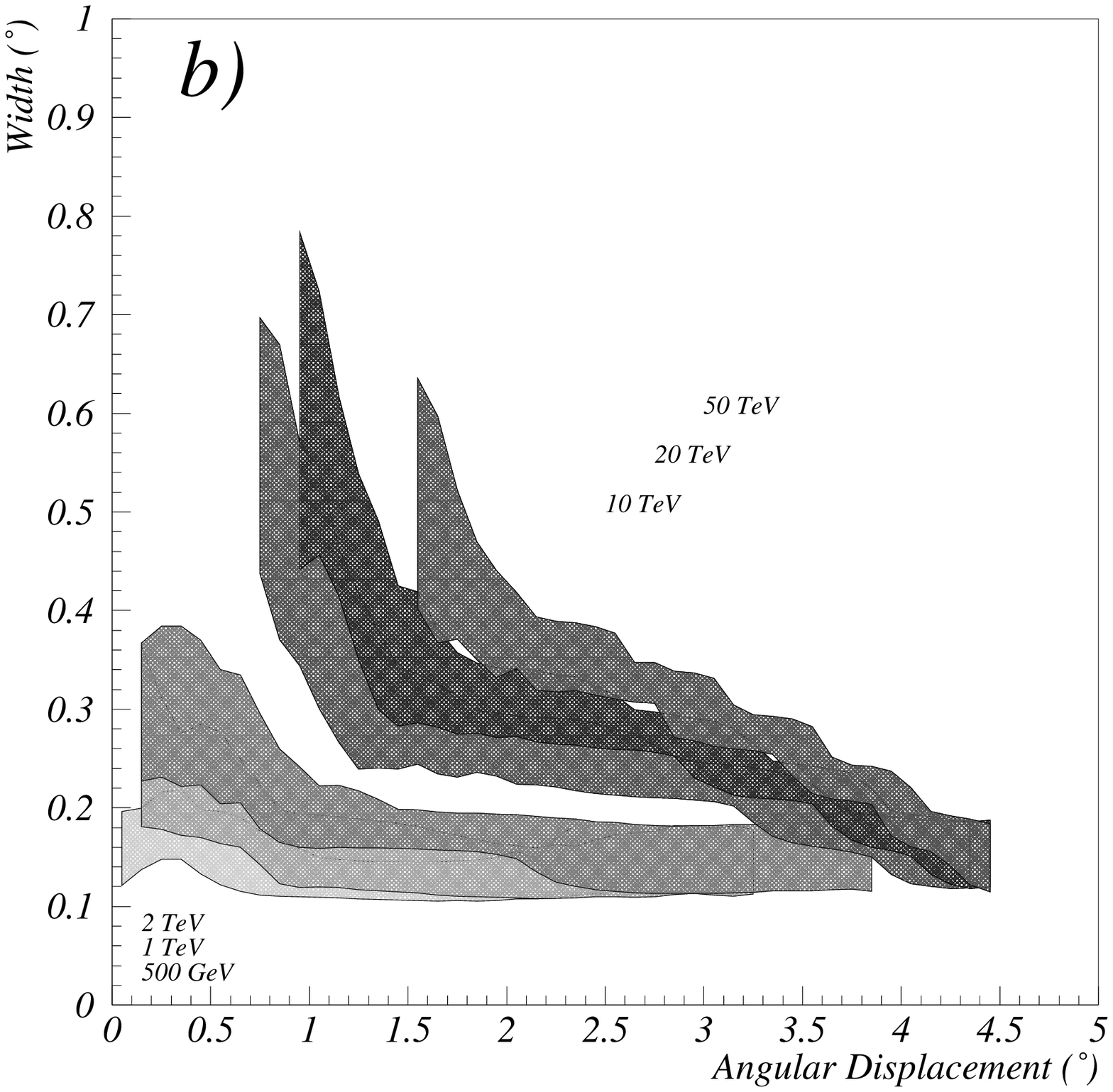,width=.45\textwidth}\\
    \epsfig{file=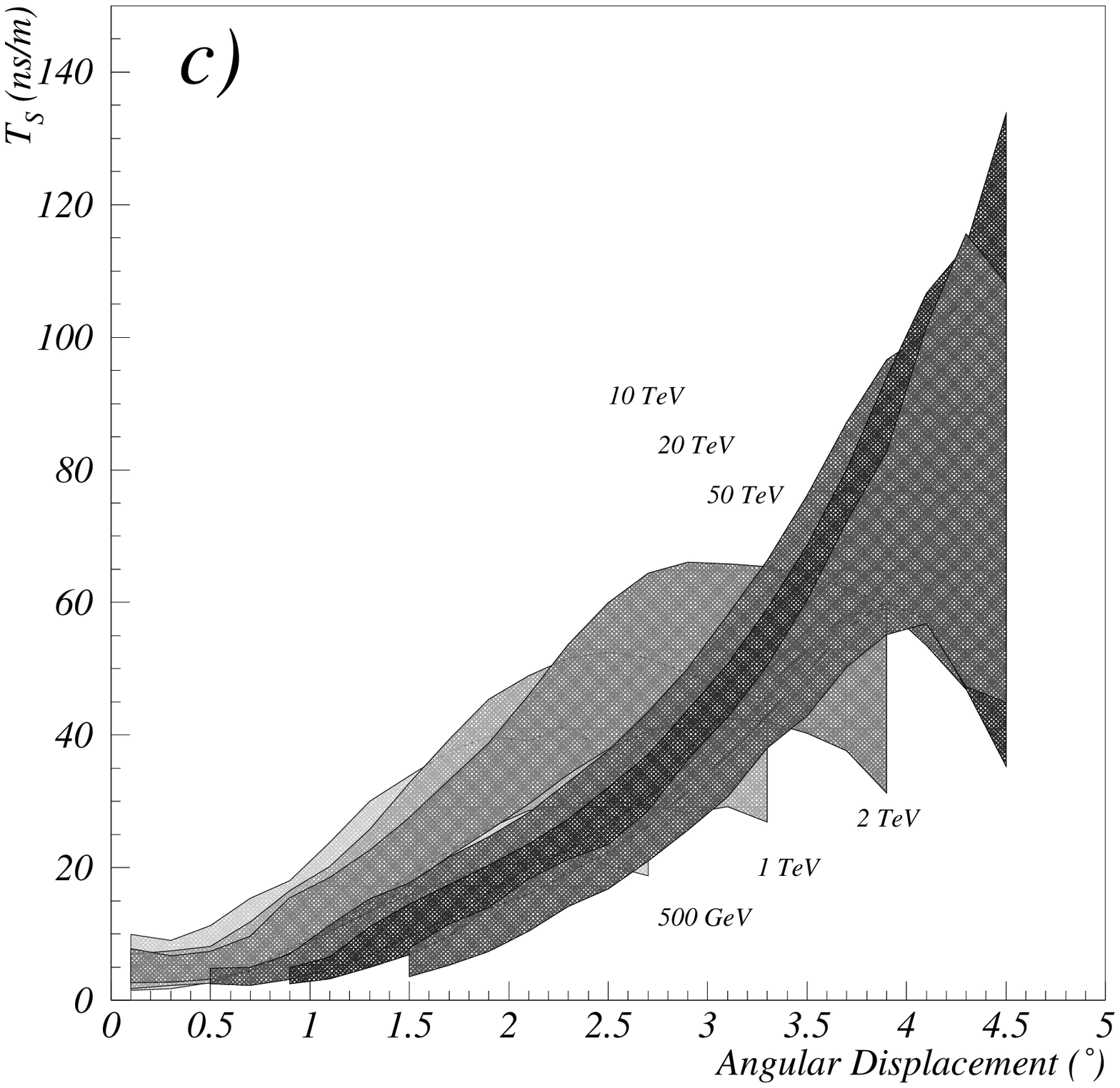,width=.45\textwidth}
    \epsfig{file=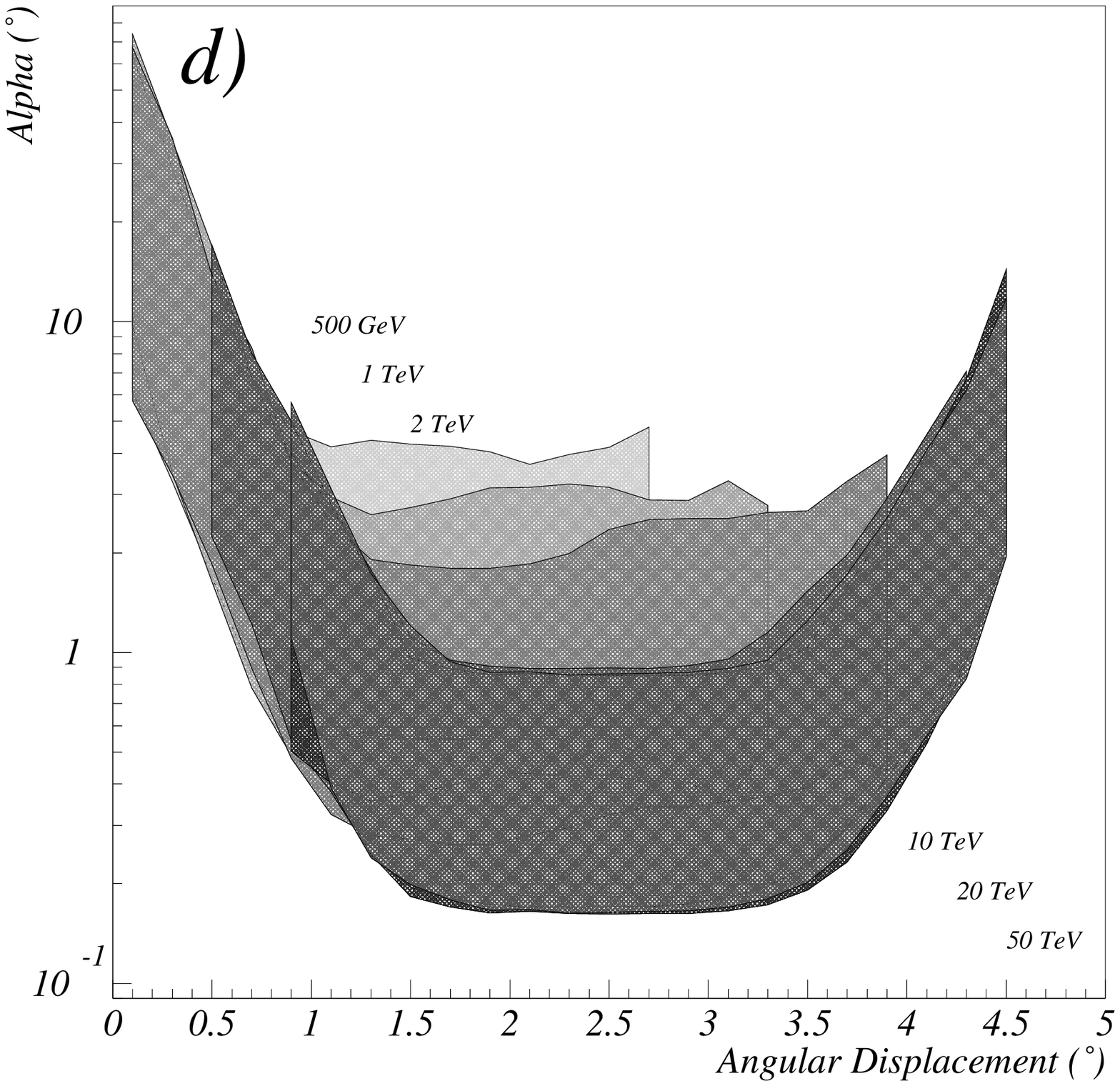,width=.45\textwidth}\\
    \caption{Image parameters as a function of angular displacement on the
      camera for triggering gamma-ray showers of 6 different
      energies. The shaded regions include 68$\%$ of the events about the
      median.}\label{gh_separation1}
  \end{center}
\end{figure}

\begin{figure}[hbt]
  \begin{center}
    \epsfig{file=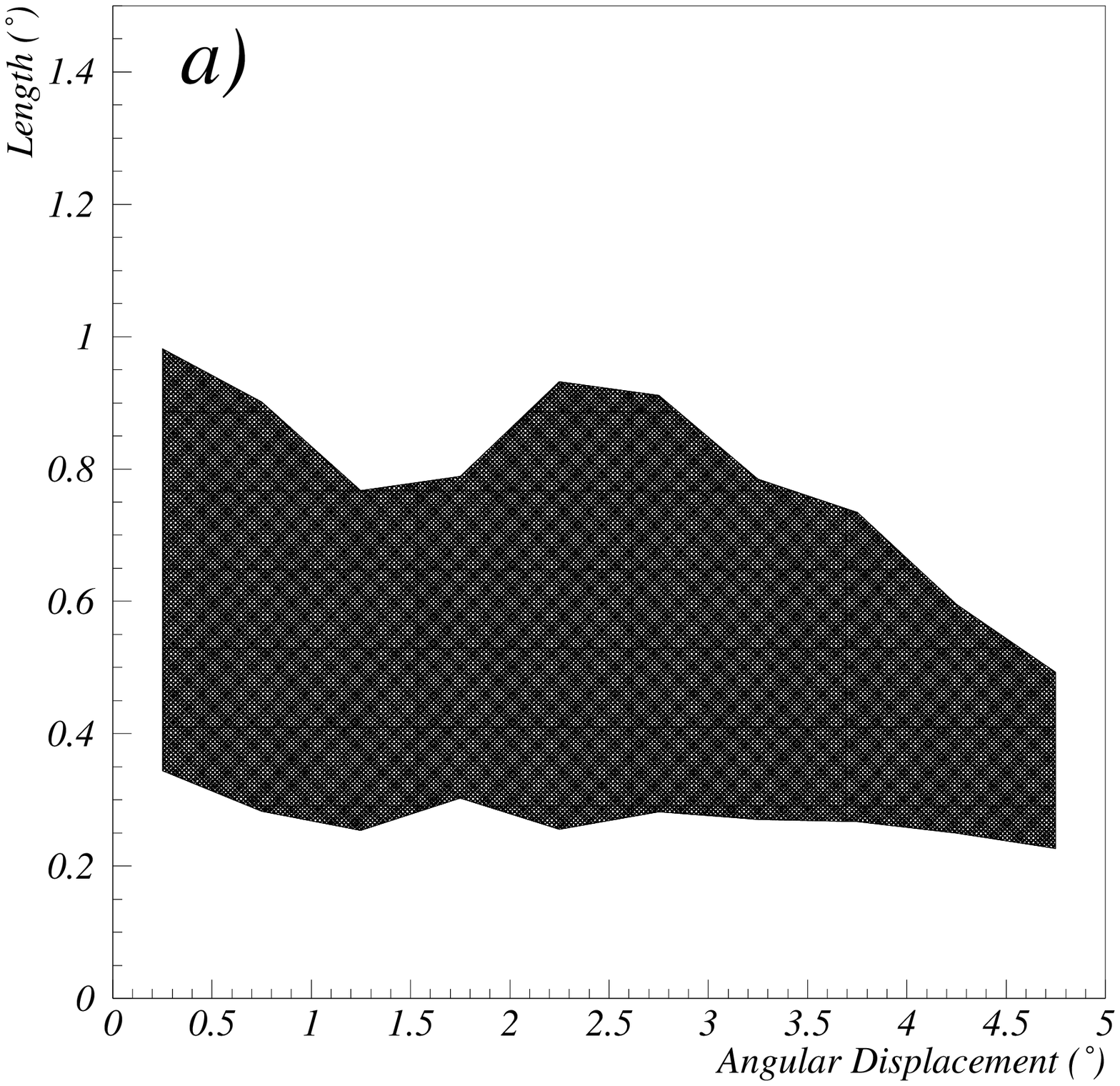,width=.45\textwidth}
    \epsfig{file=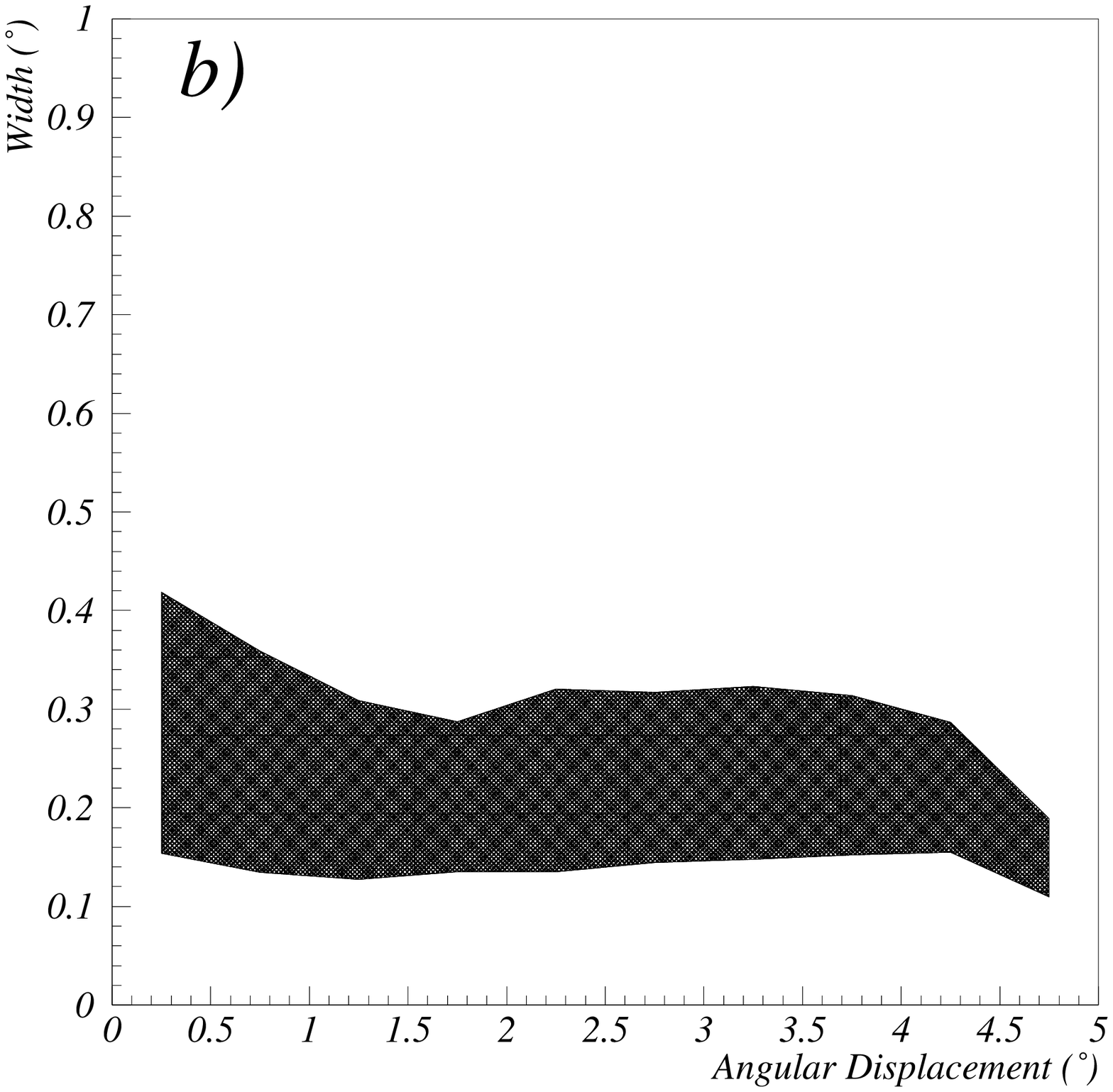,width=.45\textwidth}\\
    \epsfig{file=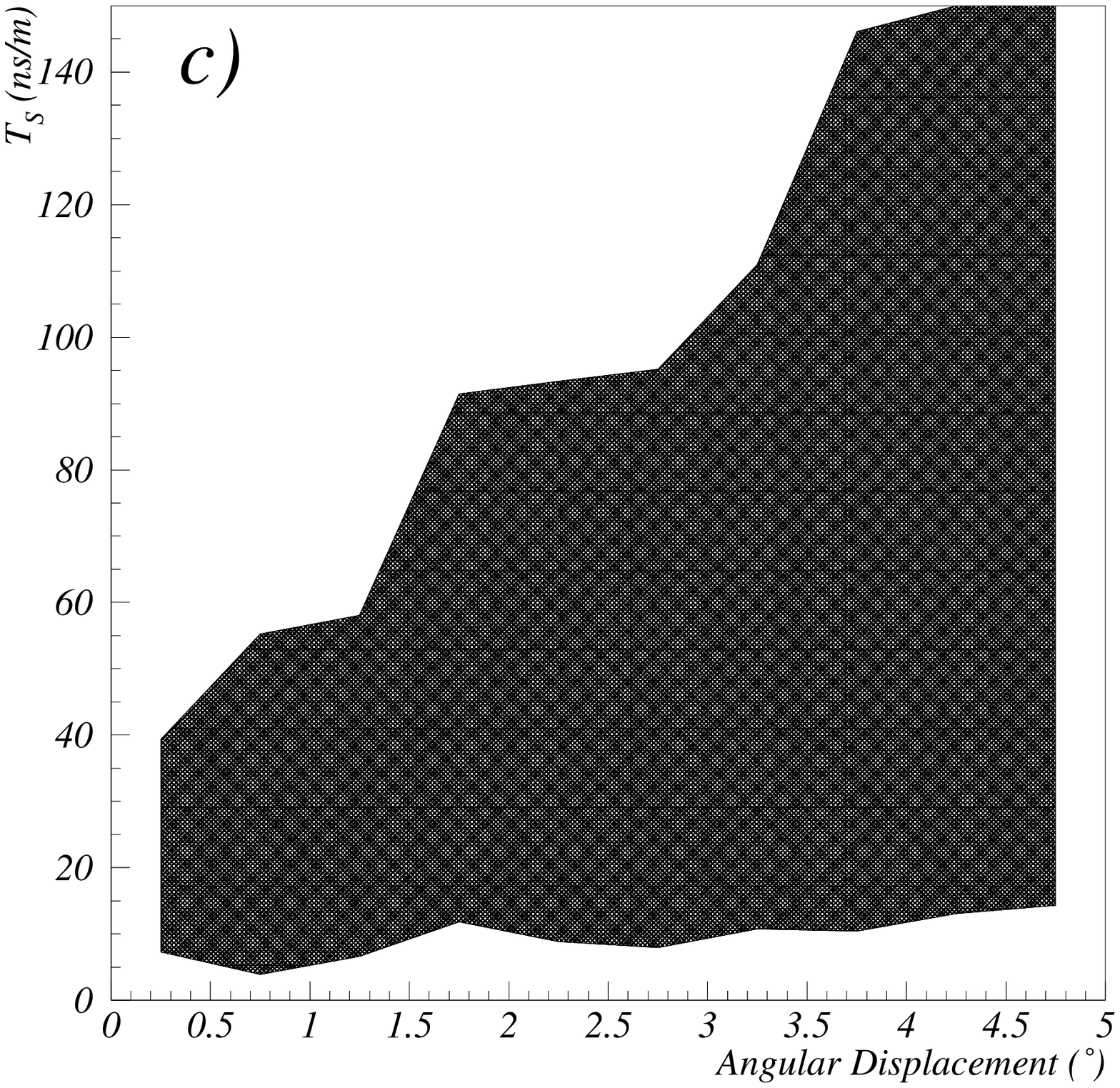,width=.45\textwidth}
    \epsfig{file=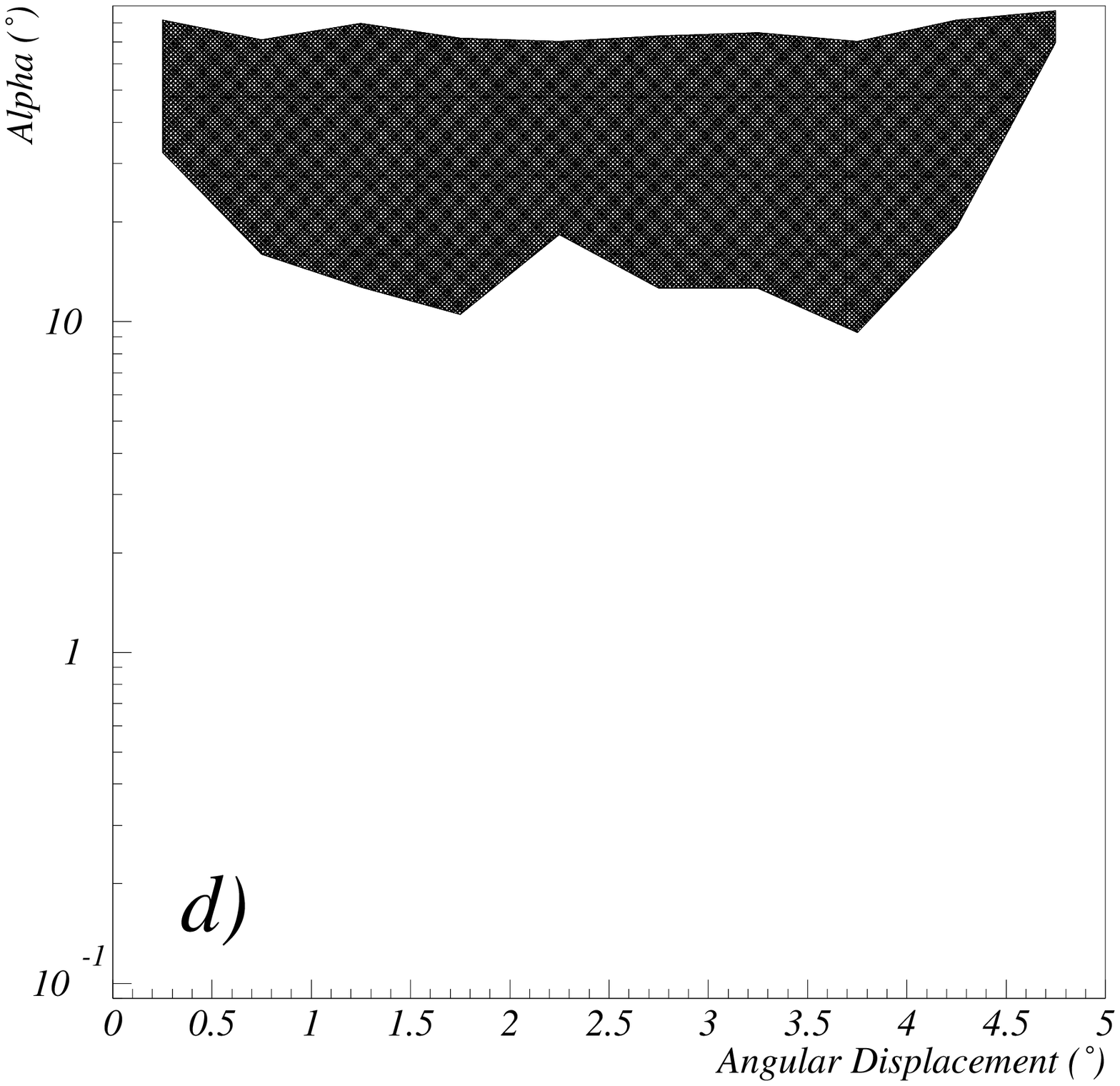,width=.45\textwidth}\\
    \caption{Image parameters as a function of angular displacement on the
      camera for triggering proton showers summed over all energies. The
      shaded regions include 68$\%$ of the events about the
      median.}\label{gh_separation2}
  \end{center}
\end{figure}

\begin{figure}[hbt]
  \begin{center}
    \epsfig{file=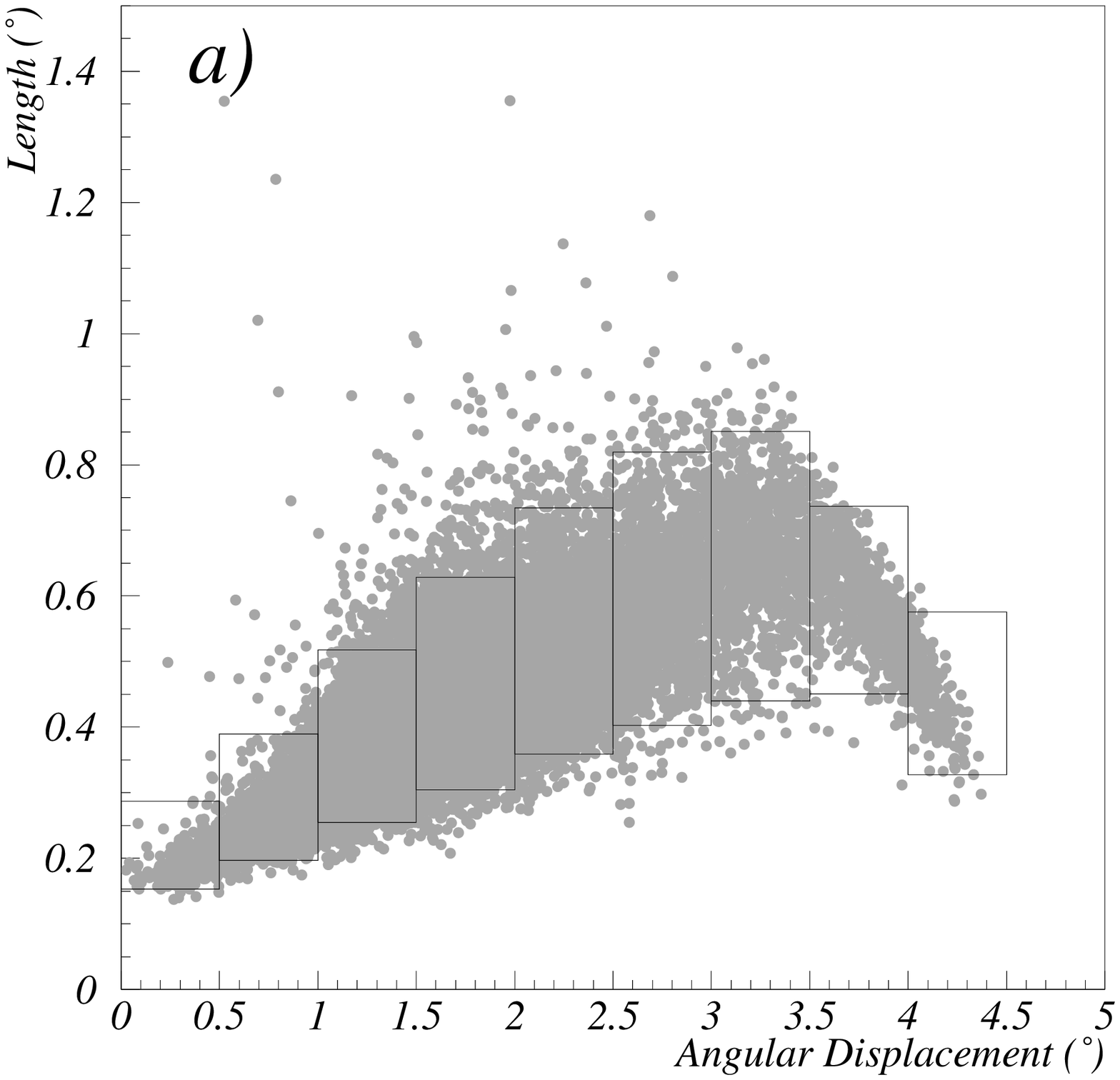,width=.45\textwidth}
    \epsfig{file=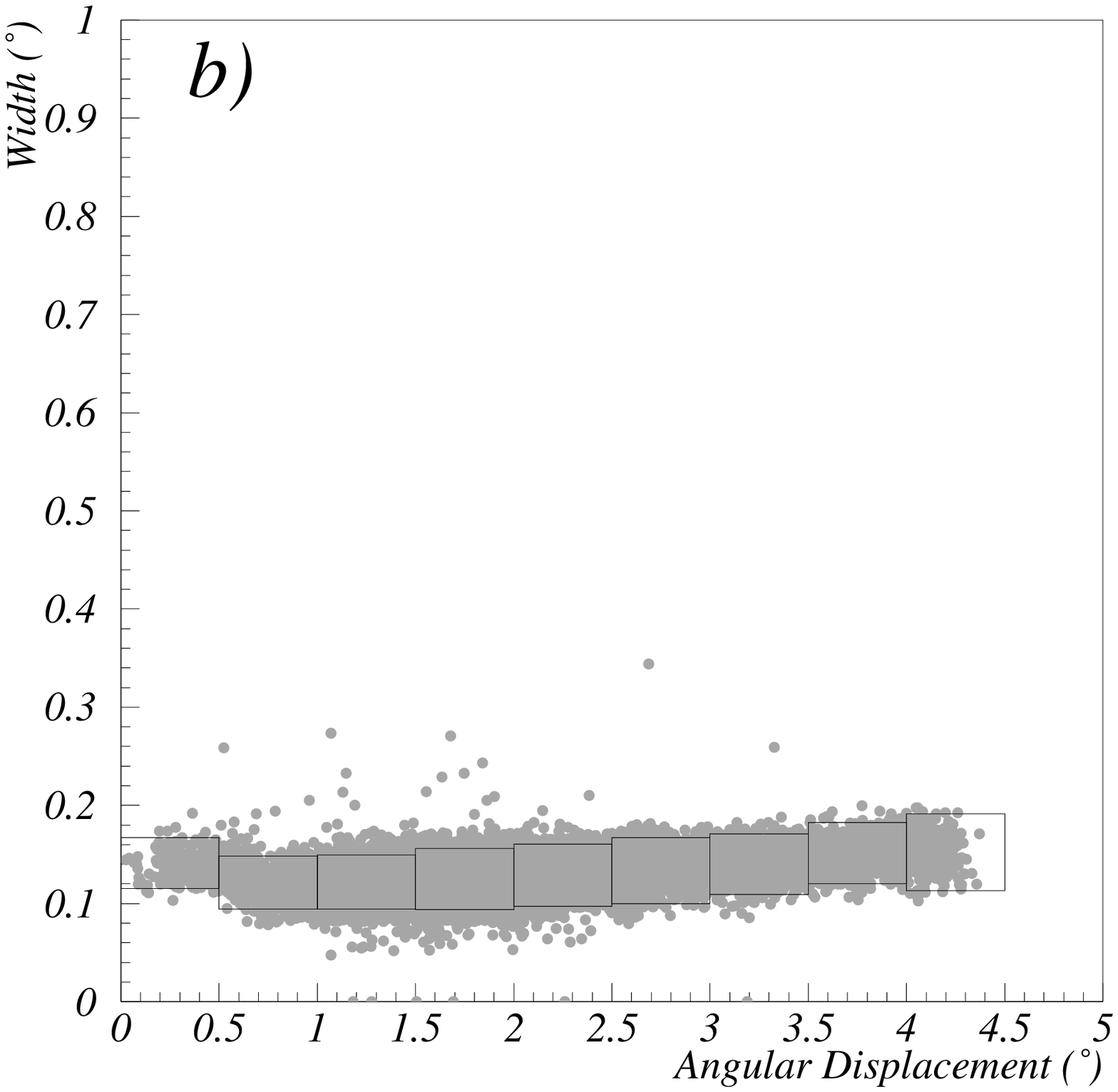,width=.45\textwidth}\\
    \epsfig{file=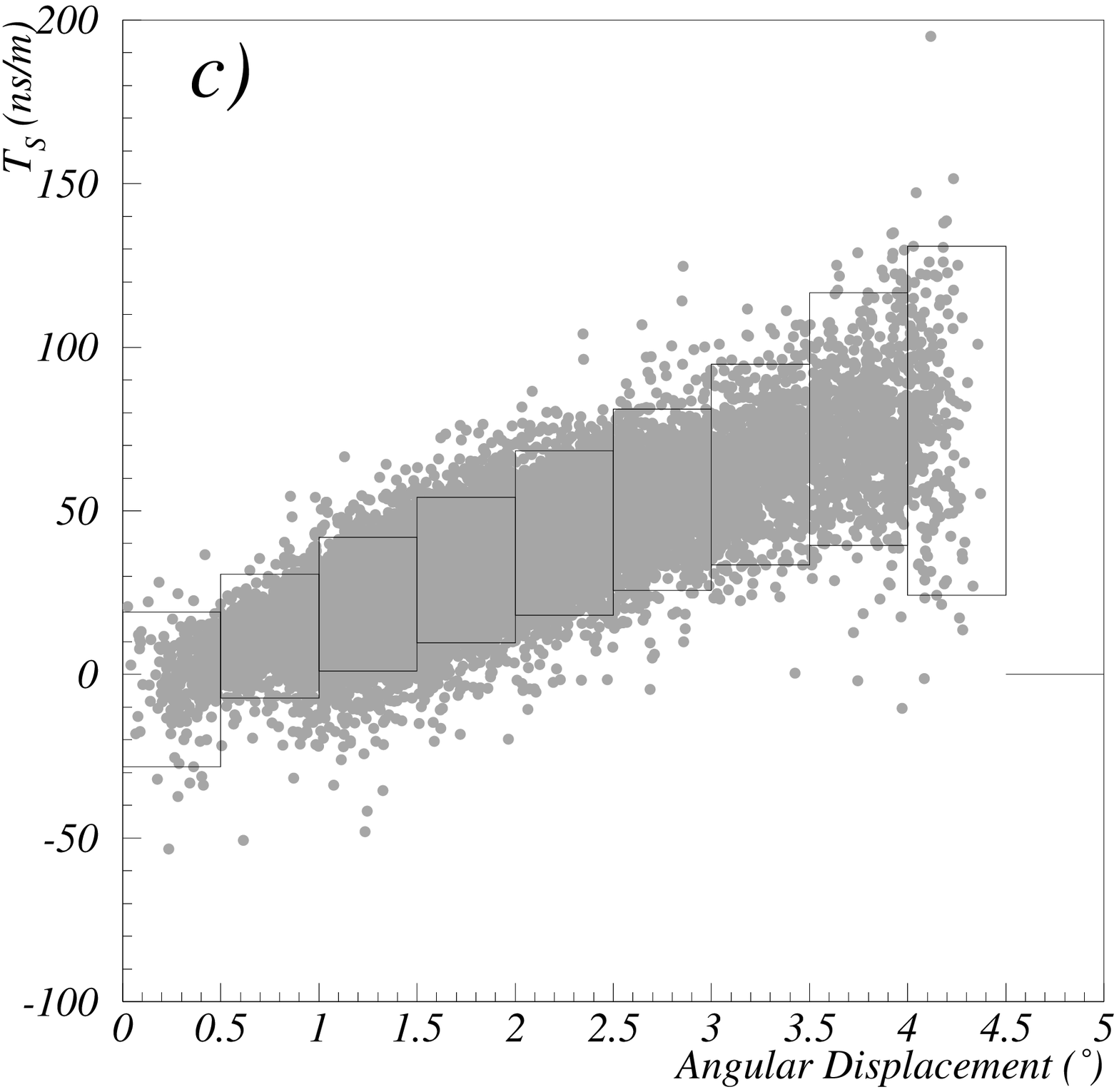,width=.45\textwidth}
    \epsfig{file=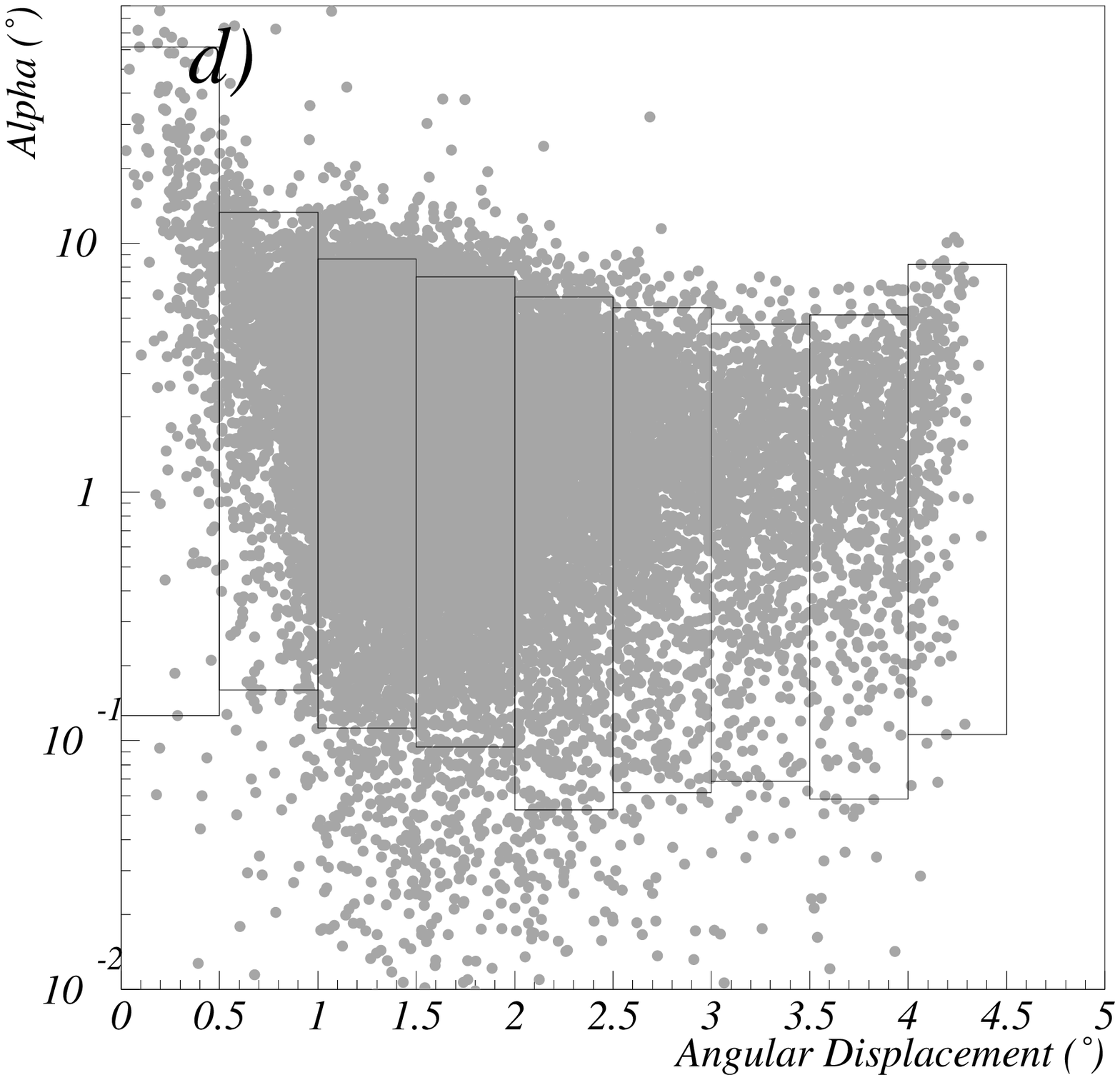,width=.45\textwidth}\\
    \caption{Acceptance intervals for each of four image parameters used in
      this work for background rejection as a function of angular
      displacement, $\mathcal{D}$, for a given range in
      $\log_{10}(\mathcal{S}$). The dots correspond to individual gamma-ray
      images. The width of each box indicates the angular displacement
      interval over which median values are obtained (see text for details),
      while the length of the box corresponds to the 95$\%$ acceptance
      interval about the median.}\label{image_cuts}
  \end{center}
\end{figure}

 In the approach followed here, the goal is to attain good background
rejection without sacrificing collection area for events over a large range of
energies and core distances from the ACT. These latter properties are most
directly related to the parameters $\mathcal{S}$ and $\mathcal{D}$. Figures
\ref{gh_separation1} and \ref{gh_separation2} show the image parameter
distribution as a function of angular displacement ($\mathcal{D}$) for
gamma-ray and background events respectively. Hence, simulated gamma-ray
images were first categorised into 10 intervals of $\mathcal{D}$ (each
spanning 0.5$^\circ$) and 20 logarithmic intervals in $\mathcal{S}$ (each
spanning 0.5 in log10($\mathcal{S}$)). For each division defined by the
resulting 2-D matrix, symmetric acceptance intervals about the median of each
of the remaining image parameters were defined so as to arbitrarily contain
95\% of the gamma-ray images. Figure \ref{image_cuts} illustrates the
definition of image selection for events within a given log10($\mathcal{S}$)
interval. The intervals for the parameters $\mathcal{L}$ and $\mathcal{W}$
define images which have {\em shapes} typical of gamma-ray images, whereas the
intervals corresponding to $\alpha$ and $\mathcal{T}_S$ define images whose
{\em orientation} is characteristic of a primary which came from the assumed
source position. In particular, $\alpha$ is related to the angular resolution
and $\mathcal{T}_S$ characterises the extent to which the value of
$\mathcal{D}$ is due to the displacement of the shower core as opposed to a
nearby shower with a larger relative inclination angle which, therefore, is
not associated with the source.  In addition, $\mathcal{T}_S$ was required to
be positive in order to remove low energy events landing very near the
detector for which the value of $\mathcal{T}_S$ is poorly defined. Table
\ref{cuts} shows the efficiency of each one of the acceptance cuts
individually applied to independent simulated gamma-ray and proton spectra.
From this, it can be seen that the introduction of $\mathcal{T}_S$ yields an
extra factor of between 2 and 3 in background rejection.

\subsection{Shower Core and Energy Resolutions}

\par
\begin{figure}[hbt]
  \begin{center}
    \epsfig{file=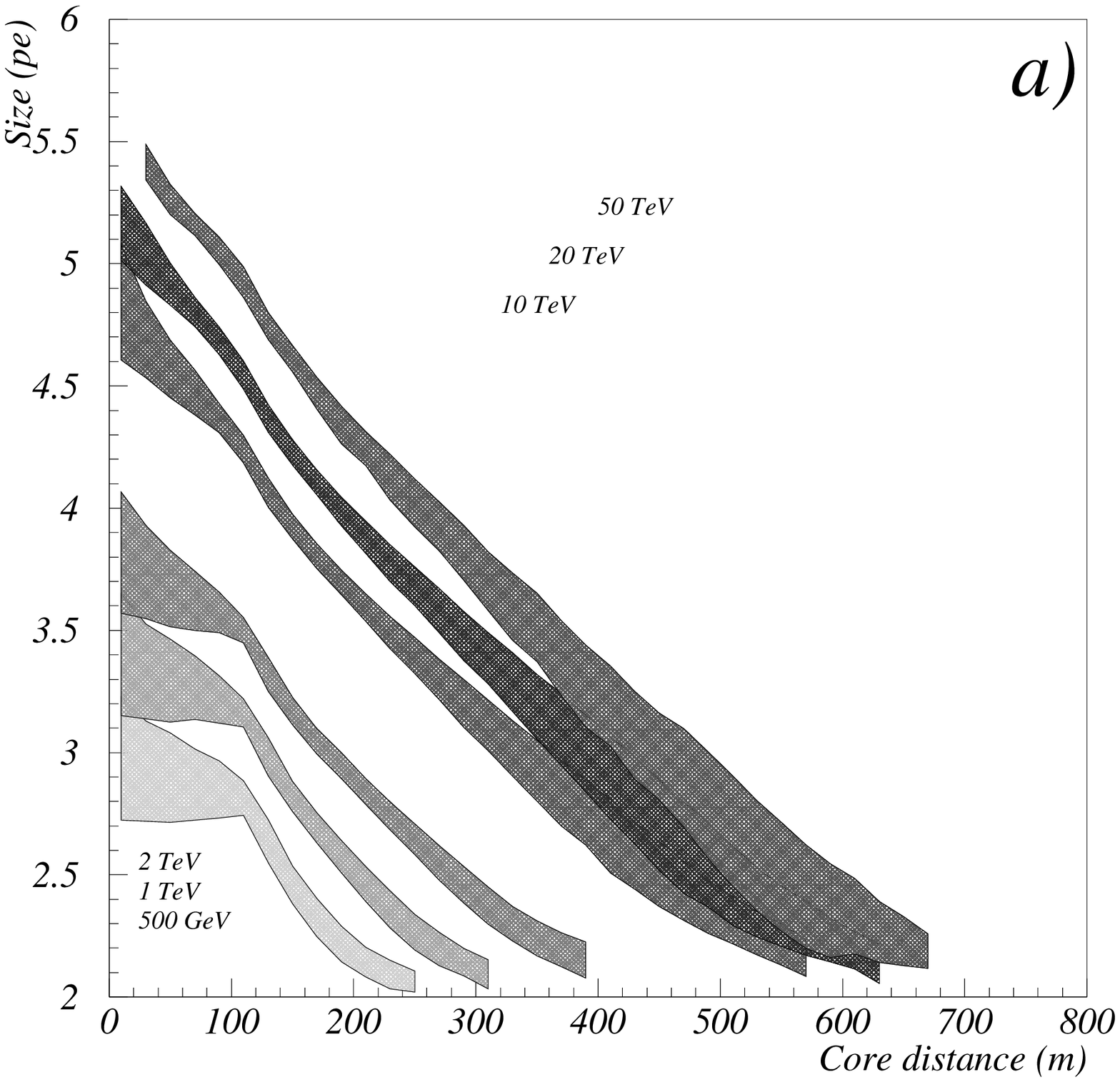,width=.33\textwidth}
    \epsfig{file=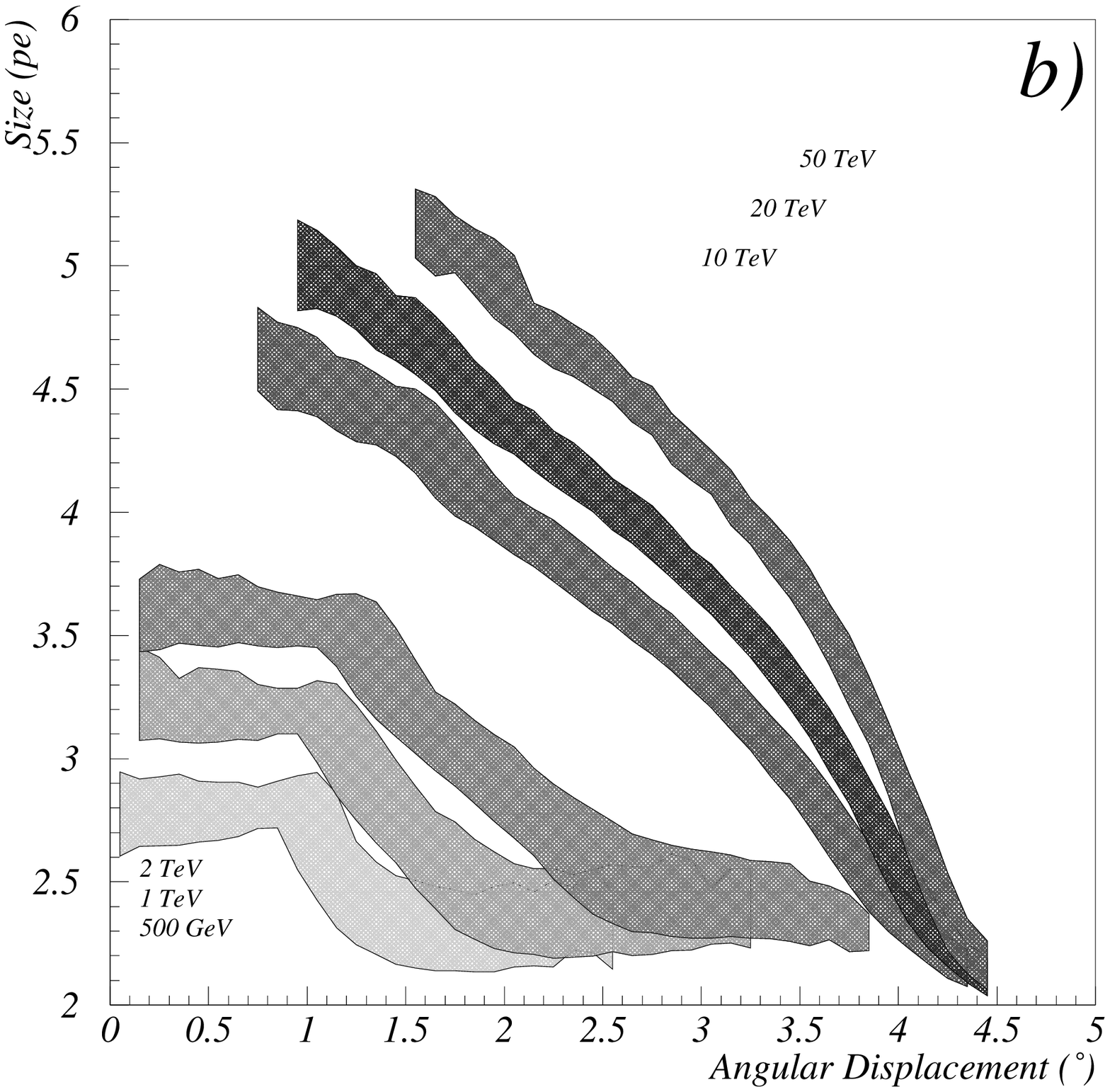,width=.33\textwidth}
    \epsfig{file=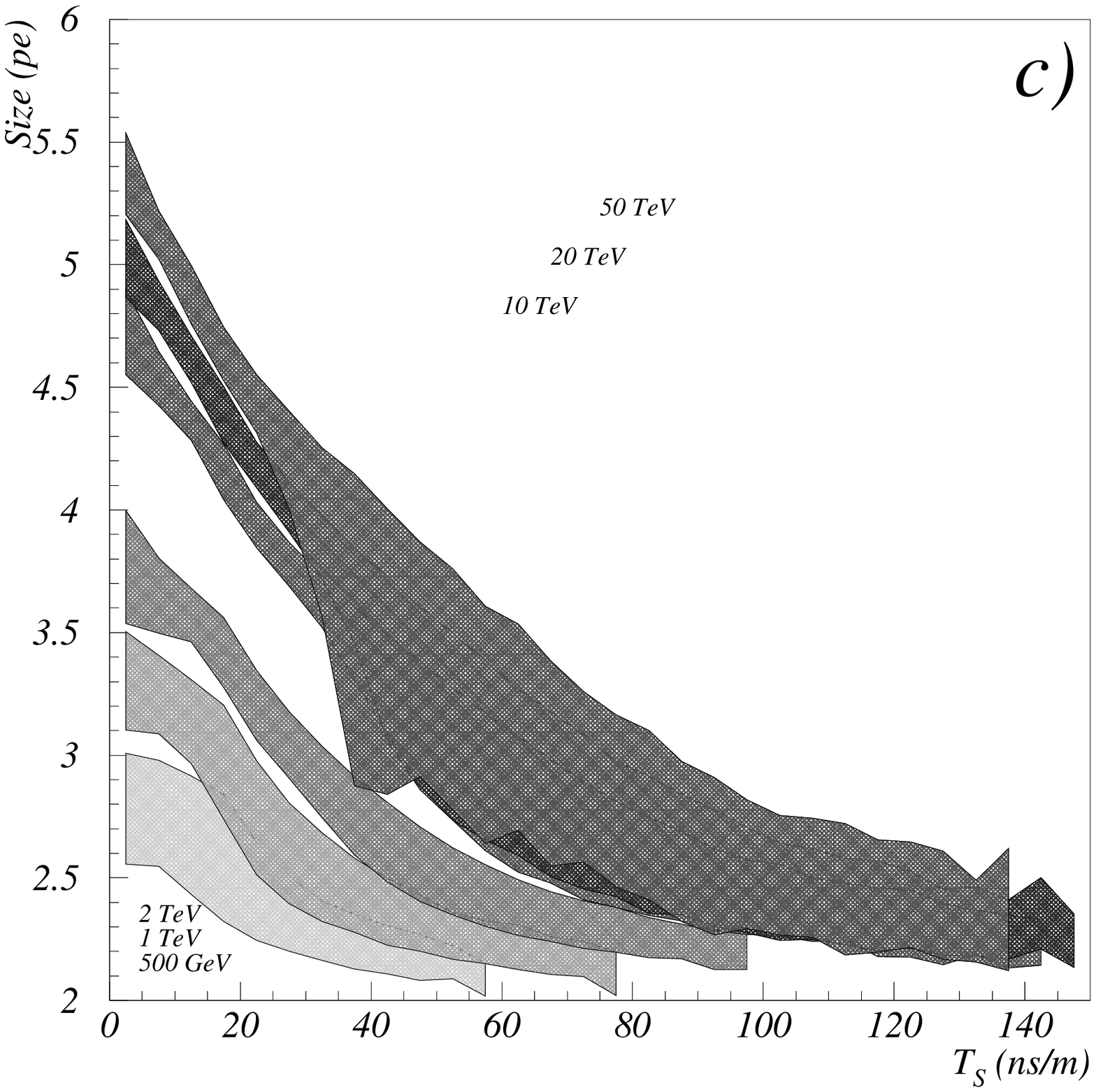,width=.33\textwidth}\\
    \caption{Image size, $\mathcal{S}$, as a function of {\it a)} core
      distance {\it b)} angular displacement $\mathcal{D}$ and {\it c)}
      $\mathcal{T}_S$ for gamma-ray showers of 6 different energies. The
      shaded regions include 68$\%$ of the events about the
      median.}\label{e_resolution}
  \end{center}
\end{figure}

\begin{figure}[hbt]
  \begin{center}
    \epsfig{file=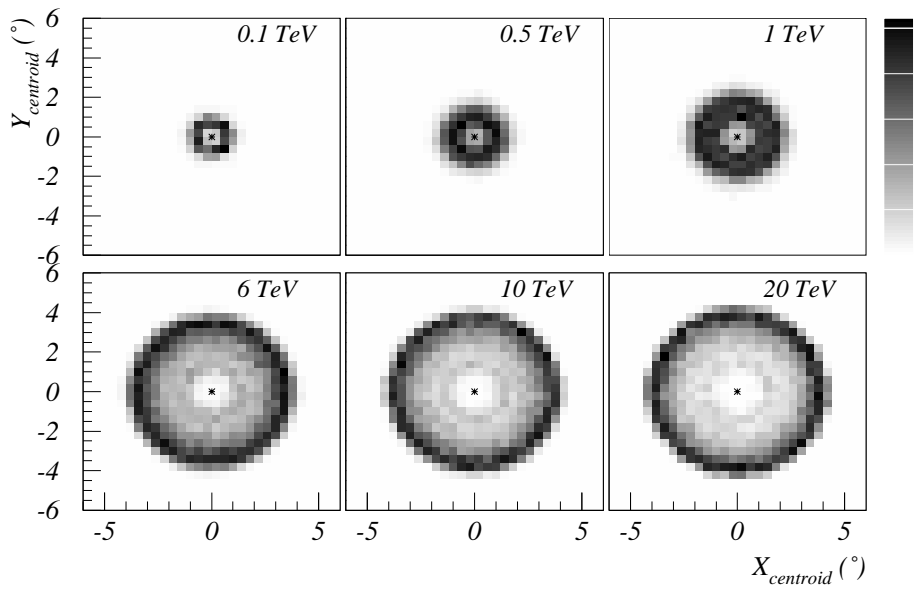,width=.9\textwidth}\\
    \caption{Distribution of image centroids on the camera plane for different
      gamma-ray energy showers.}\label{trigger}
  \end{center}
\end{figure}

 A $\chi^2$ minimisation method was employed to simultaneously reconstruct
the core position and gamma-ray energy of each event so as to explore
resolution properties of the hypothetical ACT. This approach also has the
advantage of being trivially extended for use with multiple telescopes, which
will be explored in section \ref{TandemTelescopeConfiguration}. Three image
characteristics were used in this fit (see figures \ref{e_resolution} and
\ref{trigger}): $\mathcal{D}$, $\log_{10}(\mathcal{S}$) and
$\log_{10}(\mathcal{T}_S / \mathcal{D}$).  The factor of $\mathcal{D}$ in the
latter expression removes the first-order dependence on this parameter and
allows the $\chi^2$ to be better approximated by a linear sum of independent
contributions from each of these characteristics. To define the $\chi^2$
model, simulated gamma-ray events were first categorised into 10~m intervals
of core distance and logarithmic intervals of primary gamma-ray energy using 4
bins per decade. The model was extended in energy, with respect to our
standard data set (table \ref{tab:single}), down to 30~GeV and up to 55~TeV to
reduce reconstruction biases for events near the boundaries of this set. For
each division of this 2-D matrix, the
median value of each image characteristic was determined, along with
``equivalent'' upper and lower one standard deviation values, defined as those
values which bound $\pm$34\% of images about this median. The $\chi^2$
function was then defined as:

\[\chi^2 \equiv \sum_{i=1}^{3} 
\left[\frac{m_i - m_i^0(E,R)}{\sigma_i^0(E,R)}\right]^2 \]

\noindent where $m_i$ is the measured median value of the $i$th image
characteristic for the event; and $m_i^0(E,R)$ and $\sigma_i^0(E,R)$
are the model predictions for the median and effective standard deviation
for this characteristic, respectively, as a function of primary energy, $E$, 
and distance from the shower core, $R$. Appropriate $\sigma_i^0(E,R)$ values
were used depending on whether the $m_i$ was above or below $m_i^0(E,R)$

\par
\begin{figure}[hbt]
  \begin{center}
    \epsfig{file=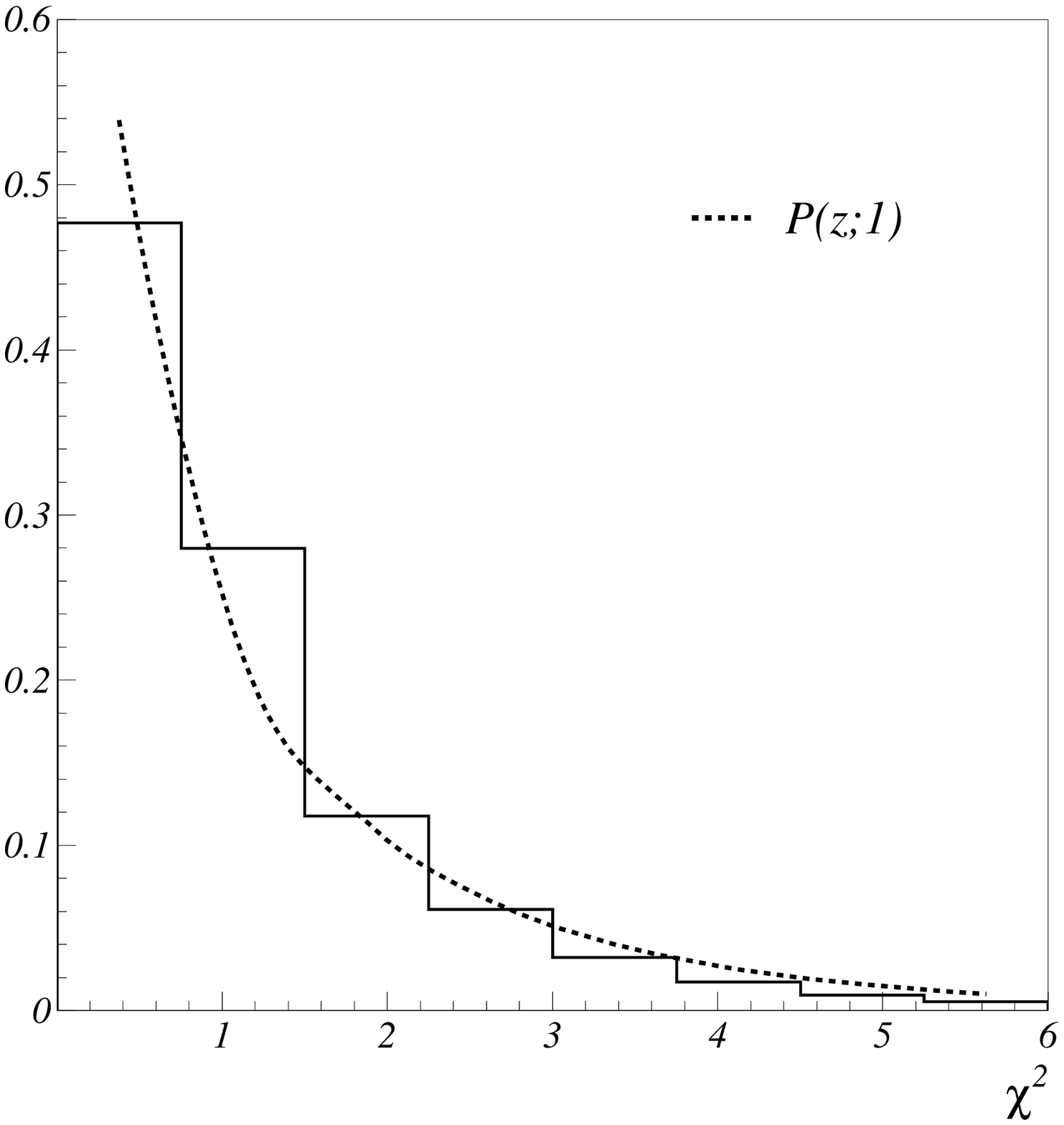,width=.9\textwidth}\\
    \caption{$\chi^2$ distribution of reconstructed single wide-angle telescope
      events. Only events which energy falls in the 300~GeV-20~TeV energy
      range are represented here. The dashed line is the probability density
      function for the $\chi^2$ distribution with 1 degree of
      freedom.}\label{1T_Chi2}
  \end{center}
\end{figure}

\begin{figure}[hbt]
  \begin{center}
    \epsfig{file=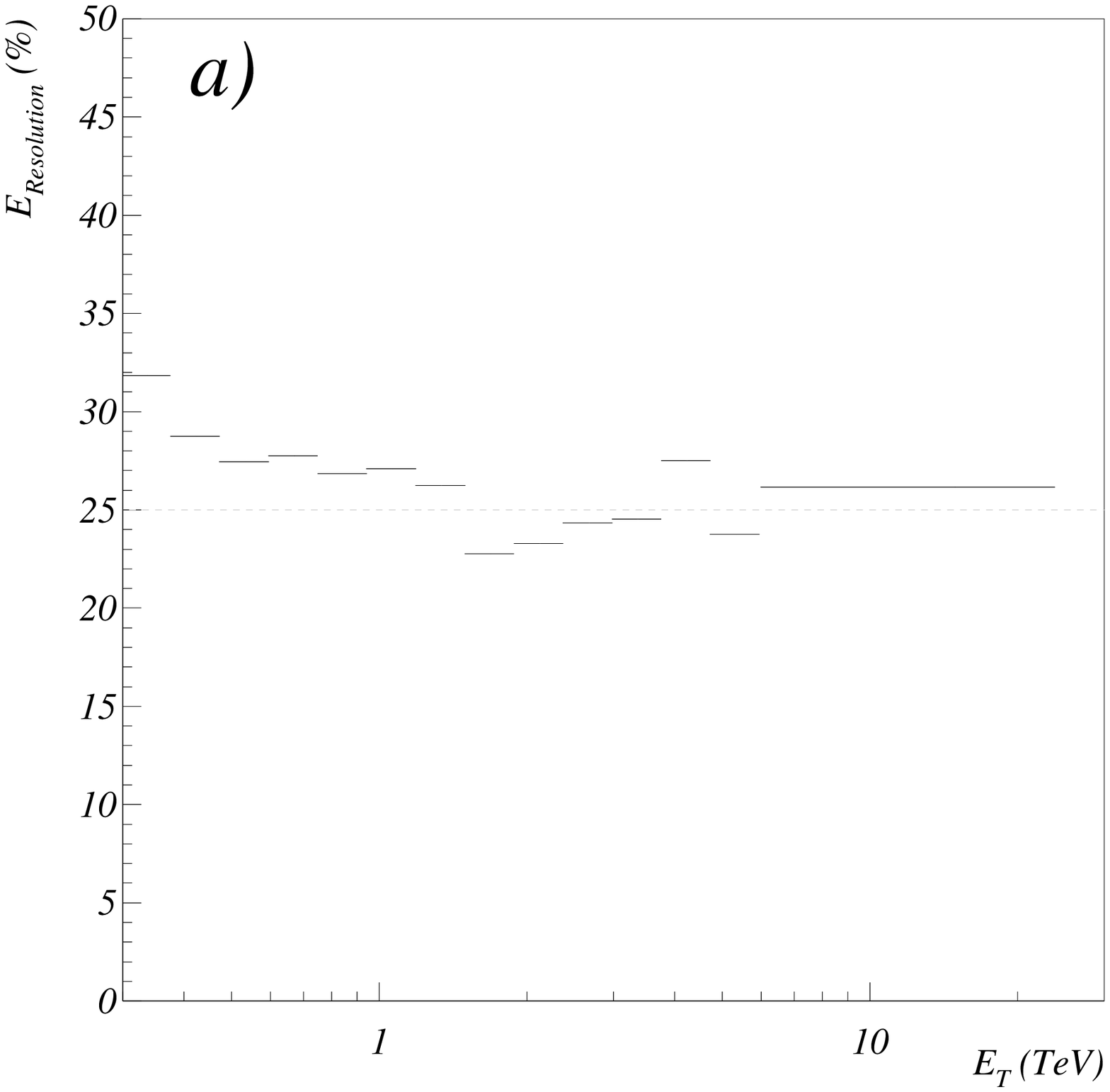,width=.33\textwidth}
    \epsfig{file=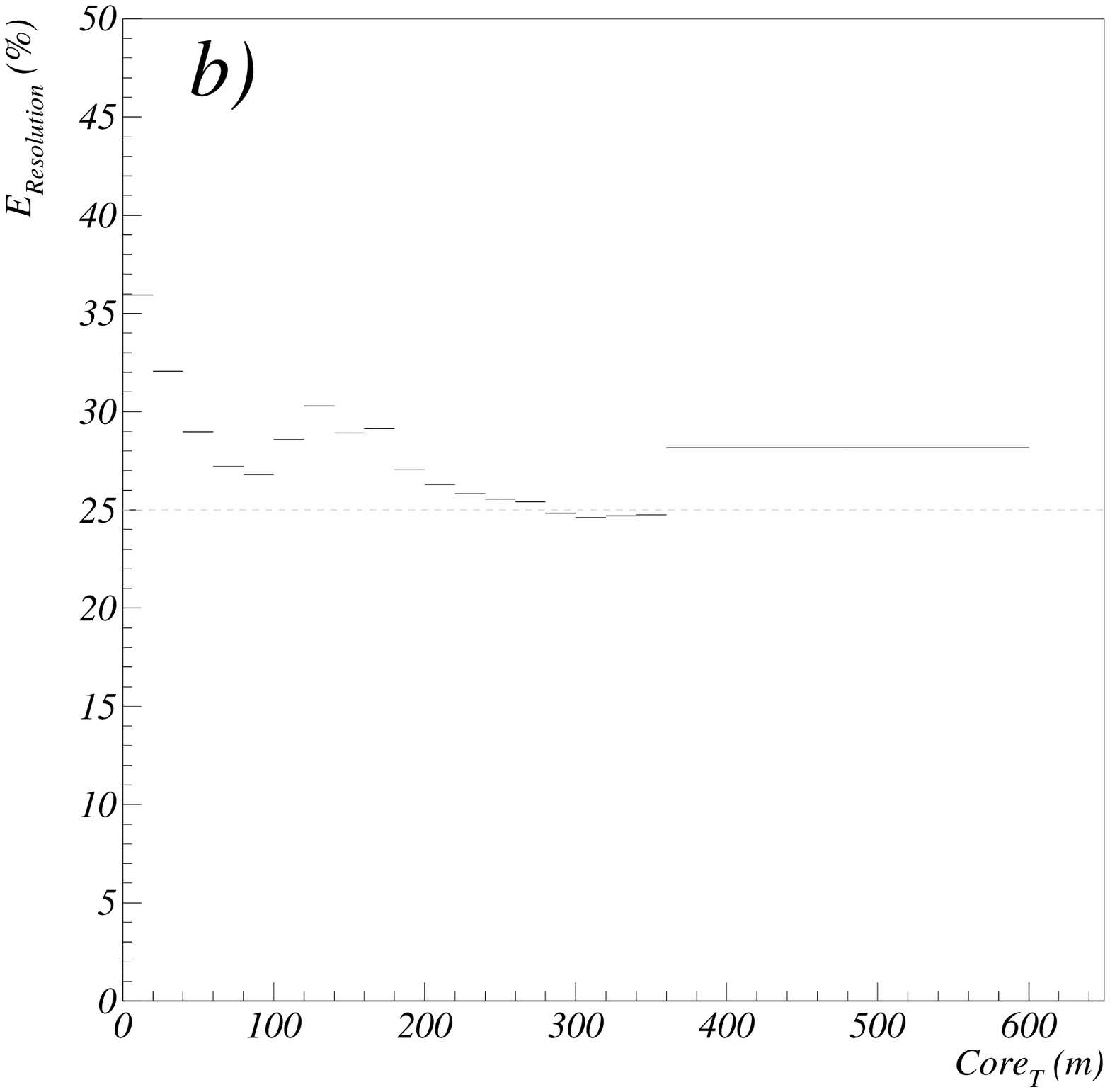,width=.33\textwidth}
    \epsfig{file=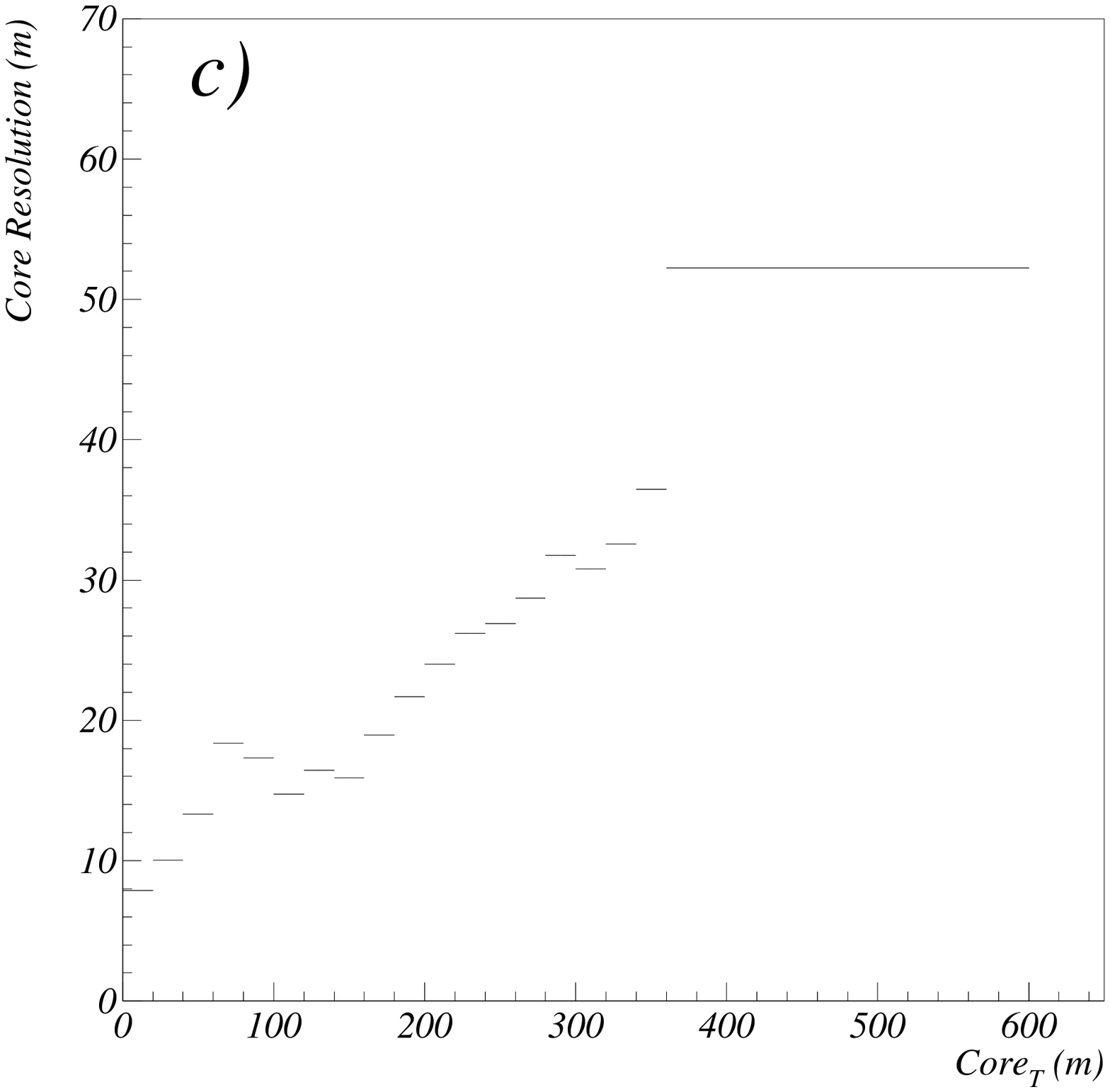,width=.33\textwidth}\\
    \caption{RMS energy and Core resolution as a function of true energy
      (E$_T$) and true core distance (Core$_T$) for a single wide-angle
      telescope.}\label{Ener_rec_1T}
  \end{center}
\end{figure}

 Using an independent simulated data set (300~GeV-20~TeV) from that used
to generate the model (30~GeV-55~TeV), this $\chi^2$ was then minimised with
respect to $E$ and $R$ for each event, using values from the 2-D matrix
described above. Despite the approximations of independent image
characteristics and normal errors, the resulting distribution of $\chi^2$,
shown in figure \ref{1T_Chi2}, is in reasonable agreement with the expected 1
degree of freedom. To remove poorly defined images, those events with values
of the $\chi^2$ probabilities less than 10\% were eliminated from the data set
as well as those with reconstructed energies outside the considered energy
range 300~GeV-20~TeV. Together, these cuts remove $\sim$27\% of the gamma-ray
events that passed image cuts and $\sim$66\% of proton events. Based on this
procedure, figure \ref{Ener_rec_1T} shows the derived core and energy
resolutions as a function of actual primary gamma-ray energy and true core
distance from the ACT. For a single telescope, an energy resolution of
$\sim$25\% is achieved above 1~TeV, essentially independent of the core
distance.

\subsection{Effective Areas}

\par
\begin{figure}[hbt]
  \begin{center}
    \epsfig{file=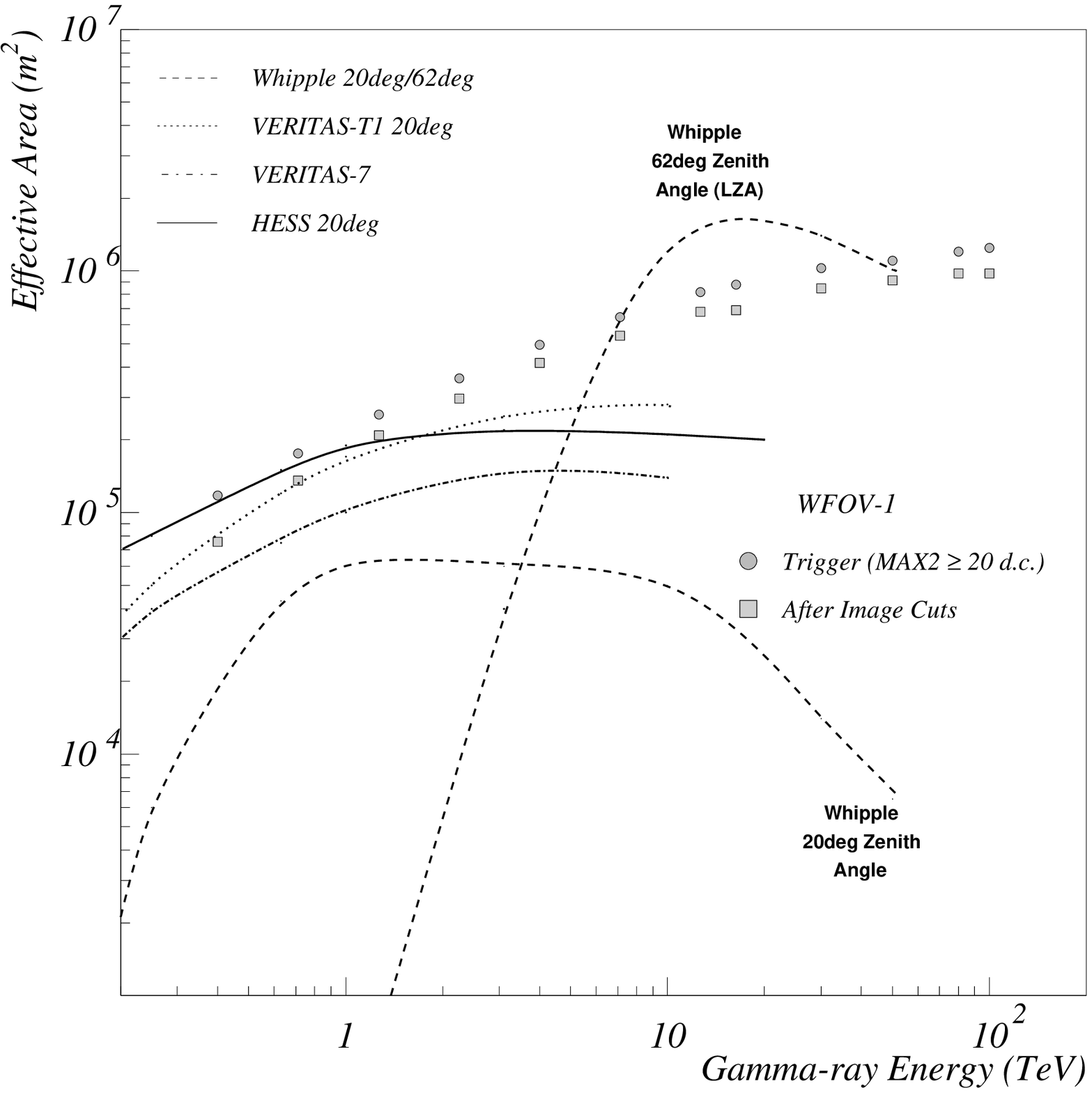,width=.45\textwidth}
    \epsfig{file=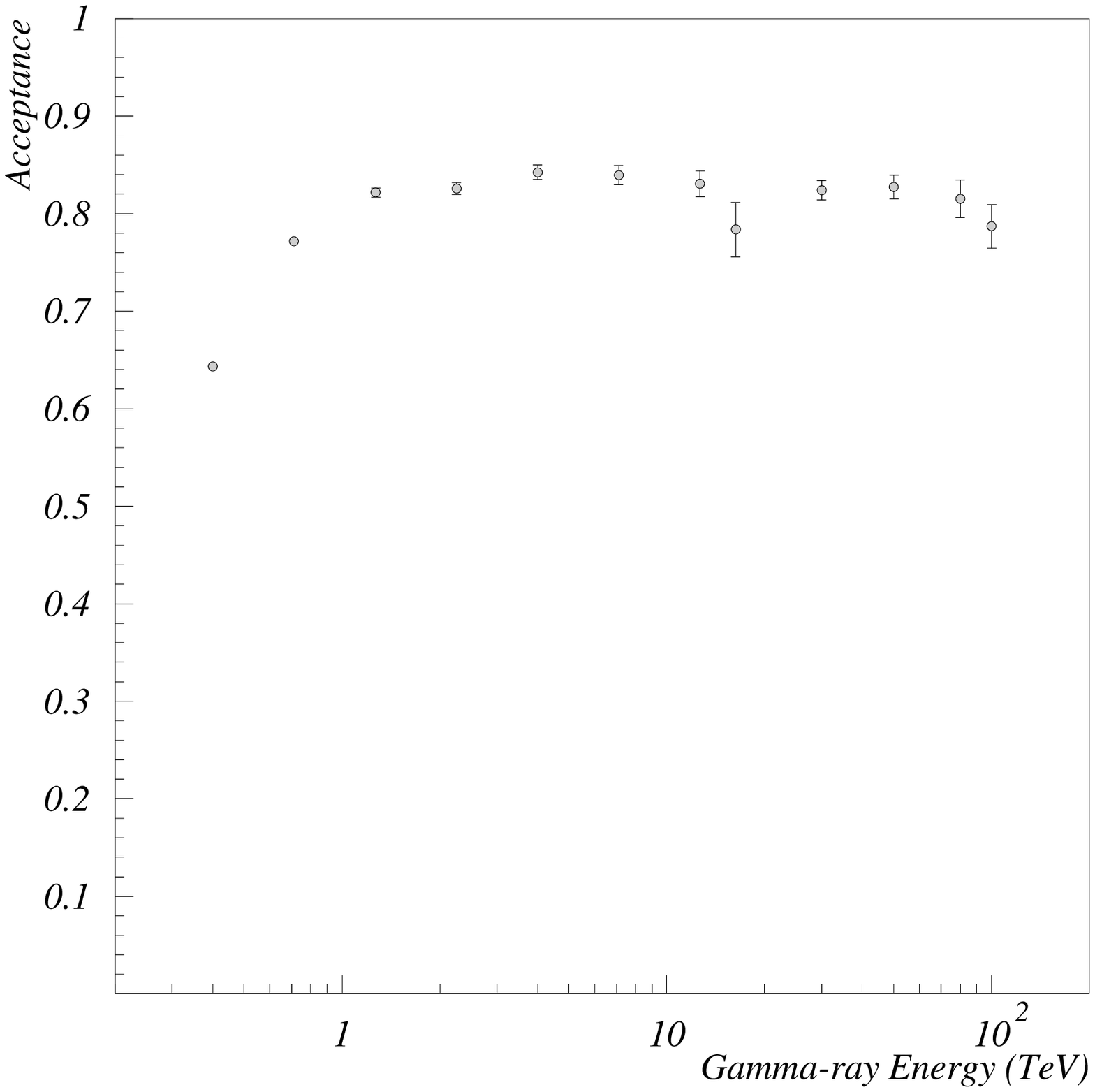,width=.45\textwidth}\\
    \epsfig{file=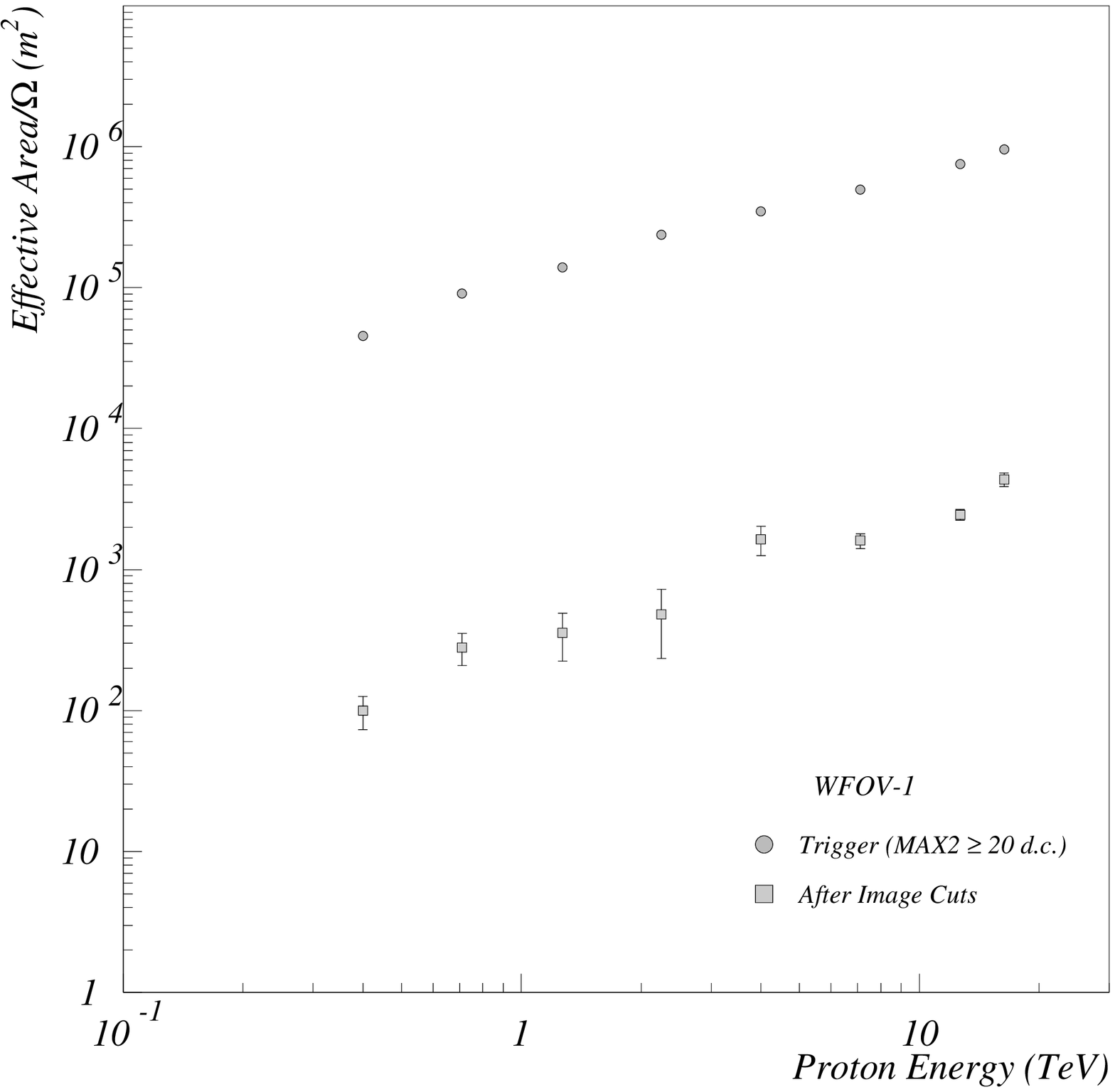,width=.45\textwidth}
    \epsfig{file=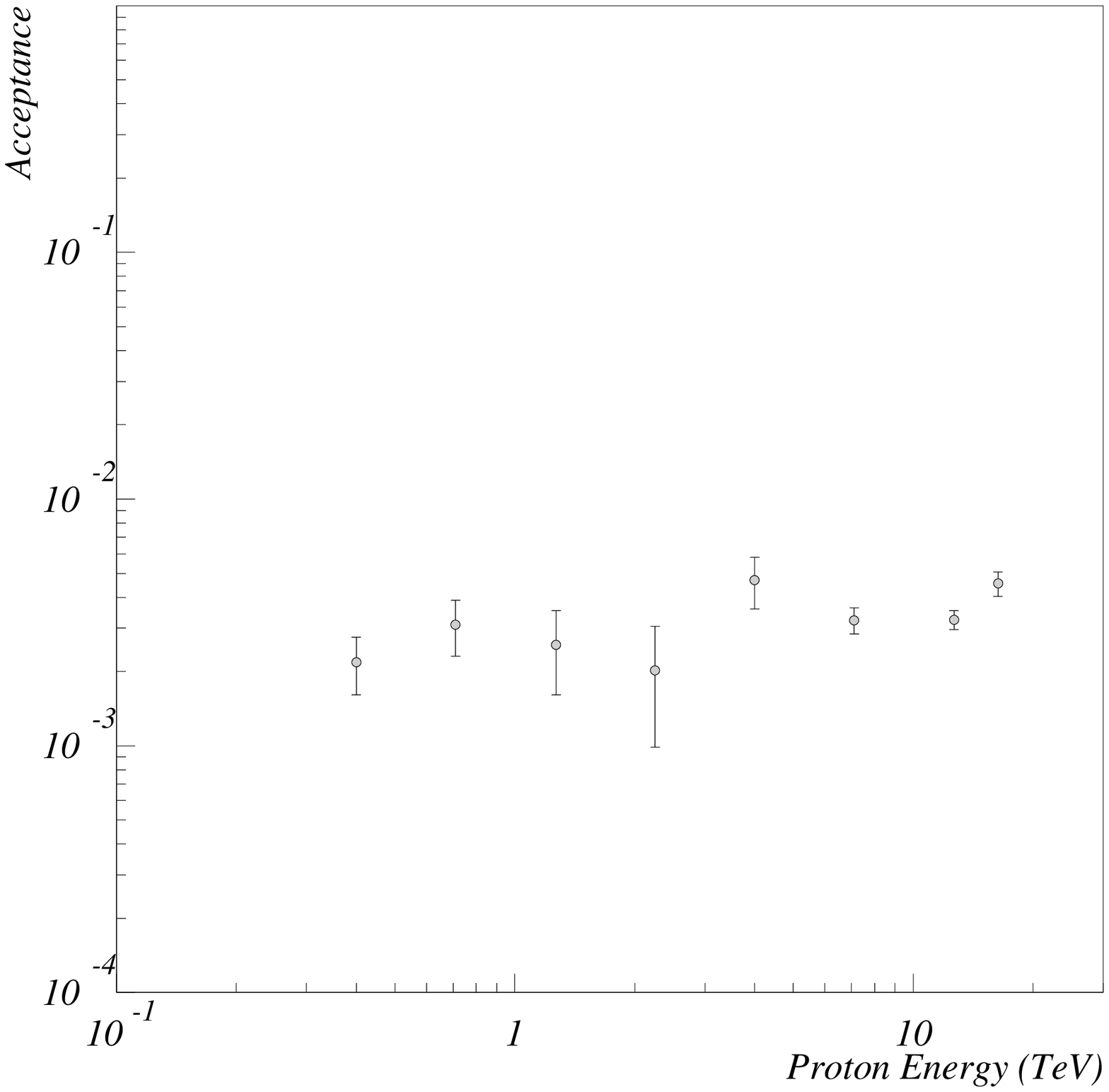,width=.45\textwidth}\\
    \caption{{\it a)} Gamma-ray collection areas, both at the trigger level
      and after image cuts, for a single wide-angle camera (WFOV-1). The
      different lines correspond to effective areas after image cuts for
      Whipple-490 and Whipple-490 at Large Zenith Angle (LZA) \cite{Petry2001}
      (dashed line), HESS \cite{Benbow2005} (continuous line) and VERITAS-T1
      \cite{Maier2005} (dotted line), VERITAS-7 \cite{Weekes2002}
      (dotted-dashed line). {\it b)} Gamma-ray acceptance as a function of
      energy. {\it c)} Background collection areas, both at the trigger level
      and after image cuts, for a single wide-angle camera (WFOV-1). The
      collection area for protons has been normalised to the solid angle
      subtended by the camera. {\it d)} Background acceptance as a function of
      energy.}\label{EffecArea}
  \end{center}
\end{figure}

 The ``effective area'' for a given class of events was computed as the
projected area on the ground over which simulated core locations are uniformly
sampled with respect to the ACT, times the fraction of such events which are
retained in the data sample. Figure \ref{EffecArea} shows the derived
effective areas for gamma-ray and proton showers as a function of the true
primary energy for triggered showers both before and after image selection. As
expected, approximately $(0.95)^4 \sim $80\% of gamma-ray images are retained
after selection based on the 4 image parameters. The effective collection area
for gamma-rays after this selection is $\sim 2\times 10^5$~m$^2$ at primary
energies of 1~TeV and exceeds $\sim 5\times 10^5$~m$^2$ above 5~TeV. This is
roughly an order of magnitude larger than a conventional, single telescope
and about twice that of the HESS or VERITAS arrays. At higher energies,
the relative gain in effective area is even greater.

\par For proton-induced showers, the effective collection areas are suppressed
by roughly a factor of $10^3$ following image selection.  However, it should
be noted that background rejection factors cannot be directly implied from
these figures since the true primary energy is not an observable. Thus, the
resolution of ``equivalent'' gamma-ray energies and their effect on assumed
background and source spectra must be taken into consideration.

\par The raw trigger rate from background showers based on an assumed proton
flux of $\Phi_p(E) = 8.73 \times 10^{-2}
E_{TeV}^{-2.7}$~m$^{-2}$~s$^{-1}$~sr$^{-1}$~TeV$^{-1}$ \cite{Horandel2003}, is
found to be $\sim 1~kHz$, using the basic trigger scheme assumed here. A simple
pattern trigger can further reduce this by a factor of $\sim$2
\cite{Bradbury1999}, making it easily manageable by relatively standard data
acquisition electronics (see e.g. \cite{Funk2004}).

\subsection{Single Telescope Flux Sensitivity}

\par
\begin{figure}[hbt]
  \begin{center}
    \epsfig{file=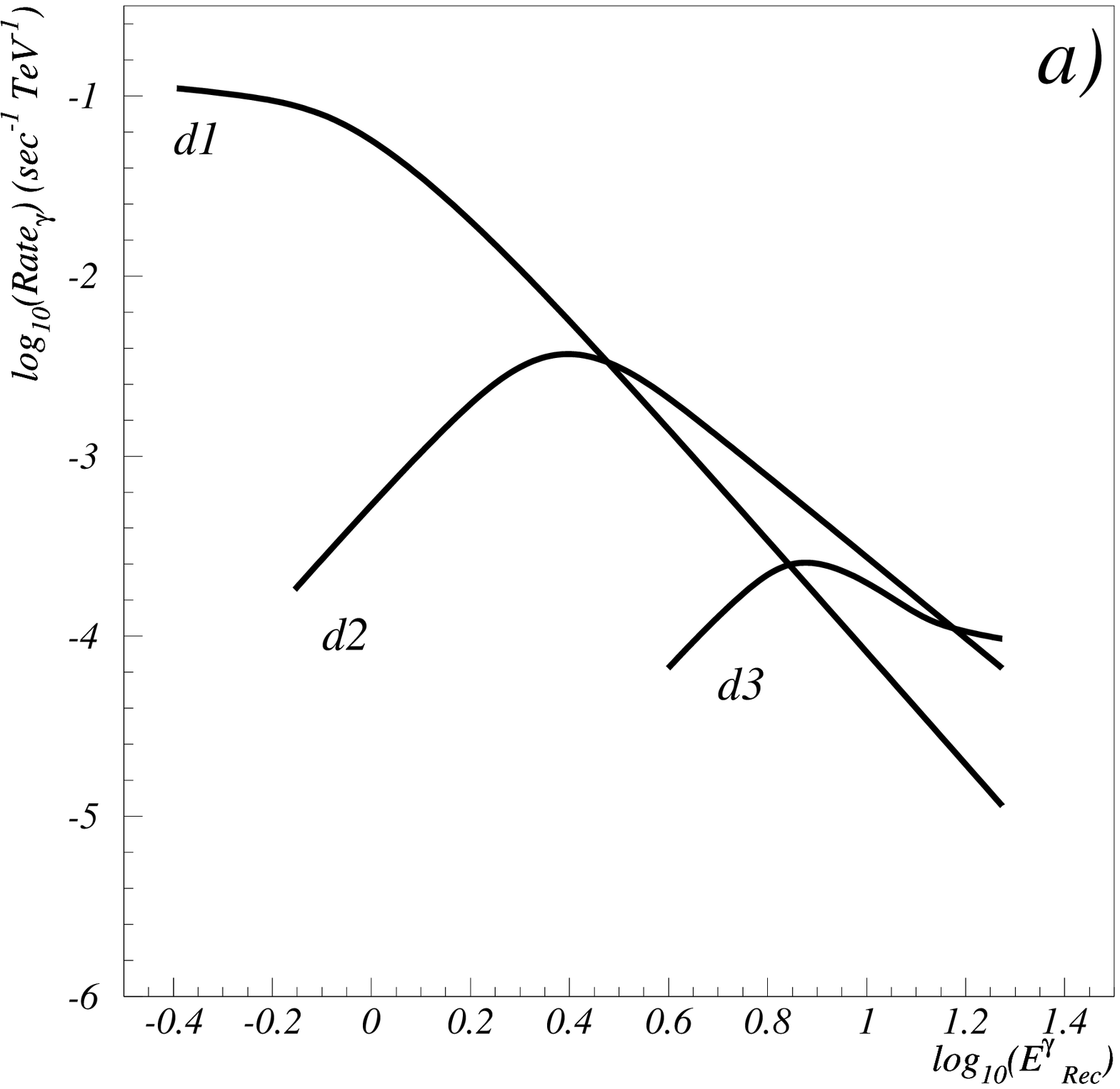,width=.45\textwidth}
    \epsfig{file=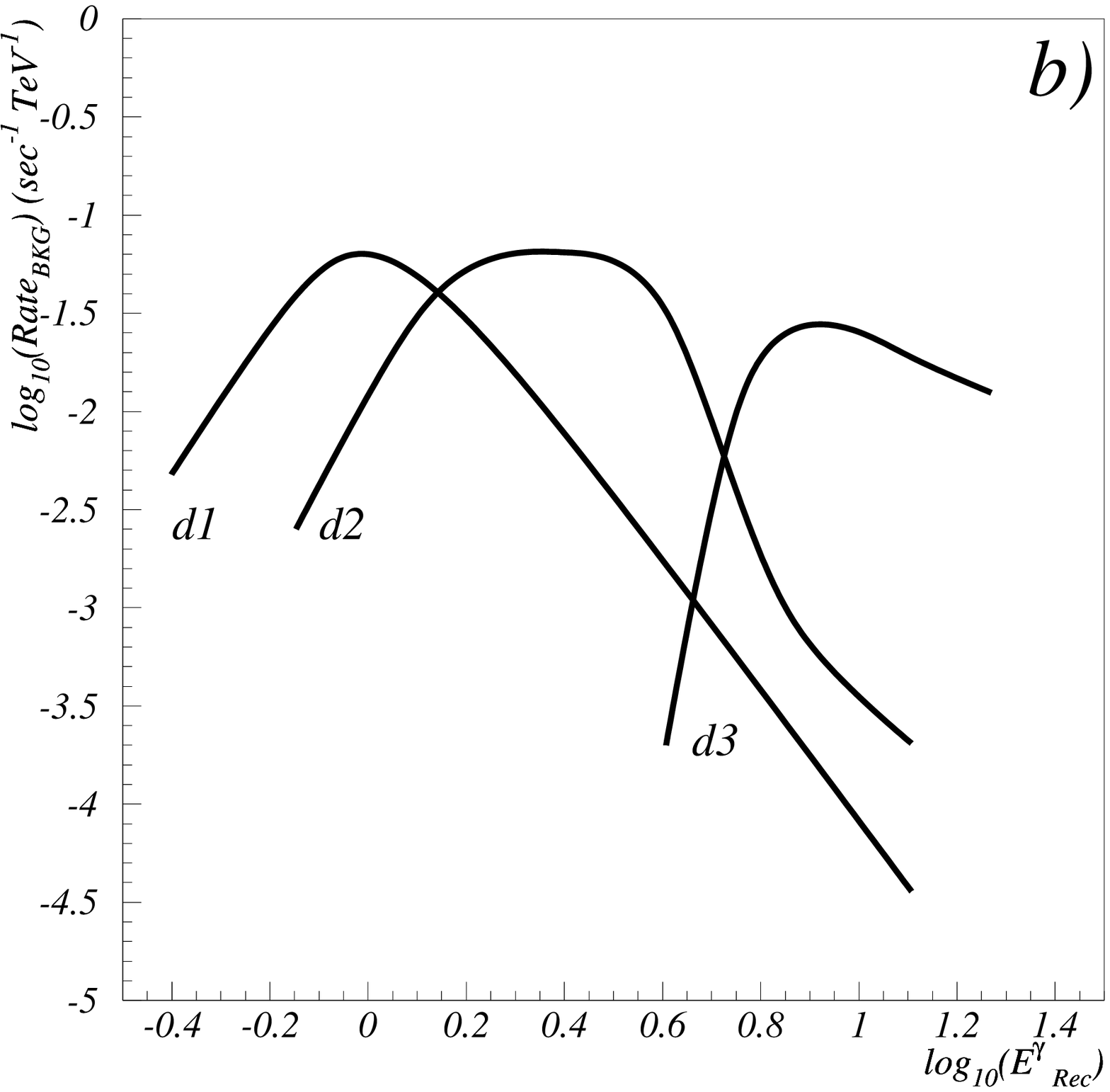,width=.45\textwidth}\\
    \caption{Differential rates for {\it a)} gamma-ray and {\it b)} background
      events as a function of reconstructed gamma-ray energy,
      E$^\gamma_{Rec}$. {\it d1}, {\it d2} and {\it d3} refer to the three
      angular displacement intervals defined in the text.}\label{WFOV_rec_rat}
  \end{center}
\end{figure}

 The image selection and energy reconstruction procedures previously
described were applied to simulated proton and gamma-ray spectra sampled from
differential power laws of the form $E^{-2.7}$ and $E^{-2.5}$, respectively.
In order to couch the derived quantities in terms of typical detection rates,
the simulated spectra were initially normalised according to $\Phi_p(E) = 8.73
\times 10^{-2} E_{TeV}^{-2.7}$ ~m$^{-2}$~s$^{-1}$~sr$^{-1}$~TeV$^{-1}$, based
on cosmic ray primary measurements \cite{Horandel2003}; and $J_\gamma(E) = 3.2
\times 10^{-7} E_{TeV}^{-2.5}$ ~m$^{-2}$~s$^{-1}$~TeV$^{-1}$, based on the
measured high energy flux from the Crab Nebula near 1~TeV
\cite{Hillas1998}. These were then convolved with the relevant effective
collection areas and used to yield differential detection rates in terms of
the reconstructed, effective gamma-ray energy, $E^{\gamma}_{Rec}$. These are
shown in figure \ref{WFOV_rec_rat}, for different ranges of $\mathcal{D}$.

\par These figures may be used to derive the detector ``sensitivity'' for such
a source spectrum by first integrating the rates over a chosen time scale,
then scaling the assumed gamma-ray flux until a given minimum significance
level for detection over some defined effective energy range is reached. For
the purposes of this paper, the sensitivity will be defined based on the
integrated signal above a given effective energy threshold which yields a
detection at a significance level of 5$\sigma$ and a minimum of 10 detected
high-energy gamma-rays. It will also be assumed that the nominal background
rate is well known. In practise, uncertainties in this background rate will
reduce the significance level to some extent, so that the curves derived here
may be more indicative of detection levels corresponding to $\sim$3-4$\sigma$.

\par Particularly for this type of camera, background levels will be highly
dependent on both the reconstructed gamma-ray energy and the apparent core
distance. To account for this, the average signal and background levels were
first computed separately for the 3 regions of $\mathcal{D}$ shown in figure
\ref{WFOV_rec_rat} (0-2.6$^\circ$, 2.6$^\circ$-3.7$^\circ$ and
3.7$^\circ$-4.5$^\circ$) and in each of 8 logarithmic energy intervals between
300~GeV and 20~TeV. A maximum likelihood ratio, $L_r$, was then constructed
relative to the zero-signal case. $-2log L_r$ is approximately distributed as
a $\chi^2$ distribution \cite{Wilks1938}, from which the average detection
significance was calculated assuming that three variable parameters of signal
strength, power-law index and spectral cut-off may be fit to the data. This
approach appropriately weights the contributions from the different
signal/background regimes.

\par
\begin{figure}[hbt]
  \begin{center}
    \epsfig{file=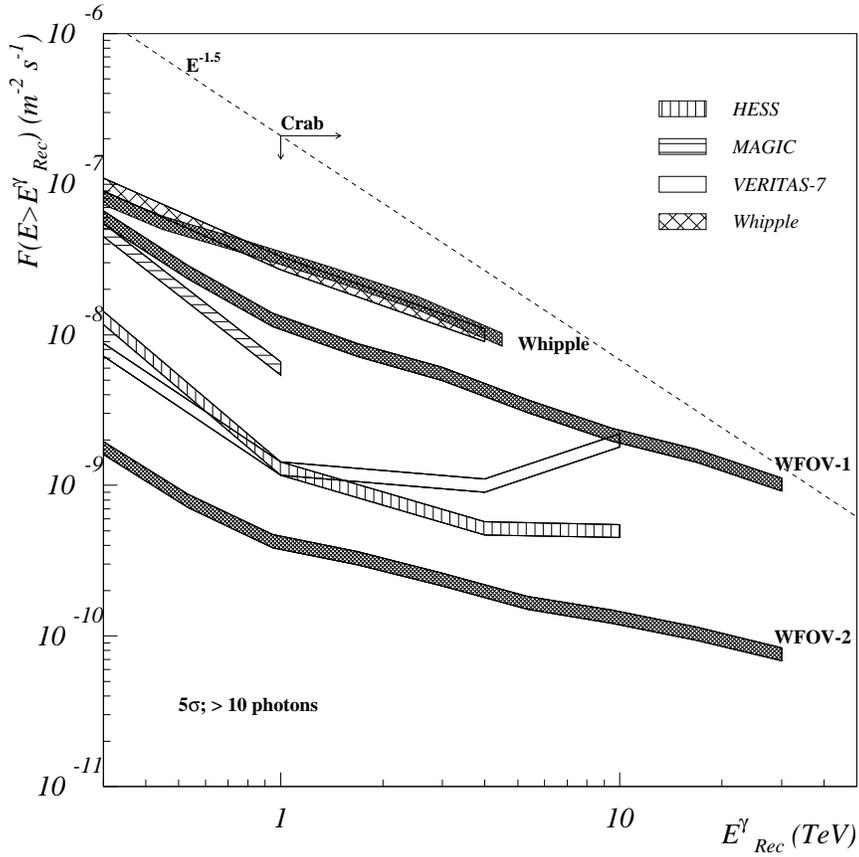,width=.9\textwidth}\\
    \caption{Integral sensitivity curve for a 50~h exposure for the detectors
      simulated in this work assuming an integral $E^{-1.5}$ source
      spectrum. WFOV-1 refers to a single detector while WFOV-2 refers to the
      tandem telescope configuration described in the text. Also shown is the
      sensitivity curve for a Whipple-like detector derived using the same
      simulation software used to explore the wide-angle telescope considered
      in this work. These are compared with the stated sensitivity curves for
      other experiments from \cite{Weekes2002} and
      \cite{Hofmann2001}.}\label{50h_sen}
  \end{center}
\end{figure}

\begin{figure}[hbt]
  \begin{center}
    \epsfig{file=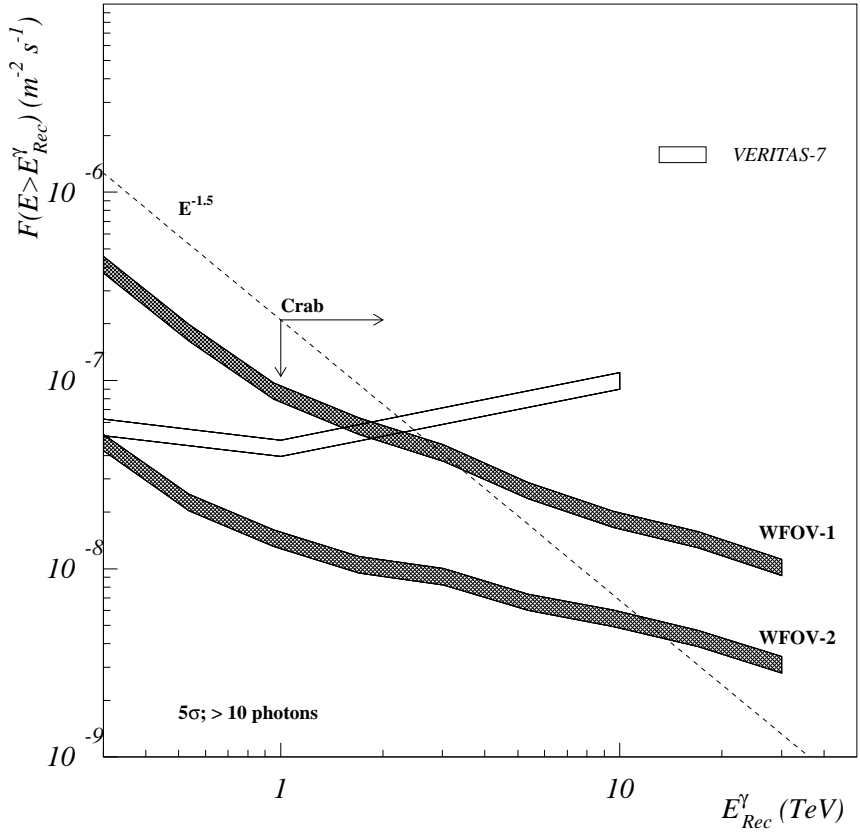,width=.9\textwidth}\\
    \caption{Scaled 1~h integral sensitivity curves for the detector simulated
      in this work assuming an integral $E^{-1.5}$ source spectrum. WFOV-1 
      refers to a single detector while WFOV-2 refers to the tandem telescope
      configuration described in the text. Also shown is the expected response
      of VERITAS-7 for a 1~h exposure. }\label{1h_sen}
  \end{center}
\end{figure}

 The resulting sensitivity curves as a function of
$E^{\gamma}_{Rec}$ are shown in figures \ref{50h_sen} and \ref{1h_sen} for
exposure times of 50 hours and 1 hour, respectively. These are compared with
similar curves for the Whipple, MAGIC, VERITAS and HESS instruments, taken
from \cite{Weekes2002} and \cite{Hofmann2001}. To verify the consistency of
the methods used to obtain the following results, the sensitivity of the
Whipple telescope was also independently estimated for 50 hours exposure time
using the same simulation software and similar analysis procedure as employed
for the wide-angle camera study. As can be seen in figure \ref{50h_sen}, the
results are in excellent agreement with published values.

\par For continuous sources based on 50 hours of exposure, the wide-angle
camera is found to be approximately a factor of two more sensitive than the
Whipple instrument, even at energies as low as 300~GeV. This stems from the
fact that, even at these low energies, the collection area for the larger
field of view is roughly a factor of two larger than the Whipple instrument
and the $\mathcal{T}_S$ selection (made more effective by the larger field of
view) suppresses the background by an addition factor of $\sim$2. As the
sensitivity is proportional to the square root of both these factors, the
overall improvement in sensitivity by a factor of two is as expected.

\par For this same exposure, the device becomes comparable to VERITAS and HESS
at energies above 10~TeV, despite backgrounds, owing to the greatly increased
collection area. For studies based on 1 hour of exposure (as might be
characteristic of transient emission from AGN), the wide-angle camera
surpasses the sensitivity of the arrays above 2-3~TeV, despite being only a
single telescope with a $\sim$30\% smaller mirror area than each element of
the arrays. This is due to the fact that, with it's much larger collection
area, the wide-angle camera tends to be background dominated rather than
statistics limited. Thus, any decrease in the relative background level
(whether through improved background rejection or higher source fluxes over
shorter exposure times) results in greater improvement relative to
conventional cameras, which become starved for signal above $\sim$1~TeV (as
indicated by the upwards turn in their sensitivity curves in figure
\ref{1h_sen}).

\section{Tandem Telescope Configuration}\label{TandemTelescopeConfiguration}

\par
\begin{figure}[hbt]
  \begin{center}
    \epsfig{file=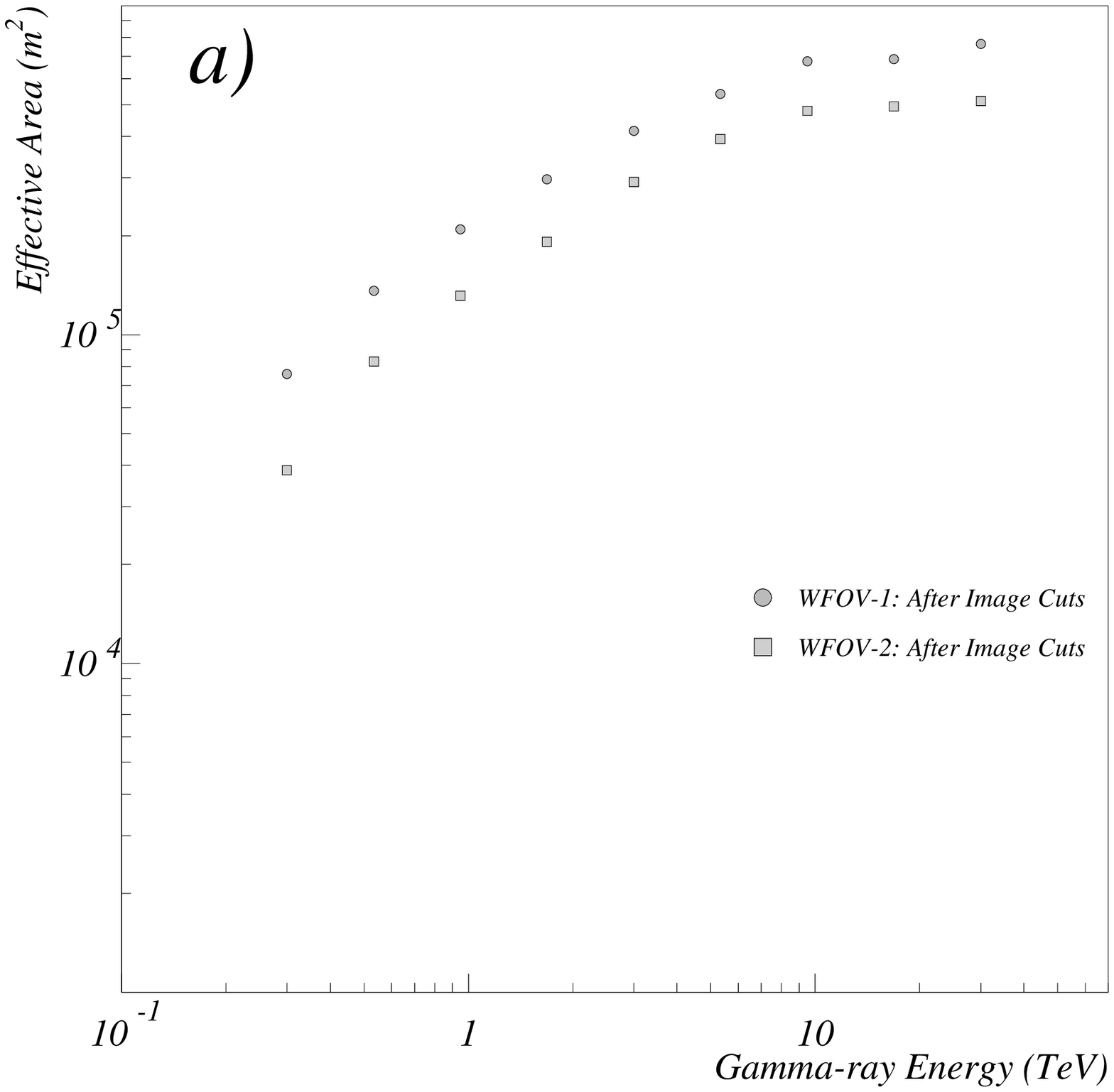,width=.45\textwidth}
    \epsfig{file=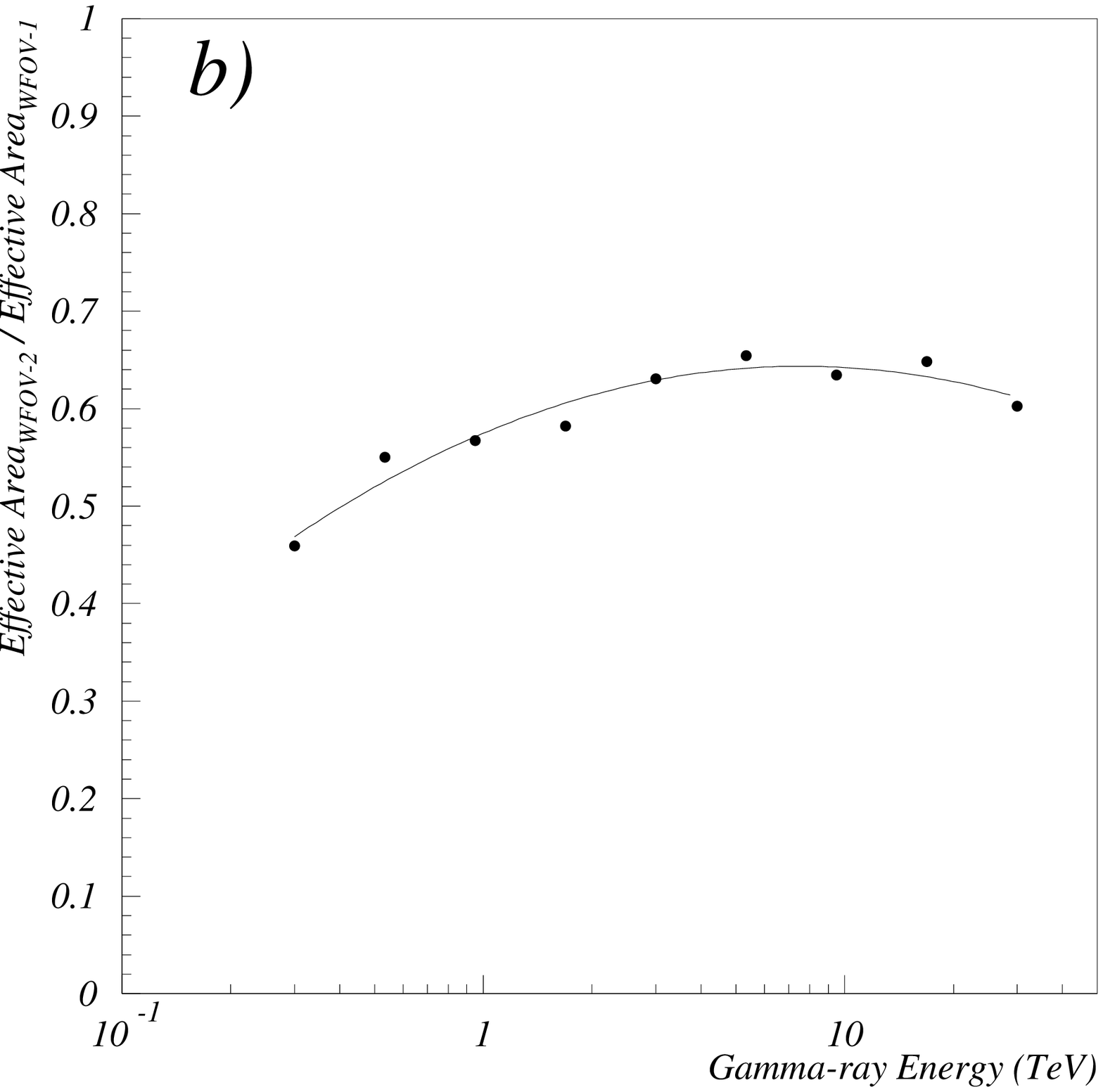,width=.45\textwidth}\\
    \caption{{\it a)} Comparison of the gamma-ray collection areas (after
      image cuts) of a single wide-angle camera detector (WFOV-1) and the
      tandem telescope configuration (WFOV-2) simulated on this work. {\it b)}
      Ratio of these effective areas.}\label{EffecArea_2T}
  \end{center}
\end{figure}

 Background rejection is clearly a key factor to instrumental sensitivity.
This is particularly true for the wide-field concept, which seeks to extend
the dynamic range by essentially swapping a statistically limited regime for
one which is more background limited, as previously mentioned. One obvious way
to gain further background rejection and also improved energy resolution is
with the use of multiple telescopes. In order to explore the potential gain in
sensitivity by this approach, we have limited our consideration here to a
tandem configuration in which identical wide-field ACTs are separated by
125~m, which is close to the optimum distance suggested by Aharonian et
al. based on trigger rates \cite{Aharonian1997}. However, note that we have
not specifically optimised this configuration in terms of background rejection
versus effective collection area for the instrument considered here, but have
only chosen this as a ``representative'' configuration to explore the basic
characteristics.  We have also not optimised the trigger for a 2-telescope
system, but have simply imposed the same single telescope trigger and image
analysis/rejection previously described to both cameras (figure
\ref{EffecArea_2T}). The simulated data set used for this study is summarised
in table \ref{tab:tandem}. As with the case of protons in the previous study,
showers were generated over several different energy intervals and joined via
an appropriate weighting to represent full spectra so as to improve the
statistical characterisation in very different energy regimes.

\subsection{Tandem Shower Core and Energy Resolution}

\par
\begin{figure}[hbt]
  \begin{center}
    \epsfig{file=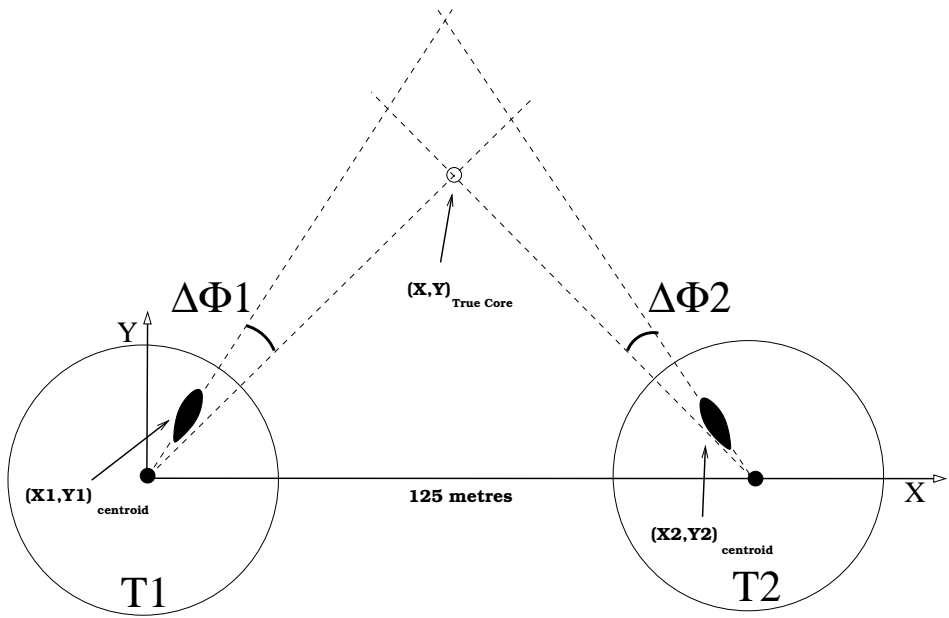,width=.9\textwidth}
    \caption{Illustration of the model parameter $\Delta \Phi$ used for core
      reconstruction with the tandem telescope configuration.}\label{2T_model}
  \end{center}
\end{figure}

 The $\chi^2$ approach for determining primary energy and core location
was modified for the 2-telescope case to simultaneously fit both camera
images.  This was done by fitting for the 2-dimensional core location (as
opposed to simply radius) and introducing a 4th image characteristic, $\Delta\Phi$,
which is sensitive to the angular orientation of the core with respect to the
coordinates of each telescope. This parameter was defined as the angular
deviation between the core direction and the direction of the image centroid,
both relative to the centre of the particular camera (which is assumed to
track the source). This definition is illustrated in figure \ref{2T_model}.
Thus, the $\chi^2$ function for the ``jth'' telescope becomes:

\[\chi^2_j \equiv \sum_{i=1}^{4} 
\left[\frac{m_i - m_i^0(E,R)}{\sigma_i^0(E,R)}\right]^2 \]

\noindent where the sum is over each of the 4 image parameters. The overall
$\chi^2$ function for an array of $n$ telescopes is therefore:

\[\chi^2_{array} = \sum_{j=1}^{n} \chi^2_j \] 

\par
\begin{figure}[hbt]
  \begin{center}
    \epsfig{file=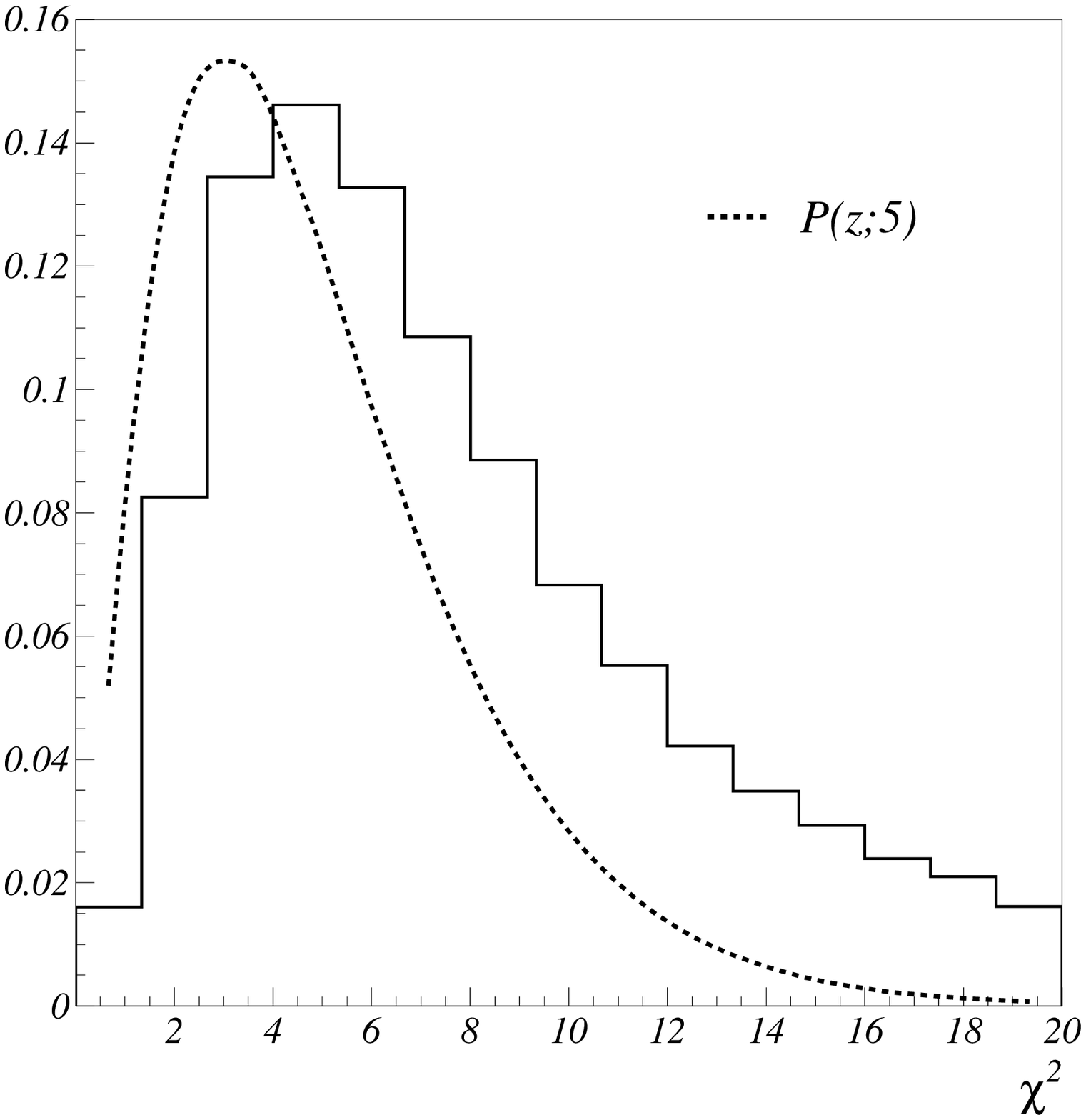,width=.9\textwidth}\\
    \caption{$\chi^2$ distribution of reconstructed tandem-telescope
      events. Only events which energy falls in the 300~GeV-20~TeV energy
      range are represented here. The dashed line is the probability density
      function for the $\chi^2$ distribution with 5 degrees of
      freedom.}\label{2T_Chi2}
  \end{center}
\end{figure}

 Given the three free fit parameters (the 2-D core location plus primary
gamma-ray energy), for the case of 2 telescopes (each involving 4 measured
characteristics), this should nominally correspond to 5 degrees of freedom.
However, this once more assumes that the image characteristics are
completely uncorrelated and that the uncertainties are normally distributed,
neither of which is strictly true. The actual $\chi^2$ distribution
is compared with the idealised case in figure \ref{2T_Chi2}.

\par
\begin{figure}[hbt]
  \begin{center}
    \epsfig{file=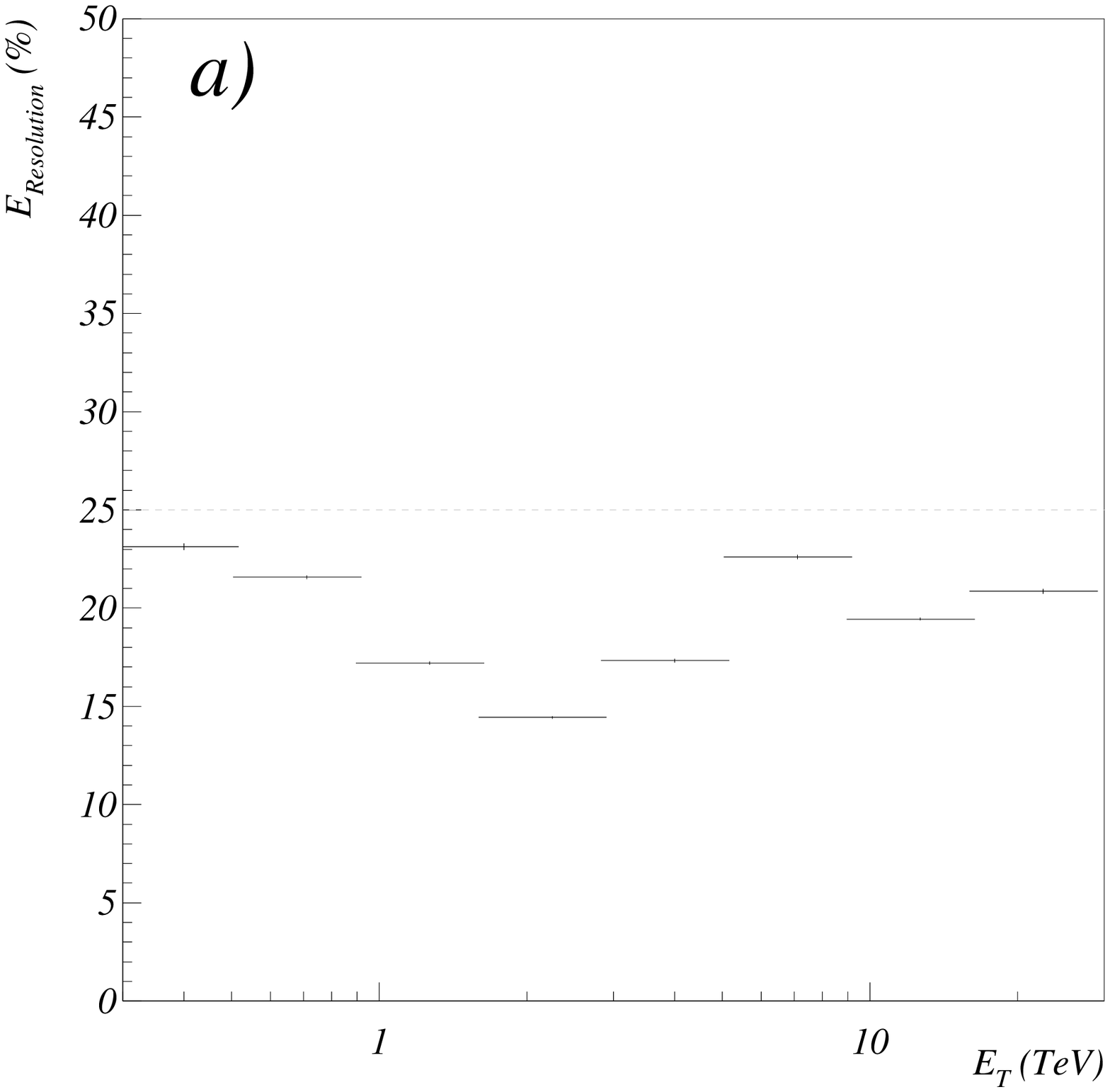,width=.33\textwidth}
    \epsfig{file=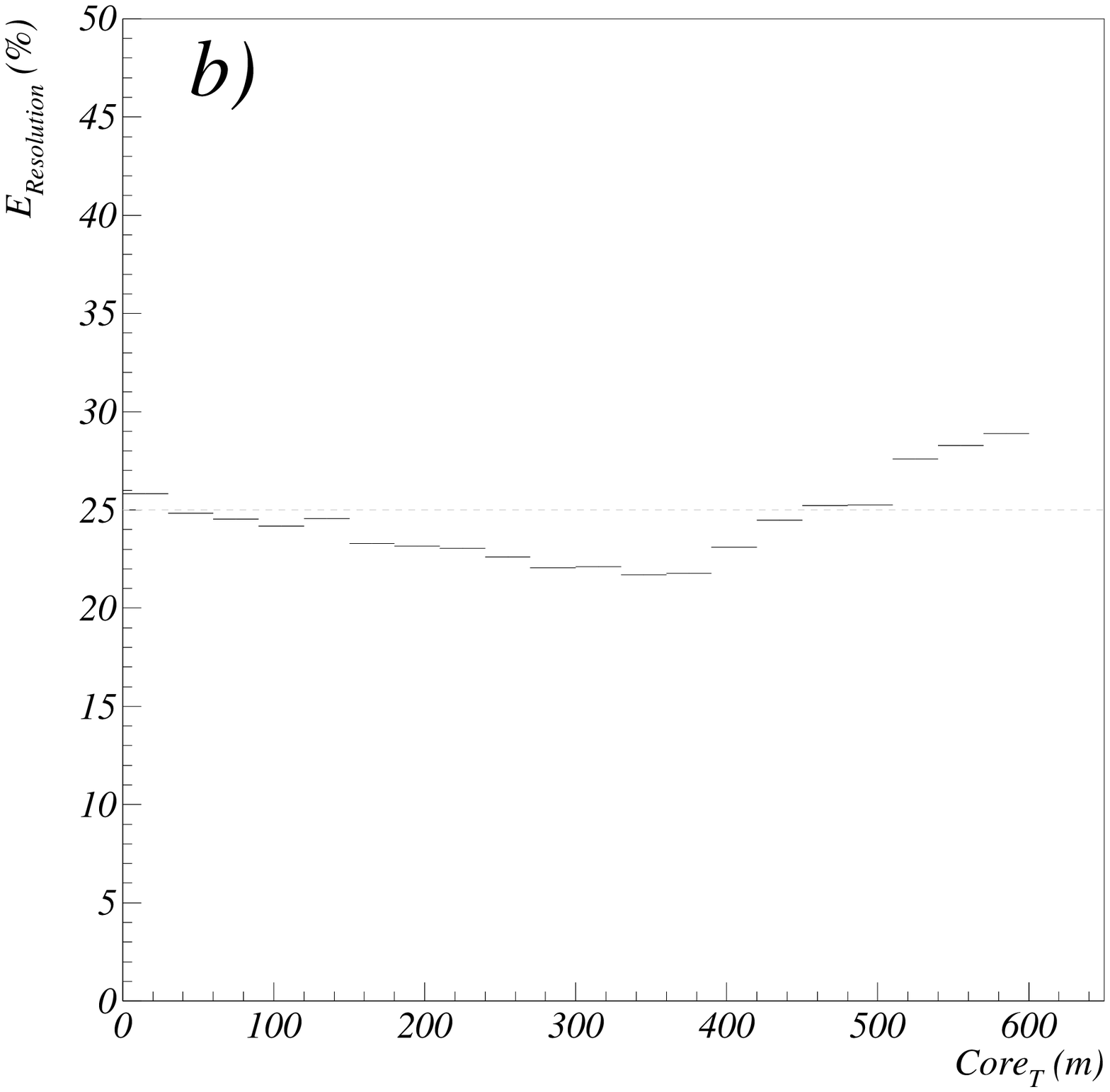,width=.33\textwidth}
    \epsfig{file=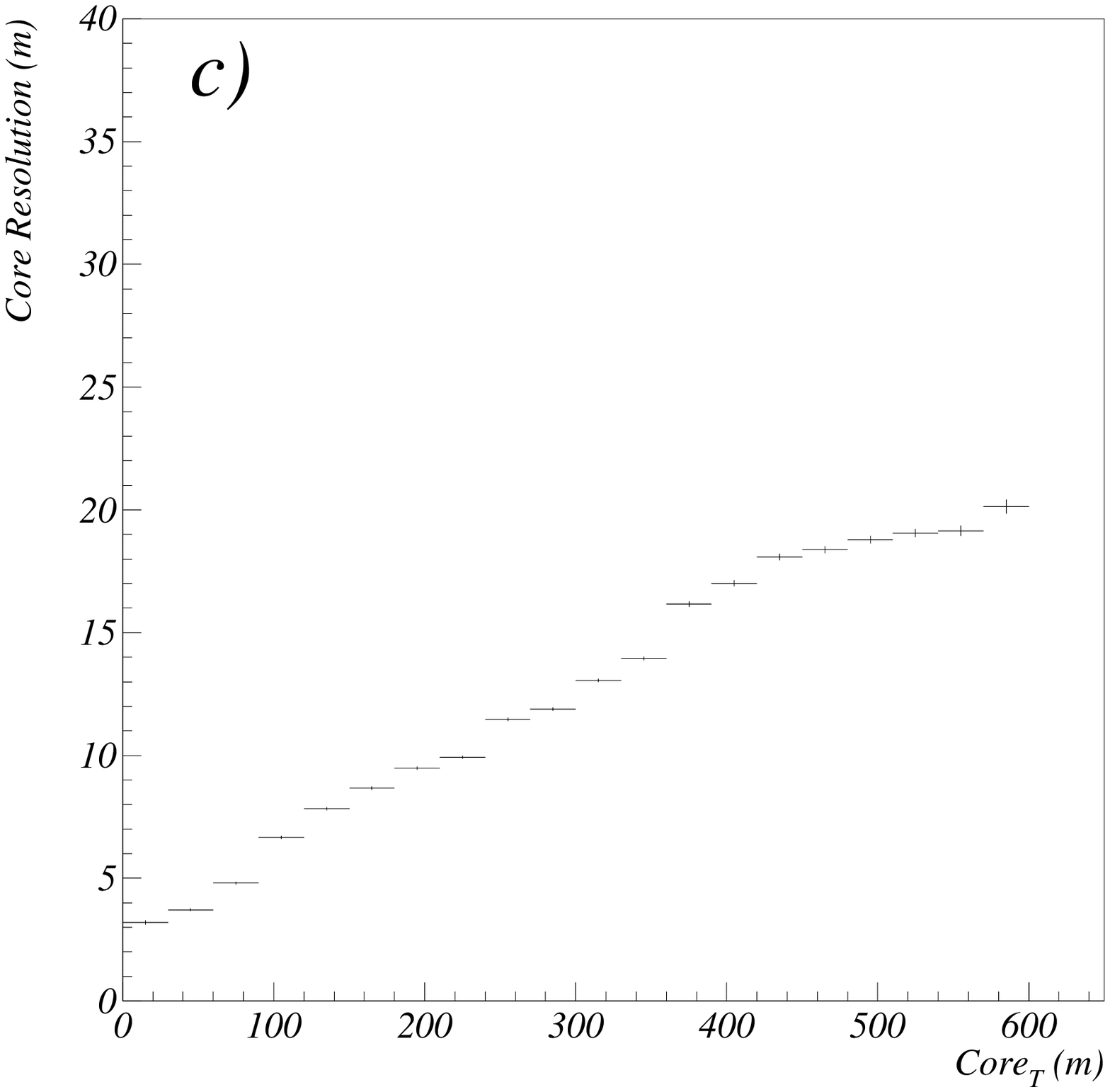,width=.33\textwidth}\\
    \caption{RMS energy and Core resolution as a function of true energy
      (E$_T$) and true core distance (Core$_T$) for the tandem telescope
      configuration.}\label{Ener_rec_2T}
  \end{center}
\end{figure}

 In applying this fitting procedure to simulated data, as in the case of a
single telescope, events with values of $\mathcal{D}$ greater than 4.5$^\circ$
or fit $\chi^2$ values less than 10\% likely were eliminated from the
data set in order to remove poorly defined images. Figure \ref{Ener_rec_2T}
shows the derived core and energy resolutions as a function of actual primary
gamma-ray energy and true core distance from the ACT. An energy resolution of
$\sim$20\% is achieved, as compared to the single telescope value of
$\sim$25\%. 

\subsection{Inter-Telescope Timing and Shower Maximum}

\par The relative arrival times of the {\Cerenkov} signal in different
telescopes can also provide valuable information. The thickness of the
{\Cerenkov} wavefront from an extensive air shower is expected to be $\sim$5-10
ns and the Davies-Cotton design considered here results in light paths to the
camera which differ by as much as 5 ns. However, a typical gamma-ray image
involves several hundred photoelectrons, such that the statistical sampling
should allow a sub-nanosecond measurement of the relative wavefront arrival
time.  For an array of telescopes spread over a baseline of $\sim$100-200 m,
this alone could provide a determination of the shower arrival direction with
an accuracy comparable to or better than that based on image orientation.
However, even for the case of two telescopes explored here, the relative
arrival times of the {\Cerenkov} signal can be combined with the reconstructed
shower core position to yield information on the height of shower maximum and
also provide another useful discriminant against background showers.

\par In particular, if it is assumed that the {\Cerenkov} light is emitted from
a height of maximum shower development, $h_{max}$, above the detector level,
for the simple case here in which the source is at zenith, the time at which
the telescope receives the light is then just given by:

\[t = \frac{n}{c}\sqrt{h_{max}^2 + R^2} \]

\noindent where n is the refractive index of air and $R$ is the distance from
that telescope to the shower core at the detection level. Thus, in the limit
where $h_{max} \gg R$ and $n=1.0$, the difference in time between two
different telescope measurements is given by:

\[\Delta t \simeq \frac{1}{2ch_{max}} (R_2^2 - R_1^2) \]

\par
\begin{figure}[hbt]
  \begin{center}
    \epsfig{file=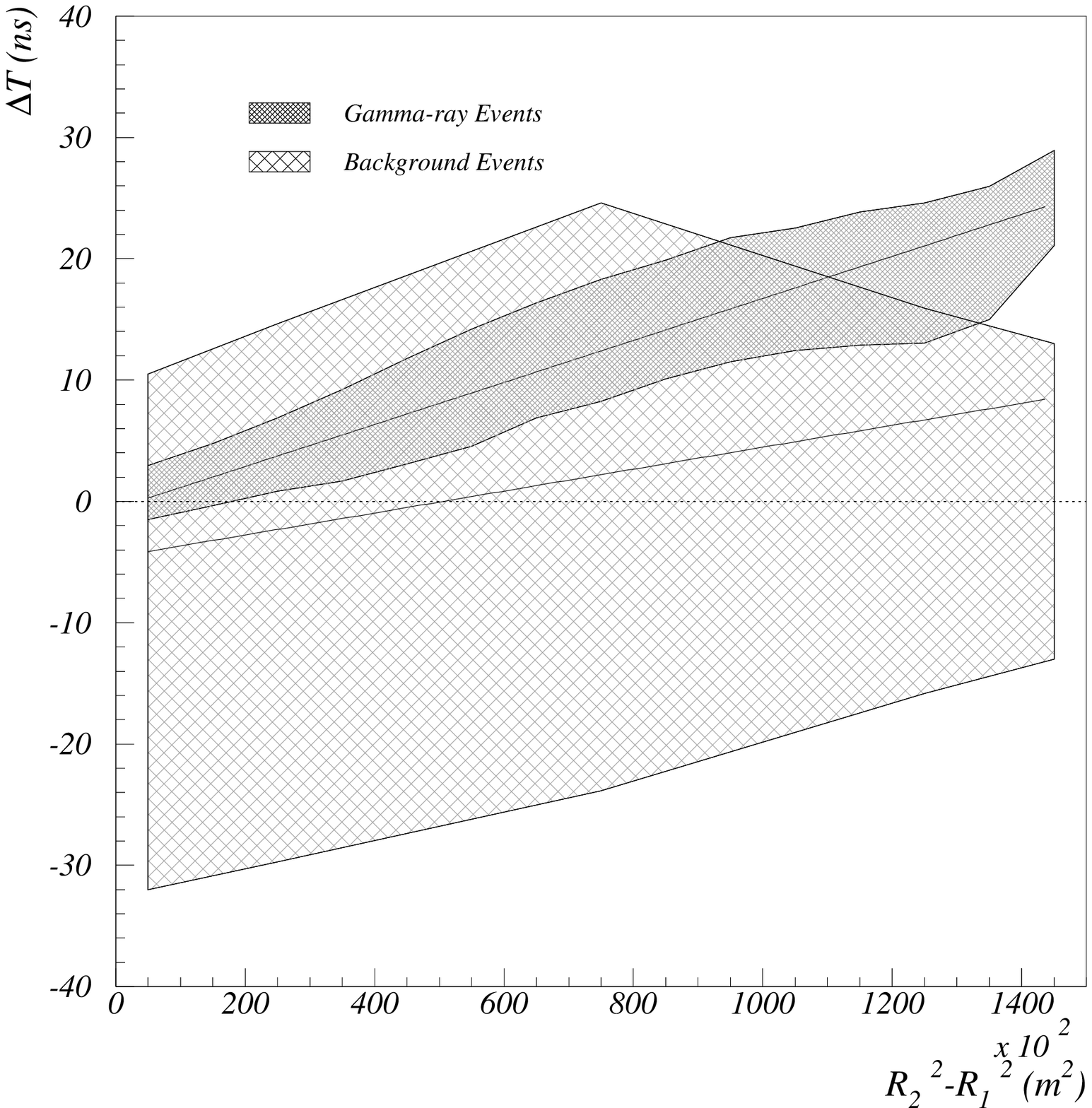,width=.9\textwidth}\\
    \caption{Relative shower arrival time as a function of relative squared
      distance to the shower core to the different telescopes. The dark-shaded
      band corresponds to the 95$\%$ gamma-ray events which have passed image
      selection. The light-shaded band is the equivalent for background
      events, where only events which have passed the trigger condition and
      have at least 5 picture pixels and $\mathcal{T}_S \ge 0.$ are
      represented.}\label{hmax}
  \end{center}
\end{figure}

 Thus, a plot of $\Delta t$ versus $(R_2^2 - R_1^2)$ should yield a
well-defined linear relation for gamma-rays, whose slope is proportional to
the height of shower maximum (figure \ref{hmax}). In practise, the lateral
extent of the gamma-ray shower at the point of maximum development means that
much of the light arrives from distances somewhat closer to the detector than
is implied by the shower core.  This causes the apparent $h_{max}$ derived from the
simplistic expression above to be overestimated. In order to apply this to a
simulated data set, the reconstructed core locations were used (as previously
described) and the relative ``image time'' in each telescope was defined as
that of the image centroid, using the linear fit parameters derived in the
determination of $\mathcal{T}_S$. Results are shown in figure \ref{hmax} for a
simulated gamma-ray source.

\par A selection criteria based on this to discriminate against background was
formed by defining a range of $\Delta t$ as a function of $(R_2^2 - R_1^2)$
which contains 95\% of the gamma-rays. This was found to reject $\sim$85\% of
proton events which would otherwise pass data selection.

\subsection{Factorised Background Rejection}

\par The very large background rejection associated with a multi-telescope
system can impose statistical limitations on the ability to extract detailed
information about the behaviour of the expected background from a finite set of
simulations. However, many of the image selection criteria are independent of
each other to a very large extent, thus allowing their individual effects to
be well determined and then combined as a product.  To explicitly verify this
approach for the current case, we first consider the background rejection of a
single camera for proton events which have triggered both telescopes and pass
the selections $\mathcal{D}<4.5^\circ$, $\mathcal{T}_S>0$ and {\it npicture}
$\ge$ 5. Table \ref{factorised_bkg_1} shows the fraction of such events which
pass different data selection criteria, both individually and in various
combinations. These numbers indicate that selections based on {\bf 1)} image
shape ($\mathcal{L}$ and $\mathcal{W}$), {\bf 2)} image orientation
{$\alpha$}, and {\bf 3)} $h_{max}$, based on relative image times (in this
case making use of information from the second telescope), are all largely
independent of each other.

\par We now consider these data selection criteria applied to the case of two
telescopes. While the effect of any given selection is generally correlated
between both cameras, the joint application of each selection will still be
independent from the joint application of another selection. Table
\ref{factorised_bkg_2} lists the two-telescope proton rejection factors for
each the three selections listed above.

\par Furthermore, the effect of a selection based on the $\chi^2$ goodness of
fit to the gamma-ray energy and core location can also be separately applied
as a factor since the parameters involved ($\mathcal{S}$, $\mathcal{D}$ and
$\mathcal{T}_S$) are independent of the three selections above: Image shape
and orientation selections are parameterised as a function of $\mathcal{S}$
and $\mathcal{D}$ to explicitly remove any dependence, while the relative
image times are, by construction, independent of individual telescope time
measurements. For the case of two telescopes, a $\chi^2$ selection chosen to
keep 90\% of gamma-ray images was found to reject $\sim$85\% of proton shower
images.

\par The combination of these factors for the case of two-telescope analysis
is shown in table \ref{factorised_bkg_2} for 2 different ranges of
reconstructed gamma-ray energy. As no clear dependence on this reconstructed
energy is seen, a single, average rejection factor will be applied to scale
the detection rates from the one telescope to the two telescope case.
  
\subsection{Tandem Telescope Sensitivity}

\par Following a similar procedure to that employed for a single telescope,
figures \ref{50h_sen} and \ref{1h_sen} show the sensitivity of a tandem,
wide-angle instrument to a hypothetical point-source with an $E^{-2.5}$
differential spectrum, both for 50 hours and 1 hour exposure times,
respectively. This is compared with similar curves for the VERITAS/HESS
instruments. The scaling relative to the single telescope case is consistent
with the factor expected from comparing the Whipple with the VERITAS/HESS
sensitivities \cite{Weekes2002} (figure \ref{50h_sen}), with an additional
factor of $\sim$2 gained from the $h_{max}$ selection.  For the 50 hour
exposure, the sensitivity of the tandem wide-angle design is a factor of $\sim$4
times more sensitive over the range above 300~GeV, with extended sensitivity
above 10~TeV. For the case of 1 hour exposure, the sensitivities are
comparable at 300~GeV and exceed an order of magnitude improvement in
sensitivity above 5 TeV.

\section{Summary and Conclusions}

\par The study presented here suggests that the sensitivity of ACT instruments
could be substantially improved in the regime above 300~GeV through the use of
cameras with much larger fields of view for the same PMT density than are
currently employed. Particularly for the tandem telescope design considered in
this paper, the ability to image showers further ``off-axis'' leads to a
significant gain in effective area with no significant loss in energy
resolution compared with current instruments, even for showers with cores
landing much further from the telescopes. The ability to suppress background
events also appears to be substantial for such a design and yields a very good
sensitivity over a remarkably large dynamic range.  This background rejection
has been enhanced by the introduction of additional imaging parameters based
on both intra- and inter- telescope timing characteristics. These parameters
may be of some use even for current instruments, though they are particularly
well suited for cameras with larger fields of view. 

To be explicit, the conclusions of this study rely on four basic premises:
{\bf 1)} That the simulations are accurate and that the analysis techniques 
applied are valid. In this regard we have verified the simulation by comparing 
with analytical calculations, alternative simulations and actual experimental
data. We have been able to reproduce trends seen in other, independent
studies and replicate parameter distributions and published sensitivity 
curves from existing experiments. {\bf 2)} That the repeated sampling of
showers employed does not lead to significant biases. To this end, we have 
explicitly tested this with regard to image selection (Table 2) and have
found no evidence of any bias to within the limits our statistical
uncertainties. Furthermore, throughout this analysis, we have specifically 
checked to insure that resulting distributions were not unduly influenced
by a handful of independent showers with unusually high ``weights.''
{\bf 3)} That the background rejection factor for the tandem WFOV design
can be factorised into groups of largely independent contributions
which can be separately assessed and combined as a product. However,
in addition for there being logical arguments for why this ought to be
the case for the parameters used, we have also explicitly verified the 
independence of these parameters to within the limits of our statistical 
uncertainties (Table 5). {\bf 4)} That the predicted rejection factors
due to newly introduced timing parameters are accurate, even though
this is yet to be experimentally tested. We find no reason to
doubt these factors given that the photon timing is largely governed by
basic air-shower development and geometry, though we certainly encourage
future experimental efforts to explicitly explore this. Consequently,
we believe the basic results of this study to be robust.

These results predict the
2-telescope design considered here to be more than $\sim$3 times
more sensitive than existing/planned arrays in the regime above 300~GeV for
continuously emitting sources; up to $\sim$10 times more sensitive for
hour-scale emission (relevant for episodic sources, such as AGN);
significantly more sensitive in the regime above 10~TeV; and possessing a sky
coverage which is roughly an order of magnitude larger than existing
instruments. This is despite the fact that the overall mirror area for the
instrument modelled here is nearly three times smaller than that of either the
current HESS or VERITAS-4 designs.  This study also suggests that the approach
could be extended even further by using larger fields of view, mirror areas
more comparable to the current devices and a more optimised array
configuration to both increase the effective area further and potentially take
better advantage of inter-telescope timing so as to improve the angular
resolution, even at lower energies. These latter issues will be explored in a
future paper.

This research has been supported by the Particle Physics and Astronomy 
Research Council (PPARC).

\begin{table}[hbt]
  \begin{center}
    \begin{tabular}{lccccc}
      \hline
      Primary    
      & Number 
      & Energy 
      & Core Distance
      & Integral 
      & Zenith \\
      & of Showers
      & Range 
      & Range
      & Spectral Index
      & Angle\\
      & 
      & (TeV)
      & (m)
      & 
      & (deg)\\
      \hline
      Gamma  &  {\bf 4.8  $\cdot$ 10$^6$} & {\bf 0.3-20} & {\bf 0.-800.} & {\bf 1.5}  & {\bf 0.}\\
      Proton &  9.56 $\cdot$ 10$^6$ & 0.2-5 & 0.-800. & 1.7  & 0.
      - 15.\\
             &  1.08 $\cdot$ 10$^6$ & 5-10 & 0.-800. & 1.7  & 0.
      - 15.\\
             &  0.70 $\cdot$ 10$^6$ & 10-15 & 0.-800. & 1.7  & 0.
      - 15.\\
             &  0.65 $\cdot$ 10$^6$ & 15-20 & 0.-800. & 1.7  & 0.
      - 15.\\
             &  {\bf 12.0 (9.59) $\cdot$ 10$^6$} & {\bf 0.2-20} & {\bf
	    0.-800.} & {\bf 1.7}  & {\bf 0. - 15.} \\
      \hline
    \end{tabular}
    \caption{Summary of principle shower simulations used in this study for
      a single telescope (WFOV-1). The numbers in parenthesis are the effective
      number of events used after normalising according to the energy spectrum.}
    \label{tab:single}
  \end{center}
\end{table}

\vskip 1.cm

\begin{table}[hbt]
\begin{center}
\begin{tabular}{ccccc}
\hline
Selection & \multicolumn{2}{c|}{$\kappa_{\gamma}$}     &  \multicolumn{2}{c}{$\kappa_{bkg}$}    \\
Parameter &  100~Throws & 1~Throw & 100~Throws & 1~Throw \\
\hline
\hline
$\mathcal{L}$     & 0.852$\pm$0.002 & 0.84$\pm$0.02 & 0.174$\pm$0.003 & 0.18$\pm$0.03  \\
$\mathcal{W}$      & 0.852$\pm$0.002 & 0.85$\pm$0.02 & 0.222$\pm$0.003 & 0.28$\pm$0.04  \\
$\alpha$      & 0.875$\pm$0.002 & 0.87$\pm$0.02 & 0.037$\pm$0.001 & 0.03$\pm$0.01  \\
$\mathcal{T}_S$     & 0.842$\pm$0.002 & 0.84$\pm$0.02 & 0.127$\pm$0.002 & 0.15$\pm$0.03  \\
($\mathcal{T}_S \ge$0   & 0.879$\pm$0.002 & 0.87$\pm$0.02 & 0.367$\pm$0.005 & 0.37$\pm$0.04)  \\
\hline
{\bf All Cuts}   & {\bf 0.750$\pm$0.002} & {\bf 0.74$\pm$0.02} & {\bf
  0.0017$\pm$0.0003}  & {\bf 0.0041$\pm$0.0041} \\

All but $\mathcal{T}_S$ & {\bf 0.795$\pm$0.002} & {\bf 0.78$\pm$0.02} & {\bf
  0.0051$\pm$0.0005}  & {\bf 0.0044$\pm$0.0041} \\
\hline
\end{tabular}
\caption{Effects of image selection as defined in the text, comparing the case
  where each shower is sampled once with that in which each is sampled 100
  times. $\kappa_{\gamma,bkg}$ are the acceptance values, with respect
  triggered events, for gamma and background events, respectively. Note that
  only events with 5 or more picture pixels are parameterised (which, by
  itself, removes about 10$\%$ of triggered gamma events and about 40$\%$ of
  triggered proton events). \label{cuts}}
\end{center}
\end{table}

\vskip 1.cm

\begin{table}[hbt]
  \begin{center}
    \begin{tabular}{lccccc}
      \hline
      Primary      
      & Number 
      & Energy 
      & Core Distance
      & Integral 
      & Zenith \\
      & of Showers
      & Range 
      & Range
      & Spectral Index
      & Angle\\
      & 
      & (TeV)
      & (m)
      & 
      & (deg)\\
      \hline
      Gamma  &  0.88 $\cdot$ 10$^6$ & 0.3-1 & 0.-400. & 1.5  &
      0.\\
             &  3.03 $\cdot$ 10$^6$ & 1-5 & 0.-500. & 1.5  &
      0.\\
             &  0.75 $\cdot$ 10$^6$ & 5-10 & 0.-600. & 1.5  &
      0.\\
             &  0.74 $\cdot$ 10$^6$ & 10-15 & 0.-600. & 1.5  &
      0.\\
             &  0.66 $\cdot$ 10$^6$ & 15-20 & 0.-700. & 1.5  &
      0.\\
             &  {\bf 6.06 (1.05) $\cdot$ 10$^6$} & {\bf 0.3-20} & {\bf 0.-700.} & {\bf 1.5}  & {\bf 0.}\\

      Proton & 0.15 $\cdot$ 10$^6$ & 0.2-5 & 0.-800. & 1.7  & 0.
      - 15.\\
             &  0.26 $\cdot$ 10$^6$ & 5-10 & 0.-800. & 1.7  & 0.
      - 15.\\
             &  0.18 $\cdot$ 10$^6$ & 10-15 & 0.-800. & 1.7  & 0.
      - 15.\\
             &  0.13 $\cdot$ 10$^6$ & 15-20 & 0.-800. & 1.7  & 0.
      - 15.\\
             &  {\bf 0.72 (0.15) $\cdot$ 10$^6$} & {\bf 0.2-20} & {\bf 0.-800.} & 1.7  & {\bf 0.- 15.}\\
      \hline
    \end{tabular}
    \caption{Summary of principle shower simulations used in this study for
      the tandem configuration (WFOV-2). The numbers in parenthesis are the
      effective number of events used after normalising according to the
      energy spectrum.}
    \label{tab:tandem}
  \end{center}
\end{table}

\vskip 1.cm

\begin{table}[hbt]
  \begin{center}
    \begin{tabular}{l||c}
      \multicolumn{2}{l}{{\bf Background rejection factors for a single
	wide-angle telescope.}}\\
      \hline
      {\bf Energy (TeV)}                       & 0.3-20 \\
      \hline
      {\bf Image Cut} & {\bf Values Relative to Trigger + $\mathcal{D}$} \\
      \hline
      $\mathcal{L,W}$ 		         & (7.1$\pm$0.4)$\cdot$10-2 \\
      $\alpha$ 			         & (1.01$\pm$0.05)$\cdot$10-1 \\
      \hline      
      \hline
      $\mathcal{L,W,\alpha}$	         & (1.1$\pm$0.1)$\cdot$10-2 \\
      \hline
      \multicolumn{1}{c||}{\bf Product of}    &  \\
      {\bf ($\mathcal{L,W,\alpha}$)} & {\bf (5.3$\pm$0.3)$\cdot$10-3} \\
      \multicolumn{1}{c||}{\bf and $\chi^2$}  &  \\
      \hline
    \end{tabular}
    \caption{Background rejection factors for a single wide-angle telescope
      (WFOV-1) and different combinations of image selection criteria. The cut in
      distance, $\mathcal{D}$, includes the conditions $\mathcal{T}_S>0$ and
      $\ge$5 picture pixels in the image. Events are only considered as having
      passed image selection if the $E^{\gamma}_{Rec}$ is within the considered
      energy range. The $\chi^2$ cut alone removes
      $\sim$25\% of the background events (see text for
      details).}\label{factorised_bkg_0}
  \end{center}
\end{table}

\vskip 1.cm

\begin{table}[hbt]
  \begin{center}
    \begin{tabular}{l||c}
      \multicolumn{2}{l}{{\bf Background rejection factors for a single}}\\
      \multicolumn{2}{l}{{\bf wide-angle telescope given a two telescope trigger.}}\\
      \hline
      {\bf Energy (TeV)}                       & 0.3-20 \\
      \hline
      {\bf Image Cut} & {\bf Values Relative to Trigger + $\mathcal{D}$} \\
      \hline
      $\mathcal{L,W}$ 		         & (1.4$\pm$0.3)$\cdot$10-1 \\
      $\alpha$ 			         & (1.4$\pm$0.3)$\cdot$10-1 \\
      h$_{max}$			         & (2.1$\pm$0.4)$\cdot$10-1 \\
      \hline      
      $\mathcal{L,W,\alpha}$	         & (4.9$\pm$1.7)$\cdot$10-2 \\
      $\mathcal{L,W}$, h$_{max}$	      & (4.4$\pm$1.6)$\cdot$10-2 \\
      $\alpha$, h$_{max}$		      & (4.3$\pm$1.5)$\cdot$10-2 \\
      \hline 
      \hline 
      {\bf $\mathcal{L,W,\alpha}$, h$_{max}$} & {\bf (2.4$\pm$1.2)$\cdot$10-2} \\
      \hline
      \multicolumn{1}{c||}{\bf Product of}    &  \\
      ({\bf $\mathcal{L,W,\alpha}$, h$_{max}$}) & {\bf (4.1$\pm$1.5)$\cdot$10-3} \\
      \multicolumn{1}{c||}{\bf and $\chi^2$}  &  \\
      \hline
    \end{tabular}
    \caption{Background rejection factors for a single wide-angle telescope
      and different combinations of image selection criteria. Note that a second
      telescope has to be used to evaluate the trigger and the {\it h$_{max}$} cut,
      but that only one telescope is used to evaluate the rest of the
      cuts. The cut in distance, $\mathcal{D}$, includes the conditions
      $\mathcal{T}_S>0$ and $\ge$5 picture pixels in the image. Events are
      only considered as having passed image selection if the $E^{\gamma}_{Rec}$ is
      within the considered energy range.}\label{factorised_bkg_1}
  \end{center}
\end{table}

\vskip 1.cm

\begin{table}[hbt]
  \begin{center}
    \begin{tabular}{l||cc|c||c}
      \multicolumn{5}{l}{{\bf Background rejection factors for the tandem telescope configuration.}}\\
      \multicolumn{4}{l|}{} & \multicolumn{1}{|c}{{\bf T2/T1}}\\
      \hline
      {\bf Energy (TeV)}                       & 0.3-3 & 3-20 & 0.3-20 & 0.3-20\\
      \hline
      {\bf Image Cut} & \multicolumn{4}{c}{{\bf Values Relative to Trigger + $\mathcal{D}$}} \\
      \hline
      $\mathcal{L,W}$ & (1.7$\pm$0.1)$\cdot$10-2 		         
      & (3.4$\pm$0.2)$\cdot$10-2 & (2.5$\pm$0.1)$\cdot$10-2 
      & (3.5$\pm$0.2)$\cdot$10-1 \\
      $\alpha$ & (8.2$\pm$1.1)$\cdot$10-3 
      & (5.9$\pm$1.0)$\cdot$10-3 & (7.1$\pm$0.7)$\cdot$10-3
      & (7.0$\pm$0.8)$\cdot$10-2 \\
      h$_{max}$	& (1.32$\pm$0.05)$\cdot$10-1 
      & (1.52$\pm$0.05)$\cdot$10-1 & (1.41$\pm$0.03)$\cdot$10-1 
      & (1.41$\pm$0.03)$\cdot$10-1 \\
      \hline     
      \hline
      \multicolumn{1}{c||}{{\bf Product of}}   & & & & \\
      ({\bf $\mathcal{L,W,\alpha}$, h$_{max}$}) & (2.8$\pm$0.4)$\cdot$10-6 
      & (4.6$\pm$0.8)$\cdot$10-6 & (3.7$\pm$0.4)$\cdot$10-6
      & {\bf (6.9$\pm$0.4)$\cdot$10-4} \\
      \multicolumn{1}{c||}{{\bf and $\chi^2$}} & & & & \\
      \hline
      \multicolumn{1}{c||}{{\bf Relative}}   & & & & \\
      Trigger + $\mathcal{D}$ & & & & {\bf (2.9$\pm$0.2)$\cdot$10-1} \\
      \hline
    \end{tabular} 
    \caption{Background rejection factors for the tandem telescope
      configuration and different combinations of image selection criteria. The
      cut in distance, $\mathcal{D}$, includes the conditions
      $\mathcal{T}_S>0$ and $\ge$5 picture pixels in the image. Events are
      only considered as having passed image selection if the $E^{\gamma}_{Rec}$ is
      within the considered energy range shown in column 1. The $\chi^2$ cut
      alone removes $\sim$85\% of the background events (see text for
      details). The relative rejection factor of WFOV-2 to WFOV-1 (table
      \ref{factorised_bkg_0}) is shown in the last column of the table
      (T2/T1).}\label{factorised_bkg_2}
  \end{center}
\end{table}

\clearpage

\clearpage

\begin{thebibliography}{999}

\bibitem{Weekes1989} Weekes, T. C., {\em et al.} 1989; ApJ {\bf 342}, 379.

\bibitem{Aharonian2001} Aharonian, F. A., {\em et al.} 2001; A$\&$A, {\bf 370},
  112.
\bibitem{Aharonian2005c} Aharonian, F. A., {\em et al.} 2005; A$\&$A {\it in press}

\bibitem{Horan2004} Horan, D., and Weekes, T. C., 2004; NewAR, {\bf 48}, 527.  

\bibitem{Aharonian2002} Aharonian, F. A., {\em et al.} 2002; A$\&$A, {\bf 393}, 37L. 

\bibitem{Aharonian2005a} Aharonian, F. A., {\em et al.} 2005; A$\&$A {\bf 442}, 1.

\bibitem{Aharonian2004a} Aharonian, F. A., {\em et al.} 2004; A$\&$A, {\bf 425},
  13L.

\bibitem{Aharonian2005b} Aharonian, F. A., {\em et al.} 2005; Science, {\bf 307},
  1839.

\bibitem{Buckley1998} Bergstr\"{o}m, L., Ullio, P. and Buckley, J. H. 1998; 
Astroparticle Physics, {\bf 9}, 137.

\bibitem{Biller1998} Biller, S. D., {\em et al.} 1998; Physical Review Letters, {\bf
80}, 14.

\bibitem{Biller1999b} Biller, S. D., {\em et al.} 1999; Physical Review Letters, {\bf
83}, 2108.

\bibitem{Aharonian2004b} Aharonian, F. A., {\em et al.} 2004; ApJ, {\bf 614},
  897.

\bibitem{ParisConference2005} Conference Proceedings of {\it Towards a Major
  Atmospheric Cherenkov Detector VII}, April 2005, Palaiseau, France ({\it in
  preparation}). Ed. Bernard Degrange ({\it
  http://polywww.in2p3.fr/actualites/cogres/cherenkov2005/}).

\bibitem{Maccarone} Maccarone, M.C., {\em et al.} 2005; Proc. 29th ICRC, 
International Cosmic Rays Conference, Pune, India, {\bf 5} 295
  
\bibitem{Biller2000} Biller, S. D., 2000; {\it Oxford internal memo}, August
  2000.

\bibitem{Rowell} Rowell, G., Aharonian, F. and Plyasheshnikov, A.  2005;
astro-ph/0512523.
   
  
\bibitem{Hillas1996} Hillas, A. M., 1996; Space Science Reviews {\bf 75}, 17. 

\bibitem{Biller1997} Biller, S. D., 1997; {\it Oxford internal memo}.

\bibitem{SLAC} Nelson, W. R., Hirayama, H. and Rogers, D. W. O., 1985; Report
  {\bf SLAC 265}, Standford Linear Accelerator Center. {\it unpublished}

\bibitem{Wrotniak} Wrotniak, A., 1985; University of Maryland Report
  No. PP.85-191. {\it unpublished}

\bibitem{Calle2002} de la Calle Per\'ez, I., {\em et al.} 2002; {\it PhD Thesis}. 

\bibitem{Finley2000} Finley, J. P., {\em et al.} 2000; AIP Conference Proceedings {\bf
  515}, 301.  

\bibitem{Bernlohr2003}  Bernlohr, K., {\em et al.} 2003; Astroparticle Physics
  {\bf 20}, 111.

\bibitem{Weekes2002} Weekes, T. C., {\em et al.} 2002; Astroparticle Physics {\bf
  17}, 221. 

\bibitem{Petry2001} Petry, D., {\em et al.} 2001; 27th ICRC, Hamburg, {\bf 7}, 2848.

\bibitem{Hillas1998} Hillas, A. M., {\em et al.} 1998; ApJ, {\bf 503}, 744.

\bibitem{Aharonian2005d} Aharonian, F. A., {\em et al.} 2005; A$\&$A {\bf 432}, L25.

\bibitem{Horandel2003} H\"{o}randel, J. R., 2003; Astroparticle Physics, {\bf
  19}, 193.

\bibitem{Reynolds1993} Reynolds, P. T., {\em et al.} 1993; ApJ, {\bf 404}, 206.

\bibitem{Hillas1985} Hillas, A. M., 1985; 19th ICRC, La Jolla, {\bf 3}, 445.

\bibitem{Aharonian1995} Aharonian, F. A., {\em et al.} 1995; J. Phys. G, {\bf 21}
  985. 

\bibitem{Bradbury1999} Bradbury, S. M., {\em et al.} 1999; 26th ICRC, Salt Lake City,
     {\bf 5}, 263. 

\bibitem{Funk2004} Funk, S., {\em et al.} 2004; Astroparticle Physics {\bf 22}, 285. 

\bibitem{Wilks1938} Wilks, S. S., 1938; Ann. Math. Stat, {\bf 9}, 60.

\bibitem{Hofmann2001} Hofmann, W., {\em et al.} 2001; 27th ICRC, Hamburg, {\bf 7},
  2785. 

\bibitem{Aharonian1997} Aharonian, F. A., {\em et al.} 1997; Astroparticle Physics
  {\bf 6}, 343. 

\bibitem{Maier2005} Maier, G., {\em et al.} 2005; 29th ICRC, Pune, {\bf 0}, 101. 

\bibitem{Mohanty1998} Mohanty, G., {\em et al.} 1998; Astroparticle Physics {\bf 9},
  15. 

\bibitem{Benbow2005} Benbow, W. {\em et al.}, 2005; 2nd High Energy Gamma-ray
  Symposium, AIP Conference Proceedings, {\bf 745}, 611.  

\end{thebibliography}
\end{document}